\documentclass[11pt,a4paper]{article}
\usepackage{jcappub}
\usepackage{tikz-feynhand}
\usepackage{graphicx}
\usepackage{xcolor}
\usepackage{caption,subcaption}
\usepackage{mathrsfs,mathtools}
\usepackage{physics,amssymb}
\usepackage{bm}
\usepackage{braket}
\usepackage{listings}
\usepackage{cases}
\usepackage{comment}
\usepackage{soul}
\usepackage{cancel}
\usepackage{cases}
\usepackage[utf8]{inputenc}
\usepackage{url}
\usepackage[normalem]{ulem}
\usepackage{xspace}
\usepackage{aas_macros}
\usepackage{acronym}
\usepackage{siunitx}

\hypersetup{colorlinks=true
,urlcolor=DARKBLUE
,anchorcolor=DARKBLUE
,citecolor=DARKBLUE
,filecolor=DARKBLUE
,linkcolor=DARKBLUE
,menucolor=DARKBLUE
,linktocpage=true
,pdfproducer=medialab
,pdfa=true
}

\definecolor{MONZA}{HTML}{CF000F}
\definecolor{DARKBLUE}{HTML}{00008b}
\definecolor{DARKMAGENTA}{HTML}{8b008b}

\newcommand{\HH}{\mathcal{H}}
\newcommand{\PP}{\mathcal{P}}

\newcommand{\ee}{\mathrm{e}}
\newcommand{\Mpl}{M_\mathrm{Pl}}

\newcommand{\pk}{\mathrm{pk}}
\newcommand{\uth}{\mathrm{th}}
\newcommand{\PBH}{\mathrm{PBH}}
\newcommand{\DM}{\mathrm{DM}}
\newcommand{\tot}{\mathrm{tot}}
\newcommand{\GW}{\mathrm{GW}}
\newcommand{\RD}{\mathrm{RD}}

\newcommand{\dk}{\frac{\dd^3k}{(2\pi)^3}}
\newcommand{\iq}[1]{\int\frac{\dd^3q_{#1}}{(2\pi)^3}}

\newcommand{\hk}[2]{h^{#2}_{\lambda^{#1}}(\tau,\bfk^{#1})}
\newcommand{\Q}[2]{Q_{\lambda_{#1}}(\bfk_{#1},\bfq_{#2})}

\newcommand{\NL}{\mathrm{NL}}
\newcommand{\probP}{\bbP}
\newcommand{\umax}{\mathrm{max}}

\newcommand{\CZ}{\mathrm{CZ}}

\acrodef{CMB}{cosmic microwave background}
\acrodef{PBH}{primordial black hole}
\acrodef{PDF}{probability density function}
\acrodef{EoM}{equation of motion}
\acrodef{GW}{gravitational wave}
\acrodef{RD}{radiation-dominated}
\acrodef{DM}{dark matter}
\acrodef{PTA}{pulsar timing array}

\newcommand{\calC}{\mathcal{C}}

\newcommand{\calF}{\mathcal{F}}
\newcommand{\scrF}{\mathscr{F}}
\newcommand{\uf}{\mathrm{f}}

\newcommand{\calI}{\mathcal{I}}
\newcommand{\ui}{\mathrm{i}}

\newcommand{\bfk}{\mathbf{k}}

\newcommand{\um}{\mathrm{m}}

\newcommand{\calO}{\mathcal{O}}

\newcommand{\calP}{\mathcal{P}}
\newcommand{\bbP}{\mathbb{P}}
\newcommand{\bfp}{\mathbf{p}}
\newcommand{\bfq}{\mathbf{q}}

\newcommand{\ur}{\mathrm{r}}
\newcommand{\calS}{\mathcal{S}}
\newcommand{\bfx}{\mathbf{x}}

\newcommand{\bae}[1]{\begin{align} #1 \end{align}}
\newcommand{\beae}[1]{\begin{equation}\begin{aligned} #1 \end{aligned}\end{equation}}
\newcommand{\bege}[1]{\begin{equation}\begin{gathered} #1 \end{gathered}\end{equation}}
\newcommand{\bme}[1]{\begin{multline} #1 \end{multline}}
\newcommand{\bmte}[1]{\begin{multlined}[t] #1 \end{multlined}}
\newcommand{\bmbe}[1]{\begin{multlined}[b] #1 \end{multlined}}
\newcommand{\bfe}[4]{
\begin{figure} 
	\centering
	\includegraphics[#1]{#2}
	\caption{#3}
	\label{#4}
\end{figure}}

\newcommand{\lr}[1]{\left( #1 \right)}
\newcommand{\bce}[1]{\begin{cases} #1 \end{cases}}

\begin{document}
\title{Primordial black holes and gravitational waves induced by exponential-tailed perturbations}

\date{\today}

\author[a]{Katsuya T. Abe,}
\emailAdd{abe.katsuya.f3@s.mail.nagoya-u.ac.jp}

\author[a]{Ryoto Inui,}
\emailAdd{inui.ryoto.a3@s.mail.nagoya-u.ac.jp}
\affiliation[a]{Department of Physics, Nagoya University, \\ 
Furo-cho Chikusa-ku, Nagoya 464-8602, Japan}

\author[a,b,c]{Yuichiro Tada,}
\emailAdd{tada.yuichiro.y8@f.mail.nagoya-u.ac.jp}
\affiliation[b]{Institute for Advanced Research, Nagoya University, \\
Furo-cho Chikusa-ku, Nagoya 464-8601, Japan}
\affiliation[c]{Theory Center, IPNS, KEK, \\
1-1 Oho, Tsukuba, Ibaraki 305-0801, Japan}

\author[d,e]{and Shuichiro Yokoyama} \emailAdd{shu@kmi.nagoya-u.ac.jp}
\affiliation[d]{%
  Kobayashi Maskawa Institute, Nagoya University, \\
  Furo-cho Chikusa-ku, Nagoya
  464-8602, Japan}%
\affiliation[e]{%
  Kavli IPMU (WPI), UTIAS, The University of Tokyo, \\
  5-1-5 Kashiwanoha, Kashiwa, Chiba 277-8583, Japan}%

\abstract{
\Acp{PBH} whose masses  
are in $\sim[10^{-15}M_\odot,10^{-11}M_{\odot}]$
have been extensively studied as a candidate of whole \ac{DM}.
One of the probes to test such a \ac{PBH}-\ac{DM} scenario
is scalar-induced stochastic \acp{GW} accompanied with the enhanced primordial fluctuations to form the \acp{PBH} with frequency
peaked in the mHz band being targeted by the LISA mission.
In order to utilize the stochastic \acp{GW} for checking the \ac{PBH}-\ac{DM} scenario, it needs to exactly relate the \ac{PBH} abundance and the amplitude of the \acp{GW} spectrum.
Recently in Kitajima et al.~\cite{Kitajima:2021fpq},
the impact of the non-Gaussianity of the enhanced primordial curvature perturbations on the \ac{PBH} abundance has been investigated based on the peak theory,
and they found that a specific non-Gaussian feature called the exponential tail significantly increases the \ac{PBH} abundance compared with the Gaussian case.
In this work, we investigate the spectrum of the induced stochastic \acp{GW} associated with \ac{PBH} \ac{DM} in the exponential-tail case.
In order to take into account the non-Gaussianity properly, we employ the diagrammatic approach for the calculation of the spectrum. 
We find that the amplitude of the stochastic \ac{GW} spectrum is slightly lower than the one for the Gaussian case, but it can still be detectable with the LISA sensitivity.
We also find that the non-Gaussian contribution can appear on the high-frequency side through their complicated momentum configurations.
Although this feature emerges under the LISA sensitivity, it might be possible to obtain information about the non-Gaussianity from \ac{GW} observation with a deeper sensitivity such as the DECIGO mission.
}

\maketitle\acresetall

\section{Introduction}

Recently, the \ac{PBH}, which could be formed in the early Universe, has been attracting much attention.
While several formation scenarios have been proposed,
one of the most extensively discussed is the formation by the gravitational collapse of over-density regions in the radiation-dominated universe after inflation~\cite{Carr:1974nx, Carr:1975qj}.
One of the interesting characteristics is that the \acp{PBH} could be formed with a wide mass range,
and \acp{PBH} heavier than $\sim 10^{15}\,\si{g}$ can exist in the present universe as \ac{DM}.
In fact, various astronomical observations have placed limits on the abundance of \acp{PBH} at various masses (see, e.g., Ref.~\cite{Carr:2020gox}). As a result, we have a small allowed parameter region for the \ac{PBH} mass called \emph{\ac{PBH} mass window}; the possibility that \ac{PBH} can be  whole \ac{DM} exists only in the case that the mass of \acp{PBH} is in $\sim[10^{-15}M_\odot,10^{-11}M_{\odot}]$.

One of the promising indirect observables to test such a remaining possibility for \ac{PBH} to be whole \ac{DM} would be scalar-induced stochastic \acp{GW}. Since large primordial scalar perturbations are necessary for the formation of \acp{PBH}, these perturbations would involve the potential to produce the large-amplitude \acp{GW} through the non-linear interactions between the scalar and tensor perturbations, which are called scalar-induced \acp{GW}. The \ac{GW} is becoming a powerful probe for cosmology along with the ongoing/future projects of ground- and space-based \ac{GW} detectors such as
LISA~\cite{LISA:2017pwj}, 
Taiji/Tianqin~\cite{TianQin:2015yph,Ruan:2018tsw}, 
DECIGO~\cite{Kawamura:2020pcg}, 
AION/MAGIS~\cite{Badurina:2019hst,MAGIS-100:2021etm}, 
LIGO/VIRGO/KAGRA~\cite{KAGRA:2013rdx},
ET/CE~\cite{Punturo:2010zz,Reitze:2019iox},
and \acp{PTA}~(see, e.g., Ref.~\cite{Verbiest:2021kmt}).
Especially, in LISA and DECIGO, the frequency ranges are corresponding to \ac{PBH} mass window scales (see, e.g., Refs.~\cite{Saito:2009jt, Bartolo:2018evs}). In Ref.~\cite{Bartolo:2018evs}, the authors suggested the detectability of \acp{GW} in LISA in the \ac{PBH} \ac{DM} model with an assumption of the Gaussian distribution for the primordial scalar perturbations. However, to utilize \acp{GW} as a probe of the \ac{PBH} \ac{DM} model, the statistical nature of the primordial perturbations such as non-Gaussianity should be taken into account properly because it highly affects the connection between the \ac{PBH} abundance and the amplitude of the scalar-induced \acp{GW} (see, e.g., Ref.~\cite{Nakama:2016gzw}).

In the standard slow-roll inflationary scenario,
the primordial curvature perturbations $\zeta$ follow the almost Gaussian distribution. On the other hand,
recently, the curvature perturbations having the
heavier tail in their distribution function
than that of the Gaussian have come to be discussed,
as the \acp{PBH} have become more actively discussed (see, e.g., Ref.~\cite{Vennin:2020kng} and references therein).
As a typical example,
the primordial curvature perturbations 
generated in the ultra slow-roll phase during inflation
could have amplitudes large enough for \ac{PBH} formations
and also are expected to have \emph{exponential tail} distribution
as $\probP(\zeta) \propto \ee^{-3 \zeta}$ in the large $\zeta$ limit~\cite{Cai:2018dkf,Atal:2019cdz,Atal:2019erb,Biagetti:2021eep}.
In Ref.~\cite{Kitajima:2021fpq},
the impact of such an exponential tail distribution on the \ac{PBH} abundance has been carefully studied based on the peak theory,
and it is found that the exponential tail significantly enhances the \ac{PBH} abundance compared with the Gaussian case. 
When the \acp{PBH} accounts for all of \ac{DM}, this fact leads to a reduction in the required amplitude of the primordial fluctuations, and then it is expected that the induced stochastic \acp{GW} associated with the \acp{PBH} should be smaller than those in the Gaussian case. Therefore, it needs to investigate whether the induced \acp{GW} in the exponential tail case can be still detected or not in the foreseeable observations.

In this work, we evaluate the stochastic \acp{GW} induced by
the primordial curvature perturbations with the exponential-tail-type non-Gaussianity. 
In addition to employing the result from Ref.~\cite{Kitajima:2021fpq}, we carefully investigate the possible spectral shape of the induced stochastic \acp{GW} by taking the non-Gaussianity of the curvature perturbations into account.
To do so, we use the diagrammatic approach
that can incorporate such a non-Gaussianity in a perturbative
and systematic manner. There are several works that
discuss the stochastic \acp{GW} induced by the primordial curvature perturbations with the perturbative non-Gaussianities
characterized by the so-called non-linearity parameters, $f_{\rm NL}$ and $g_{\rm NL}$~\cite{Cai:2018dig, Unal:2018yaa, Adshead:2021hnm,Yuan:2020iwf,Atal:2021jyo,Garcia-Saenz:2022tzu} (and Ref.~\cite{Domenech:2021ztg}
as a recent review).
By making use of this diagrammatic approach, we show that all contributions generally can be summarized into nine topologically-independent diagrams and there are still new contributions at the fourth-order of the amplitude of the primordial power as those examined in previous studies.

This paper is organized as follows. In Sec.~\ref{sec: exptail}, we will briefly review the generation of the exponential-tailed curvature perturbations and their effect on the abundance of \acp{PBH}
studied in Ref.~\cite{Kitajima:2021fpq}. In Sec.~\ref{sec:inducedGWs}, we provide a systematic perturbative approach for the 
calculation of the stochastic \acp{GW} induced by the non-Gaussian curvature perturbations, by making use of the diagrams.
Then, in Sec.~\ref{sec: result&diss}, based on the approach
given in Sec.~\ref{sec:inducedGWs}, we investigate the spectrum
of the induced \acp{GW} in the exponential tail
case
and discuss the observability in LISA. Section~\ref{sec:conclusion}
is devoted to the conclusion.
We adopt the natural unit, $c=\hbar=1$, throughout this work.

\section{Exponential-tailed curvature perturbations and primordial black holes\label{sec: exptail}}

In the standard scenario of inflation, primordial perturbations originate from the quantum vacuum fluctuation of the inflaton fields. It is therefore expected to well follow the Gaussian distribution at the leading order, which is in fact confirmed with high accuracy in the observation of the \ac{CMB}~\cite{Planck:2019kim}.
While this is natural because the \ac{CMB} scale perturbation is well in the perturbative range as its amplitude is the order of $10^{-5}$, it is however non-trivial whether the Gaussian assumption is valid for \acp{PBH}, the object related to the order-unity perturbation. In this section, we review the significant non-Gaussian feature called \emph{the exponential tail} and its effect on the \ac{PBH} abundance.

In order to deal with the nonlinear feature of gravity on the primordial metric perturbation, the so-called $\delta N$ formalism is useful~\cite{Starobinsky:1982ee,Starobinsky:1985ibc,Sasaki:1995aw,Wands:2000dp}.
Under the assumptions of the separate universe and the energy conservation, the superHubble inflaton perturbation $\delta\phi$ can be non-perturbatively converted to the conserved curvature perturbation $\zeta$ on the uniform density slice as the difference $\delta N$ in the e-foldings $N$ from the initial flat slice (no perturbation in the spatial curvature) to the final uniform density slice~\cite{Lyth:2004gb}.
One then imagines that an extremely large $\zeta$ or $\delta N$ can be realized in a so-called reproductive region.
In the eternal inflation for example~\cite{Linde:1982ur,Steinhardt:1982kg,Vilenkin:1983xq,Linde:1986fc,Linde:1986fe,Goncharov:1987ir}, the low probability of the large $N$ is compensated by the volume factor $a^3\propto\ee^{3N}$~\cite{Barenboim:2016mmw}, which means that such a probability decays only exponentially $\propto\ee^{-3N}$ rather than the Gaussian in that case.
Such a slow decay of the large-$\zeta$ probability may happen in a wider class of inflation.
If the decay of the large-$\zeta$ probability is slower than the Gaussian, the estimation of the \ac{PBH} abundance can be significantly altered from the one under the Gaussian assumption.

Though the precise probability should be calculated taking all quantum noise into account in, e.g., the stochastic approach (see, e.g., Refs.~\cite{Starobinsky:1982ee,Starobinsky:1986fx,Nambu:1987ef,Nambu:1988je,Kandrup:1988sc,Nakao:1988yi,Nambu:1989uf,Mollerach:1990zf,Linde:1993xx,Starobinsky:1994bd} for the first papers on this approach, and also Refs.~\cite{Pattison:2017mbe,Ezquiaga:2019ftu,Pattison:2021oen,Figueroa:2020jkf,Figueroa:2021zah,Tada:2021zzj,Jackson:2022unc,Ahmadi:2022lsm} for its application to the exponential tail),
qualitative features are often extracted by a simple assumption that only one noise gives a dominant contribution and the other dynamics is well approximated by the one without noise~\cite{Cai:2018dkf,Hooshangi:2021ubn,Cai:2021zsp,Cai:2022erk}.
Let us also focus on the extremely-flat-potential region, i.e., the ultra slow-roll phase to make perturbations large.
There, the \ac{EoM} for the background homogeneous mode of inflaton $\phi_0$ is approximated as
\bae{
    \dv[2]{\phi_0}{N}+3\dv{\phi_0}{N}\simeq0 \qc H\simeq\text{const.},
}
where $H$ is the Hubble parameter, and we used e-foldings $N(t)=\int^t_{t_\ui}H\dd{t}$ from some initial time $t_\ui$ to $t$ as the time variable.
It can be easily solved as
\bae{
    \phi_0(N)=\phi_\ui+\frac{\pi_\ui}{3}\qty(1-\ee^{-3N}), \quad \Leftrightarrow \quad N(\phi_0\mid\phi_\ui)=-\frac{1}{3}\ln\pqty{1-3\frac{\phi_0-\phi_\ui}{\pi_\ui}},
}
with the initial field value $\phi_\ui=\eval{\phi_0}_{t_\ui}$ and momentum $\pi_\ui=\eval{\dd\phi_0/\dd N}_{t_\ui}$. $N(\phi_0\mid\phi_\ui)$ denotes the e-foldings from $\phi_\ui$ to $\phi_0$.
Let us then assume that inflation ends or it is rapidly followed by the ordinary slow-roll phase at the end point $\phi_\uf$ and the total curvature perturbation is mainly given by the time difference between $\phi_\ui$ and $\phi_\uf$ due to the shift $\delta\phi_\ui$ at $\phi_\ui$ keeping the momentum intact.
That is, the curvature perturbation is simply given by
\bae{
    \zeta=\delta N=N(\phi_\uf\mid\phi_\ui+\delta\phi_\ui)-N(\phi_\uf\mid\phi_\ui)=-\frac{1}{3}\ln\pqty{1+3\frac{\delta\phi_\ui}{\pi_\uf}},
}
where $\pi_\uf=\pi_\ui\ee^{-3N(\phi_\uf\mid\phi_\ui)}=\pi_\ui-3(\phi_\uf-\phi_\ui)$ is the momentum at $\phi_\uf$ without noise.
Supposing the inflaton's noise $\delta\phi_\ui$ follows the Gaussian distribution and defining the Gaussian part of the curvature perturbation by $\zeta_g= - \delta\phi_\ui/\pi_\uf$, the full curvature perturbation is understood as a nonlinear transformation of the Gaussian field in this case~\cite{Cai:2018dkf,Atal:2019cdz,Atal:2019erb,Biagetti:2021eep}:
\bae{\label{eq: zeta in zetag}
    \zeta(\bfx)=-\frac{1}{3}\ln\pqty{1-3\zeta_g(\bfx)}.
}

In the small perturbation region $\abs{\zeta}\ll1$, the full curvature perturbation is well approximated by the Gaussian part $\zeta_g$ with perturbative corrections as can be seen in its series expansion, 
\bae{\label{eq: expansion of zeta}
    \zeta=-\frac{1}{3}\ln\pqty{1-3\zeta_g}=\zeta_g+\frac{3}{2}\zeta_g^2+3\zeta_g^3+\frac{27}{4}\zeta_g^4+\frac{81}{5}\zeta_g^5+\calO(\zeta_g^6).
}
However, it is obviously non-Gaussian essentially for a large enough value $\abs{\zeta}\gtrsim1$.
In fact, the probability density function of $\zeta$ can be inferred from that of $\zeta_g$ with the chain rule as
\bae{
    \probP_\zeta(\zeta)=\abs{\dv{\zeta_g}{\zeta}}\probP_{\zeta_g}(\zeta_g)=\ee^{-3\zeta}\probP_{\zeta_g}(\zeta_g),
}
making use of the inverse relation $\zeta_g=(1-\ee^{-3\zeta})/3$ of Eq.~\eqref{eq: zeta in zetag}.
In the large value limit $\zeta\to+\infty$ or equivalently $\zeta_g\to1/3$, the probability only decays exponentially as $\probP_\zeta(\zeta)\propto\ee^{-3\zeta}$ contrary to the Gaussian $\propto\ee^{-\zeta^2/2\braket{\zeta^2}}$.\footnote{Note that the probability $\probP_\zeta$ is not normalized to unity, $\int\probP_\zeta\dd{\zeta}<1$, as $\zeta=-\frac{1}{3}\ln(1-3\zeta_g)$ is defined only for $\zeta_g\in(-\infty,1/3)$. $\zeta_g\geq1/3$ corresponds to an eternally inflating baby universe in the current setup and can be also seen as a \ac{PBH} from the outer universe~\cite{Atal:2019erb}. The proper renormalization might be done by taking account of the cumulative noise in the stochastic formalism.
We simply neglect such a contribution in this work as it is probabilistically suppressed.}
This is a simple example of the exponential-tailed curvature perturbation. The decay rate $\Lambda\coloneqq-\dd{\ln \probP_\zeta}/\dd{\zeta}$ ($=3$ in this case) can depend on the details of the model, such as the potential smoothness around $\phi_\uf$ for example (see Refs.~\cite{Cai:2018dkf,Atal:2019cdz,Atal:2019erb}).
Heavier tails such that $\lim_{\zeta\to+\infty}\Lambda=0$ have been also proposed~\cite{Hooshangi:2021ubn,Cai:2021zsp,Cai:2022erk}.

\medskip

If the large-$\zeta$ probability is much amplified than the Gaussian one due to the exponential/heavy tail, the \ac{PBH} abundance can be significantly altered from the prediction under the Gaussian assumption.
The proper abundance taking account of the exponential tail can be calculated in, e.g., the so-called peak theory~\cite{Bardeen:1985tr} (see Refs.~\cite{Yoo:2018kvb,Yoo:2019pma,Yoo:2020dkz,Kitajima:2021fpq,Escriva:2022pnz} for its application to the \ac{PBH} mass function).
While we refer readers to Ref.~\cite{Kitajima:2021fpq} for details, let us briefly review the approach.

Once the functional form of $\zeta$ is fixed as Eq.~\eqref{eq: zeta in zetag}, all phenomena caused by the perturbations are statistically deterministic in principle.
In particular, it is understood that the profile around a very high peak of the Gaussian field $\zeta_g$ is typically spherical-symmetric and given by 
\bae{
    \hat{\zeta}_g=\tilde{\mu}_2\bqty{\frac{1}{1-\gamma_3^2}\pqty{\psi_1(r)+\frac{1}{3}R_3^2\Delta\psi_1(r)}-\frac{\tilde{k}_3^2}{\gamma_3(1-\gamma_3^2)}\pqty{\gamma_3^2\psi_1(r)+\frac{1}{3}R_3^2\Delta\psi_1(r)}}+\zeta_g^\infty,
}
with three dimensionless combined-Gaussian variables $\tilde{\mu}_2$, $\tilde{k}_3$, and $\zeta_g^\infty$, and characteristics
\bae{
    \sigma_n^2=\int\frac{\dd{k}}{k}k^{2n}\calP_g(k) \qc \psi_n(r)=\frac{1}{\sigma_n^2}\int\frac{\dd{k}}{k}k^{2n}\frac{\sin(kr)}{kr}\calP_g(k) \qc
    \gamma_3=\frac{\sigma_3^2}{\sigma_{2}\sigma_{4}} \qc R_3=\frac{\sqrt{3}\sigma_3}{\sigma_{4}},
}
determined by $\zeta_g$'s power spectrum
\bae{
    \calP_g(k)=\frac{k^3}{2\pi^2}\int\dd[3]{x}\ee^{-i\bfk\cdot\bfx}\Braket{\zeta_g\qty(\frac{\bfx}{2})\zeta_g\qty(-\frac{\bfx}{2})}.
}
Roughly speaking, three variables $\tilde{\mu}_2$, $\tilde{k}_3$, and $\zeta_g^\infty$ indicate the peak height, width, and overall offset, respectively.
The (comoving) number density of such a peak is statistically given by
\bae{
    \!\!n_\pk\dd{\tilde{\mu}_2}\dd{\tilde{k}_3}\dd{\zeta_g^\infty}
    =\frac{2\cdot3^{3/2}}{(2\pi)^{3/2}}\frac{\sigma_2^2\sigma_4^3}{\sigma_1^4\sigma_3^3}\tilde{\mu}_2\tilde{k}_3f\pqty{\frac{\sigma_2}{\sigma_1^2}\tilde{\mu}_2\tilde{k}_3^2}\probP_1^{(3)}\pqty{\frac{\sigma_2}{\sigma_1^2}\tilde{\mu}_2,\frac{\sigma_2}{\sigma_1^2}\tilde{\mu}_2\tilde{k}_3^2}\probP_\infty(\zeta_g^\infty)\dd{\tilde{\mu}_2}\dd{\tilde{k}_3}\dd{\zeta_g^\infty}\!,
}
where
\beae{
    &f(\xi)=\bmte{
    \frac{1}{2}\xi(\xi^2-3)\pqty{\erf\bqty{\frac{1}{2}\sqrt{\frac{5}{2}}\xi}+\erf\bqty{\sqrt{\frac{5}{2}}\xi}} \\
    +\sqrt{\frac{2}{5\pi}}\Bqty{\pqty{\frac{8}{5}+\frac{31}{4}\xi^2}\exp\bqty{-\frac{5}{8}\xi^2}+\pqty{-\frac{8}{5}+\frac{1}{2}\xi^2}\exp\bqty{-\frac{5}{2}\xi^2}},
    } \\
    &\probP_1^{(3)}(\nu,\xi)=\frac{1}{2\pi\sqrt{1-\gamma_3^2}}\exp\bqty{-\frac{1}{2}\pqty{\nu^2+\frac{(\xi-\gamma\nu)^2}{1-\gamma_3^2}}}, \\
    &\probP_\infty(\zeta_g^\infty)=\pqty{\frac{1-\gamma_3^2}{2\pi D\sigma_0^2}}^{1/2}\exp\bqty{-\frac{1-\gamma_3^2}{2D\sigma_0^2}{\zeta_g^\infty}^2},
}
with
\bae{
    D=1-\gamma_1^2-\gamma_2^2-\gamma_3^2+2\gamma_1\gamma_2\gamma_3 \qc \gamma_1=\frac{\sigma_1^2}{\sigma_0\sigma_2} \qc \gamma_2=\frac{\sigma_2^2}{\sigma_0\sigma_4}.
}
The peak profile of the full $\zeta$ is of course given by
\bae{
    \hat{\zeta}(r)=-\frac{1}{3}\ln\pqty{1-3\hat{\zeta}_g(r)},
}
with the same number density.

Whether such a peak collapses into a black hole or not can be judged by the mean compaction function, backed by several numerical works~\cite{Atal:2019erb,Escriva:2019phb}.
The compaction function is defined by
\bae{
    \calC(r)=\frac{2}{3}\bqty{1-(1+r\hat{\zeta}^\prime(r))^2},
}
and the (maximal) mean compaction is given by
\bae{
    \bar{\calC}_\um=\left.\pqty{4\pi\int^{R(r_\um)}_0\calC(r)R^2(r)\dd{R(r)}}\middle/\pqty{\frac{4\pi}{3}R^3(r_\um)}\right.,
}
with the areal radius $R(r)=a\ee^{\zeta(r)}r$ and the radius $r_\um$ corresponding to the (innermost) maximum of $\calC(r)$.
If the mean compaction $\bar{\calC}_\um$ exceeds the threshold value $\bar{\calC}_\uth=2/5$, the corresponding peak is supposed to form a \ac{PBH}.

The mass of the resultant \ac{PBH} is assumed to follow the scaling relation:
\bae{\label{eq: MPBH}
    M_\PBH(\tilde{\mu}_2,\tilde{k}_3,\zeta_g^\infty)=K\qty(\tilde{\mu}_2-\tilde{\mu}_{2,\uth}(\tilde{k}_3,\zeta_g^\infty))^pM_H(\tilde{\mu}_2,\tilde{k}_3,\zeta_g^\infty),
}
with the universal power $p\simeq0.36$~\cite{Choptuik:1992jv,Evans:1994pj,Koike:1995jm,Niemeyer:1997mt,Niemeyer:1999ak,Hawke:2002rf,Musco:2008hv}.
$K$ is the slightly-profile-depending order-unity coefficient and we uniformly approximate it as $K\simeq1$ for simplicity in this paper.
$\tilde{\mu}_{2,\uth}(\tilde{k}_3,\zeta_g^\infty)$ is the $\tilde{\mu}_2$ value on the threshold, i.e. $\bar{\calC}_\um(\tilde{\mu}_{2,\uth},\tilde{k}_3,\zeta_g^\infty)=\bar{\calC}_\uth$, which depends on the other variables $\tilde{k}_3$ and $\zeta_g^\infty$.
$M_H$ is the Hubble mass at the Hubble reentry of the maximal radius, $R(r_\um)H=1$.

With use of this expression of the mass, the \ac{PBH} number density $n_\PBH(M)\dd{\ln M}$ within the mass range of $[M,M\ee^{\dd{\ln M}}]$ is computed as
\bae{
    n_\PBH(M)=\int_{\bar{\calC}_\um(\tilde{\mu}_2,\tilde{k}_3,\zeta_g^\infty)>\bar{\calC}_\uth}n_\pk(\tilde{\mu}_2,\tilde{k}_3,\zeta_g^\infty)\delta\qty(\ln M_\PBH(\tilde{\mu}_2,\tilde{k}_3,\zeta_g^\infty)-\ln M)\dd{\tilde{\mu}_2}\dd{\tilde{k}_3}\dd{\zeta_g^\infty}.
}
The current density ratio $f_\PBH(M)$ of \acp{PBH} to total dark matters in this mass bin then reads
\bae{
    f_\PBH(M)\dd{\ln M}=\frac{Mn_\PBH(M)}{3\Mpl^2H_0^2\Omega_\DM}\dd{\ln M},
}
with the current Hubble parameter $H_0$ and the dark matter density parameter $\Omega_\DM$. $\Mpl=1/\sqrt{8\pi G}$ is the reduced Planck mass.
The total \ac{PBH} abundance is given by
\bae{\label{eq: fPBHtot}
    f_\PBH^\tot=\int f_\PBH(M)\dd{\ln M}.
}

\bfe{width=0.7\hsize}{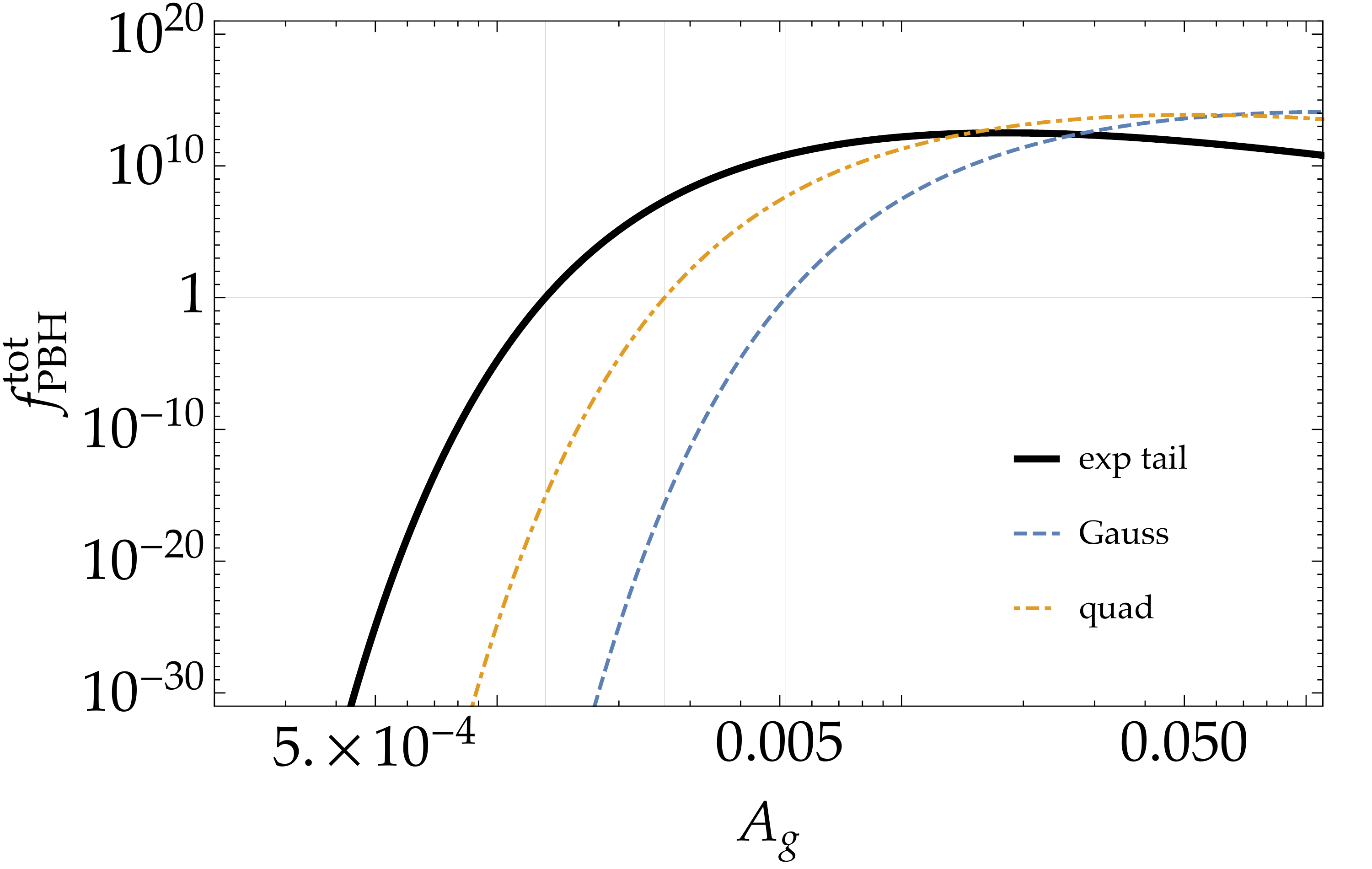}{the total \ac{PBH} abundance $f_\PBH^\tot$~\eqref{eq: fPBHtot} as a function of the perturbation amplitude $A_g$ in the monochromatic power spectrum case~\eqref{eq: monochromatic P} for $k_*=\SI{1.56e12}{Mpc^{-1}}$.
The black thick line is the result of the exponential-tailed perturbation, while the blue dashed and orange dot-dashed correspond to the Gaussian $\zeta\simeq\zeta_g$ and the quadratic approximation $\zeta\simeq\zeta_g+(3/2)\zeta_g^2$, respectively.
The vertical thin lines indicate the required amplitude
$A_g=1.32\times10^{-3}$, $2.59\times10^{-3}$, and $5.17\times10^{-3}$ for $f_\PBH^\tot=1$ in each case. The exponential tail feature amplifies the \ac{PBH} abundance and thus reduces the required perturbation amplitude for given $f_\PBH^\tot$.}{fig: fPBHtot}

If one simply assumes the monochromatic power spectrum for $\zeta_g$,
\bae{\label{eq: monochromatic P}
    \calP_g(k)=A_g\delta(\ln k-\ln k_*),
}
one finds that the variables $\tilde{k}_3$ and $\zeta_g^\infty$ are fixed to $1$ and 0 respectively as
\bae{
    \probP_1^{(3)}\pqty{\frac{\sigma_2}{\sigma_1^2}\tilde{\mu}_2,\frac{\sigma_2}{\sigma_1^2}\tilde{\mu}_2\tilde{k}_3^2}\to\frac{A_g}{2\tilde{\mu}_2}\frac{1}{\sqrt{2\pi A_g}}\ee^{-\tilde{\mu}_2^2/(2A_g)}\delta(\tilde{k}_3-1) \qc
    \probP_\infty(\zeta_g^\infty)\to\delta(\zeta_g^\infty).
}
Then only $\tilde{\mu}_2$ remains and the analysis is much simplified.
In this case, the PBH mass is sharply distributed around the mass scale corresponding to $k_*$ (see, e.g., Ref.~\cite{Tada:2019amh}):
\bae{\label{eq: Mk}
	M_{k}(k_*)=10^{22}\left(\frac{g_{\ast}}{106.75}\right)^{-1/6}\left(\frac{k_*}{\SI{1.56e12}{Mpc^{-1}}}\right)^{-2}\,\si{g},
}
where $g_*$ is the effective degrees of freedom for energy at the horizon reentry of the mode $k_*$.
An example result is shown in Fig.~\ref{fig: fPBHtot} for $k_*=\SI{1.56e12}{Mpc^{-1}}$ (or $M_k(k_*)=10^{22}\,\si{g}$), where we show the total \ac{PBH} abundance $f_\PBH^\tot$ as a function of the perturbation amplitude $A_g$. While $A_g=5.17\times10^{-3}$ is required for $f_\PBH^\tot=1$ if the curvature perturbation is Gaussian, the required amplitude is reduced to $A_g=1.32\times10^{-3}$ for the exponential-tailed perturbation as expected. The quadratic approximation $\zeta\simeq\zeta_g+(3/2)\zeta_g^2$ (orange dot-dashed) is better than the simple Gaussian assumption but one sees it is far from enough.
Note that, the ratio of the $A_g$ value between the Gaussian and the exponential tail, $\left(\frac{1.32\times10^{-3}}{5.17\times10^{-3}}\right)\sim 
0.25$, would universally hold when $f_{\PBH}\lesssim 1$. Since the density parameter of the induced GWs is, at the leading order, proportional to $A_g^2$, $\Omega_\GW\propto A_g^2$, as we will see below, this universal reduction of $A_g$ gives the universal relation that the induced GW amplitude in the exponential tail case is 
$\sim6\%$ of the one in the Gaussian tail.

\begin{figure}
    \centering
    \begin{tabular}{c}
        \begin{minipage}{0.514\hsize}
            \centering
            \includegraphics[width=0.95\hsize]{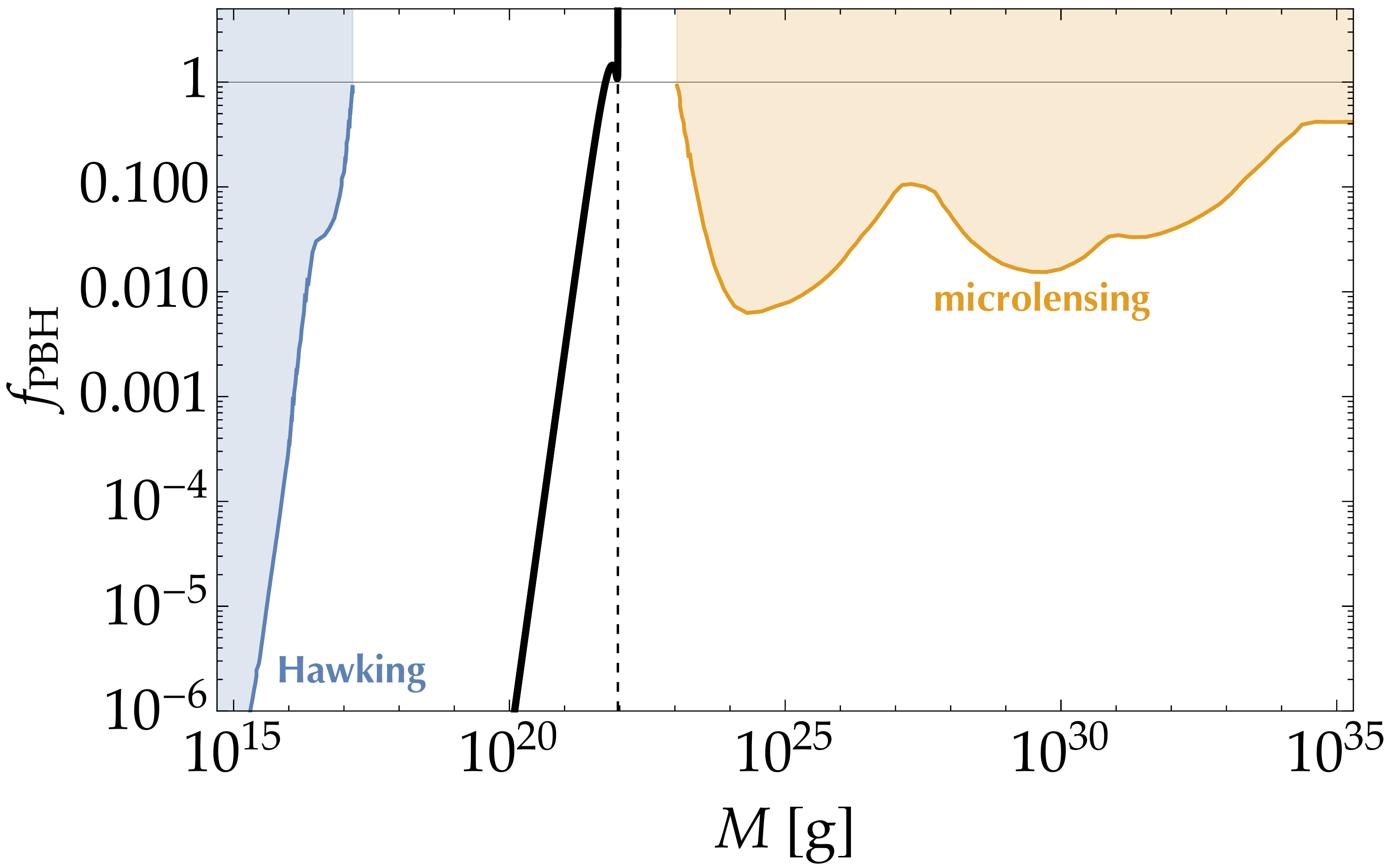}
        \end{minipage}
        \begin{minipage}{0.486\hsize}
            \centering
            \includegraphics[width=0.95\hsize]{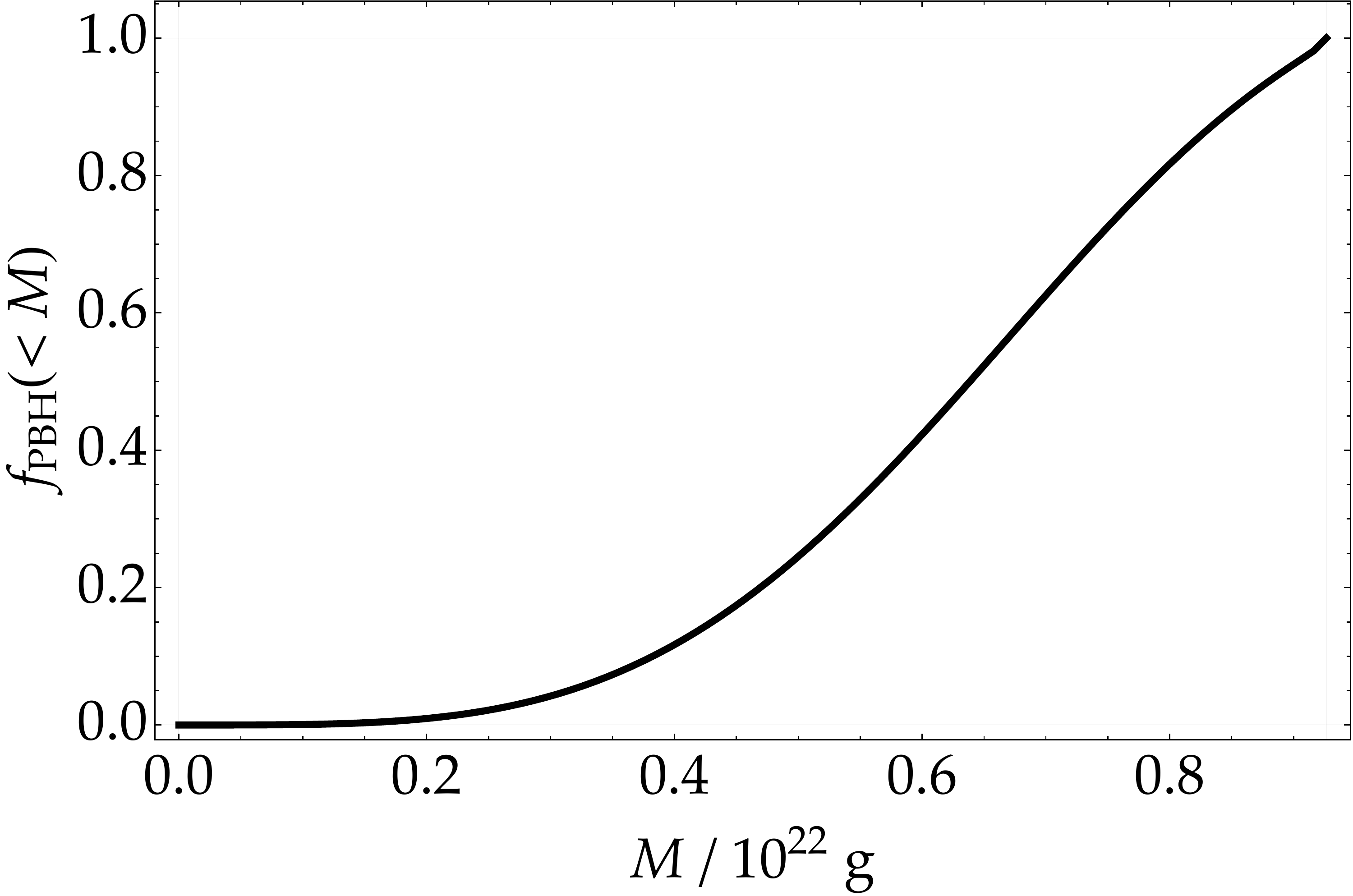}
        \end{minipage}
    \end{tabular}
    \caption{\emph{Left}: the corresponding \ac{PBH} mass function (black) for $f_\PBH^\tot=1$ in the exponential tail case with the observational constraints due to the Hawking radiation (blue) and the gravitational microlensing (orange), taken from Ref.~\cite{Carr:2021bzv} (see also references therein).
    The mass function is diverging at the maximal mass $M_\umax$ ($\simeq\SI{0.926e22}{g}$ in this setup; the vertical black dashed line) due to a characteristic feature of the exponential tail but its integral is converging healthily (see the text). \emph{Right}: the cumulative mass spectrum~\eqref{eq: fPBH cum} for intuitive understanding of the \ac{PBH} mass distribution. Gridlines denote $M=0$, $M=M_\umax$, $f_\PBH(<M)=0$, and $f_\PBH(<M)=1$, respectively.
    One finds $f_\PBH^\tot=f_\PBH(<M_\umax)=1$ indeed.}
    \label{fig: mass function}
\end{figure}

We show the 
corresponding \ac{PBH} mass spectrum in the 
left panel of Fig.~\ref{fig: mass function} with several observational constraints. Interestingly, the mass function has a hard cut and is divergent at $M_\umax\simeq\SI{0.926e22}{g}$ shown by the black dashed line as discussed in Ref.~\cite{Kitajima:2021fpq}. This is because the \ac{PBH} mass~\eqref{eq: MPBH} is not monotonic in the perturbation amplitude $\tilde{\mu}_2$ but has a maximum value $M_\umax=M(\tilde{\mu}_{2,\umax})$ and hence the Jacobian $\abs{\dd{\ln M}/\dd{\tilde{\mu}_2}}^{-1}$ from the distribution of $\tilde{\mu}_2$ to that of $\ln M$ is divergent at that mass (see the bottom panel of Fig.~6 of Ref.~\cite{Kitajima:2021fpq}).
Noting that the \ac{PBH} mass behaves quadratically around that point as $M_\umax-M\propto(\tilde{\mu}_2-\tilde{\mu}_{2,\umax})^2$, one finds that the divergence is as slow as $f_\PBH\propto(M_\umax-M)^{-1/2}$ and hence its integral is healthily convergent.
In the right panel of Fig.~\ref{fig: mass function}, we plot the cumulative spectrum
\bae{\label{eq: fPBH cum}
    f_\PBH(<M)\coloneqq\int^Mf_\PBH(M')\frac{\dd{M'}}{M'},
}
to show the \ac{PBH} mass distribution more intuitively.

\section{ 
Gravitational waves induced by scalar perturbations}
\label{sec:inducedGWs}

Let us move on to
\acp{GW} induced by scalar perturbations $\zeta$. 
We first note that the series expansion of $\zeta$ given by Eq.~\eqref{eq: expansion of zeta} is expected to work well for the calculation of \acp{GW} contrary to the \ac{PBH} abundance. This is because the amplitude of \acp{GW} is mainly determined by the $\zeta$'s typical behavior with high probability, i.e., $\zeta\sim0$, while \acp{PBH} is associated with the rare high peaks $\zeta\gtrsim1$. Therefore, we develop the \ac{GW} calculation method with use of this series expansion in this section. We will see in the next section that the result indeed converges well practically even in the exponential
tail case.  

We begin with the conformal Newtonian gauge (see Refs.~\cite{Matarrese:1997ay,Boubekeur:2008kn,Arroja:2009sh,Hwang:2017oxa,Domenech:2017ems,Gong:2019mui,Tomikawa:2019tvi,Inomata:2019yww,Yuan:2019fwv,Chang:2020iji,Chang:2020mky,Domenech:2020xin} for the gauge choice issue). 
With the assumption that the vector perturbations and the anisotropic stress are negligible, the perturbed metric is defined by
\bae{
    \dd{s}^2 = -a(\tau)^2(1+2\Phi)\dd{\tau} +a(\tau)^2\left((1-2\Phi)\delta_{ij}+\frac{1}{2}h_{ij}\right)\dd{x}^i\dd{x}^j,
}
where $\tau$ is the conformal time, $\Phi$ is the scalar gravitational potential, and $h_{ij}$ is the transverse traceless tensor perturbation. 

We below consider the tensor perturbation $h$ generated by the second-order effect of the scalar perturbation $\Phi$.\footnote{\label{footnote: Phi3}
The higher order contributions such as $h\sim\Phi^3$ and $\Phi^4$ have been recently discussed in Refs.~\cite{Yuan:2019udt,Zhou:2021vcw,Chang:2022nzu}. We will touch on them again later in Sec. \ref{sec: result&diss}.}

We expand the tensor
perturbation with the Fourier modes as 
\bae{
     h_{ij}(\tau,\bfx)=\sum_{\lambda = +,\times}\int\dk \ee^{i\bfk\cdot\bfx}e^{\lambda}_{ij}(k)\hk{}{},
}
where the two time-independent transverse traceless polarization tensors are defined by
\beae{
    e^{+}_{ij}(\bfk)&=\frac{1}{\sqrt{2}}(e_i(\bfk)e_j(\bfk)-\bar{e}_i(\bfk)\bar{e}_j(\bfk)), \\
    e^{\times}_{ij}(\bfk)&=\frac{1}{\sqrt{2}}(e_i(\bfk)\bar{e}_j(\bfk)+\bar{e}_i(\bfk)e_j(\bfk)),
}
with
the two normalized vectors $e_i(\bfk)$ and $\bar{e}_i(\bfk)$ orthogonal to each other and to the wave vector $\bfk$.
 
The tensor power spectrum $P_{\lambda \lambda^{\prime}}(\tau,k)$ is defined as
\bae{
    \braket{\hk{}{}\hk{\prime}{}}=(2\pi)^3 \delta^3(\bfk+\bfk^{\prime})P_{\lambda \lambda^{\prime}}(\tau,k),
}
and the dimensionless power spectrum $\PP_{\lambda\lambda^{\prime}}(\tau,k)$ 
is given by
\bae{
   \PP_{\lambda\lambda^{\prime}}(\tau,k) = \frac{k^3}{2\pi^2}P_{\lambda \lambda^{\prime}}(\tau,k).
}
The energy density of the scalar-induced GWs on the subhorizon scales 
is evaluated as
\bae{
    \rho_{\text{GW}}(\tau) =\int \dd {\ln{k}}\rho_{\text{GW}}(\tau,k)= \frac{\Mpl}{16a^2(\tau)}\overline{\left<h_{ij,k}h_{ij,k}\right>},
}
where
$h_{ij,k}=\partial_{x^k}h_{ij}$ and the overline stands for the oscillation average.
The \ac{GW} density parameter per logarithmic wavenumber 
reads
\bae{\label{eq: OmegaGW}
    \Omega_{\text{GW}}(\tau,k) 
    =\frac{\rho_{\text{GW}}(\tau,k)}{\rho_{\text{tot}}(\tau)}
    =\sum_{\lambda,\lambda^\prime=+,\times}\Omega_{\GW,\lambda\lambda^\prime}
    =\frac{1}{48}\lr{\frac{k}{a(\tau)H(\tau)}}^2\sum_{\lambda,\lambda' = +,\times}\overline{\PP_{\lambda\lambda'}(\tau,k)}.
}
Note that the contribution of $\lambda\neq\lambda^\prime$ will vanish  
in the parity-conserving universe as we will check either analytically or numerically (see also Appendix~\ref{sec: appendix}).
Since the energy density dilution of GWs is the same as the one of the radiation,  
i.e. $\rho\propto a^{-4}$,  
unless energy injection by decay or annihilation of particles, the \ac{GW} density parameter converges to a constant in the deep subhorizon limit during the \ac{RD} era. We will below calculate this limit value $\Omega_\GW^\RD(k)\coloneqq\Omega_\GW(\tau\to\infty,k)$.
The current density parameter $\Omega_{\GW,0}(k)$ can be 
simply estimated by multiplying 
it by the current radiation parameter $\Omega_{\ur,0}h^2\simeq4.2\times10^{-5}$ as $\Omega_{\GW,0}(k)h^2\simeq\Omega_\GW^\RD(k)\Omega_{\ur,0}h^2$.

\subsection{Gravitational waves induced by the second-order scalar perturbations}

We here review the formulation of \acp{GW} induced by the second-order scalar perturbations (see, e.g., Refs.~\cite{Kohri:2018awv, Espinosa:2018eve} for the details).
Note that  
we only focus on the induced \acp{GW} and neglect
the primordial tensor perturbations caused by the vacuum fluctuations in this work. 
In Fourier space, the \ac{EoM} for 
\acp{GW} including the quadratic 
terms of $\Phi$ is given 
by
\bae{
    \partial_\tau^2 \hk{}{} + 2\HH \partial_\tau \hk{}{} +k^2\hk{}{}=4S_\lambda(\tau,\bfk),
    \label{eq:h}
}
where $S_\lambda(\tau,\bfk)$ is the source term, 
$\HH=a(\tau)H(\tau)$ is the conformal Hubble parameter.
If one adopts the linear relation between the gravitational potential $\Phi$ and the primordial curvature perturbation $\zeta$ with the transfer function $\Phi(k\tau)$ as
\bae{\label{eq: relation_phi_zeta}
    \Phi(\tau,\bfk)=\frac{2}{3}\Phi(k\tau)\zeta(\bfk) \quad \text{in the \ac{RD} era},
}
the source term $S_\lambda(\tau,\bfk)$ can be written in terms of $\zeta$ as
\bae{\label{eq: source}
    S_\lambda(\tau,\bfk)=\iq{}\Q{}{}f(\abs{\bfk-\bfq},q,\tau)\zeta(\bfq)\zeta(\bfk-\bfq).
}
Here, the projection factor $\Q{}{}$ is given by
\bae{
    Q_\lambda(\bfk,\bfq)=e^\lambda_{ij}(\bfk)q^iq^j=\frac{q^2}{\sqrt{2}}\sin^2\theta\times\bce{
        \cos(2\phi) & (\lambda=+) \\
        \sin(2\phi) & (\lambda=\times)
    },
}
for the spherical coordinate expression $\bfq=q(\sin\theta\cos\phi, \sin\theta\sin\phi, \cos\theta)^T$ with $\bfk$ in the $z$-direction,
and the source factor $f(p,q,\tau)$ is
\bae{
    f(p,q,\tau)&=\frac{4}{9}\Bigl[3\Phi(p\tau)\Phi(q\tau)+ \Phi^\prime(p\tau)\Phi^\prime(q\tau) +\pqty{\Phi(p\tau)\Phi^\prime(q\tau)+\Phi^\prime(p\tau)\Phi(q\tau)}\Bigr], 
}
in the \ac{RD} era where $\Phi'(x) = \dd{\Phi(x)}/\dd{\ln x}$.
The transfer function in the \ac{RD} era is given by\footnote{Here we consider the adiabatic scalar perturbations. In the case of the isocurvature perturbations, see e.g. Ref.~\cite{Domenech:2021and} about the transfer function.}
\bae{
    \Phi(x)=-\frac{9}{x^2}\pqty{\frac{\sin(x/\sqrt{3})}{x/\sqrt{3}}-\cos(x/\sqrt{3})}.
}
Adopting the Green's function method to solve Eq.~(\ref{eq:h}), the particular solution of the induced GWs is formally solved as
\bae{
    \hk{}{}=\frac{4}{a(\tau)}\int^\tau\dd{\tilde{\tau}}G_\bfk(\tau,\tilde{\tau})a(\tilde{\tau})S_\lambda(\tilde{\tau},\bfk),
}
with the Green's function $G_\bfk(\tau,\tilde{\tau})$ satisfying
\bae{
    \partial_\tau^2G_\bfk(\tau,\tilde{\tau})+\pqty{k^2-\frac{\partial_\tau^2a(\tau)}{a(\tau)}}G_\bfk(\tau,\tilde{\tau})=\delta(\tau-\tilde{\tau}).
} 
It is solved as
\bae{
    G_\bfk(\tau,\tilde{\tau})=\frac{\sin k(\tau-\tilde{\tau})}{k},
}
in the \ac{RD} era.
Combining the above equations, the two-point function of induced GWs is given by
\bme{\label{eq:hh}
    \braket{h_{\lambda_1}(\tau,\bfk_1)h_{\lambda_2}(\tau,\bfk_2)}=\iq{1}\iq{2}\Q{1}{1}\Q{2}{2} \\
    \times I_k(\abs{\bfk_1-\bfq_1},q_1,\tau)I_k(\abs{\bfk_2-\bfq_2},q_2,\tau) 
    \braket{\zeta(\bfq_1)\zeta(\bfk_1-\bfq_1)\zeta(\bfq_2)\zeta(\bfk_2-\bfq_2)},
}
with the kernel function
\bae{
    I_k(p,q,\tau) = 4\int^\tau\dd{\tilde{\tau}}G_\bfk(\tau,\tilde{\tau})\frac{a(\tilde{\tau})}{a(\tau)}f(p,q,\tilde{\tau}).
}
As we have mentioned,  
we calculate $\Omega_{\rm GW} (\tau, k)$ in the deep subhorizon limit during the RD era with $\tau \to \infty$,
where the asymptotic form of this kernel function is simply given by
\bae{
    k\tau I_k(p,q,\tau
    )
    \underset{\tau\to\infty}{\sim}\calF_k(p,q)\bqty{\calS_k(p,q)\sin(k\tau)+\calC_k(p,q)\cos(k\tau)},
}
with
\beae{
    \calF_k(p,q)&=\frac{3(p^2+q^2-3k^2)}{p^3q^3}, \\
    \calS_k(p,q)&=-4pq+(p^2+q^2-3k^2)\ln\abs{\frac{3k^2-(p+q)^2}{3k^2-(p-q)^2}}, \\
    \calC_k(p,q)&=-\pi(p^2+q^2-3k^2)\Theta(p+q-\sqrt{3}k).
}
$\Theta(x)$ is the step function. Therefore, the oscillation average of their cross-correlation reads
\bae{\label{eq: I2 average}
    \overline{
    J_k^2(p_1,q_1;p_2,q_2)}\coloneqq{}&\lim_{\tau\to\infty}(k\tau)^2\overline{I_k(p_1,q_1,\tau
    )I_k(p_2,q_2,\tau
    )} \nonumber \\
    ={}&\frac{1}{2}\calF_k(p_1,q_1)\calF_k(p_2,q_2)\bqty{\calS_k(p_1,q_1)\calS_k(p_2,q_2)+\calC_k(p_1,q_2)\calC_k(p_2,q_2)}.
}
In order to evaluate the spectrum of induced GWs, we need to 
specify the remaining trispectrum of the primordial curvature perturbations.

\subsection{Diagrammatic approach}\label{sec_diag}

Let us turn next to introduce our approach to take account of the primordial non-Gaussianity in the trispectrum of the curvature perturbations.
The curvature perturbation with the local-type non-Gaussianity (i.e., $\zeta(\bfx)$ given by some function $\scrF_\NL(\zeta_g(\bfx))$ of the Gaussian field $\zeta_g(\bfx)$ at the same spatial point) can be expanded in general as
\bae{\label{eq: perturbexpand}
    \zeta(\bfx)= F_\NL^{(0)} \zeta_g(\bfx) + F_\NL^{(1)} \zeta_g^2(\bfx) +   F_\NL^{(2)} \zeta_g^3(\bfx) + F_\NL^{(3)} \zeta_g^4(\bfx)+ F_\NL^{(4)} \zeta_g^5(\bfx) + \cdots,
}
with the expansion coefficient $F_\NL^{(n)}$.
We assume the Gaussianity at the leading order as $F_\NL^{(0)}=1$. We also use specific characters for the first several coefficients as
$F_\NL\coloneqq F_\NL^{(1)}$, $G_\NL\coloneqq F_\NL^{(2)}$, $H_\NL\coloneqq F_\NL^{(3)}$, $I_\NL\coloneqq F_\NL^{(4)}$, 
$\cdots$, following the convention. 
Based on this expression, one can obtain the perturbative expression for the trispectrum of the curvature perturbations, and calculate the tensor power spectrum perturbatively in the power spectrum of $\zeta_g$ (specifically in the amplitude parameter $A_g$ given by Eq.~\eqref{eq: monochromatic P} in our monochromatic case).
As direct computations would be tedious,
we employ the helpful diagrammatic approach advocated in Ref.~\cite{Unal:2018yaa} first and organized by Adshead et al.~\cite{Adshead:2021hnm}.

\begin{figure}
	\centering
    \begin{tabular}{lcc}
		i) & 
		\begin{minipage}{0.5\hsize}
			\centering
			\includegraphics[width=0.4\hsize]{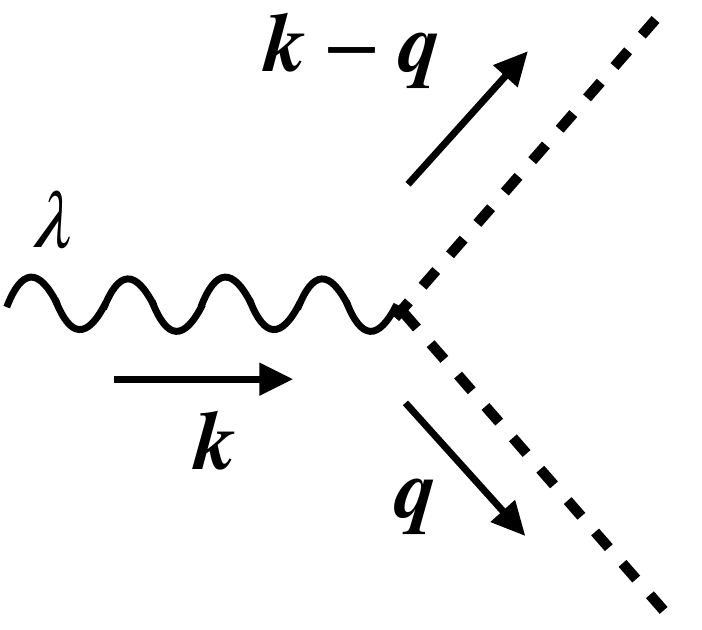}
		\end{minipage} \smallskip &
		\begin{minipage}{0.3\hsize}
			\centering
			$\displaystyle I_k(\abs{\bfk-\bfq},q,\tau)Q_\lambda(\bfk,\bfq)$
		\end{minipage} \cr\cr 
		ii) & 
		\begin{minipage}{0.5\hsize}
			\centering			\includegraphics[width=0.5\hsize]{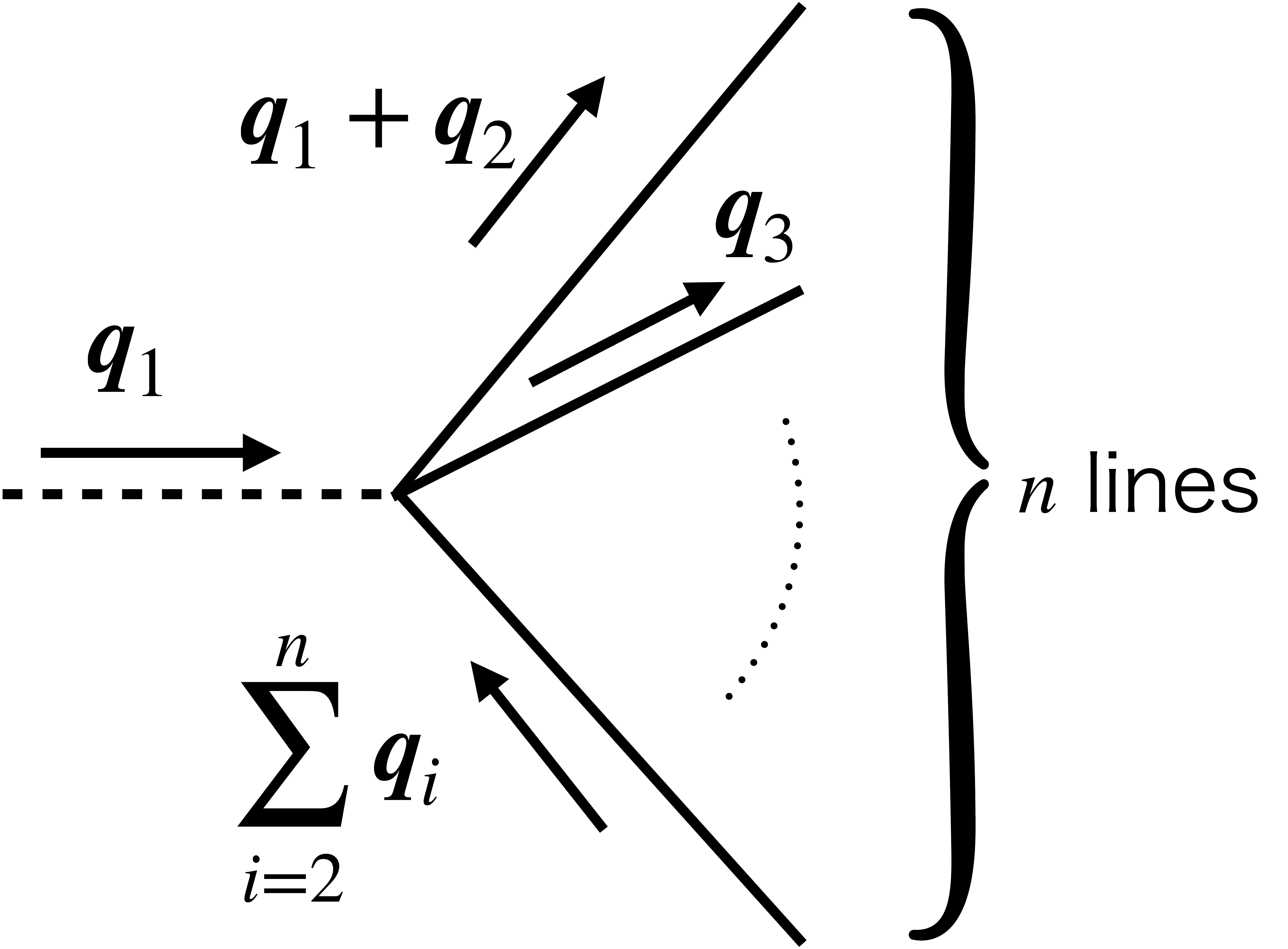}
		\end{minipage} \smallskip &
		\begin{minipage}{0.3\hsize}
			\centering
			$\displaystyle n!F_\NL^{(n-1)}$
		\end{minipage} \cr\cr
		iii) & 
		\begin{minipage}{0.5\hsize}
			\centering
			\includegraphics[width=0.3\hsize]{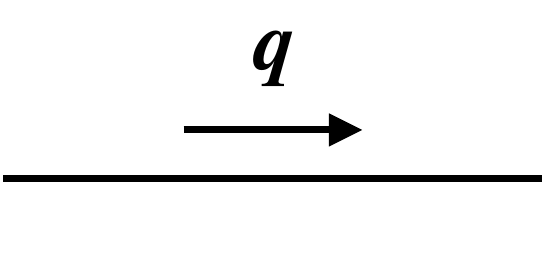}
		\end{minipage} &
		\begin{minipage}{0.3\hsize}
			\centering
			$\displaystyle P_g(q)$
		\end{minipage} \cr\cr
    	iv) & 
		\begin{minipage}{0.5\hsize}
			\centering			
			$\displaystyle \text{ Integrate over each undetermined momentum }$
		\end{minipage} &
		\begin{minipage}{0.3\hsize}
			\centering
			$\displaystyle \int\frac{\dd[3]q}{(2\pi)^3}$
		\end{minipage} \cr\cr 
		v) & 
		\begin{minipage}{0.45\hsize}
			\centering			
			$\displaystyle \text{ 
			Divide by the symmetric factor}$
		\end{minipage} &
		\begin{minipage}{0.3\hsize}
		\end{minipage} \cr\cr
	\end{tabular}
	\caption{Feynman rules as the building blocks.}
	\label{fig: feynman rule}
\end{figure}

Including the transformation from the curvature perturbation to the tensor one, all the relevant Feynman rules are summarized in Fig.~\ref{fig: feynman rule}.
Making use of them, we calculate the two-point function of tensor modes sourced by the scalar perturbations. 
That is, we first set two external tensor lines (wave lines) with the same momentum $\bfk$ and the polarization $\lambda$ as otherwise, the contributions will trivially vanish (see discussion in Appendix~\ref{sec: appendix} particularly for the polarization). These two tensor lines are connected through 
i) the coupling $I_k(\abs{\bfk-\bfq},q,\tau)Q_\lambda(\bfk,\bfq)$ between one tensor
$h_\lambda(\tau,\bfk)$ and two scalar curvature perturbations $\zeta(\bfk-\bfq)$ and $\zeta(\bfq)$ (dotted lines), ii) the coupling $n!F_\NL^{(n-1)}$ between one curvature perturbation $\zeta(\bfq_1)$ and $n$ Gaussian fields $\zeta_g(\bfq_1+\bfq_2)$, $\zeta_g(\bfq_3)$, $\cdots$, $\zeta_g(-\sum_{i=2}^n\bfq_i)$ (plane lines), satisfying the momentum conservation, and iii) the propagator $P_g(q)$ of the Gaussian field. 
Then iv)  
one has to integrate it over each undetermined momentum $\bfq_i$.

\begin{figure}
 \centering
  \begin{tabular}{ccc}
		\begin{minipage}{0.5\hsize}
		\centering
			\includegraphics[width=0.7\hsize]{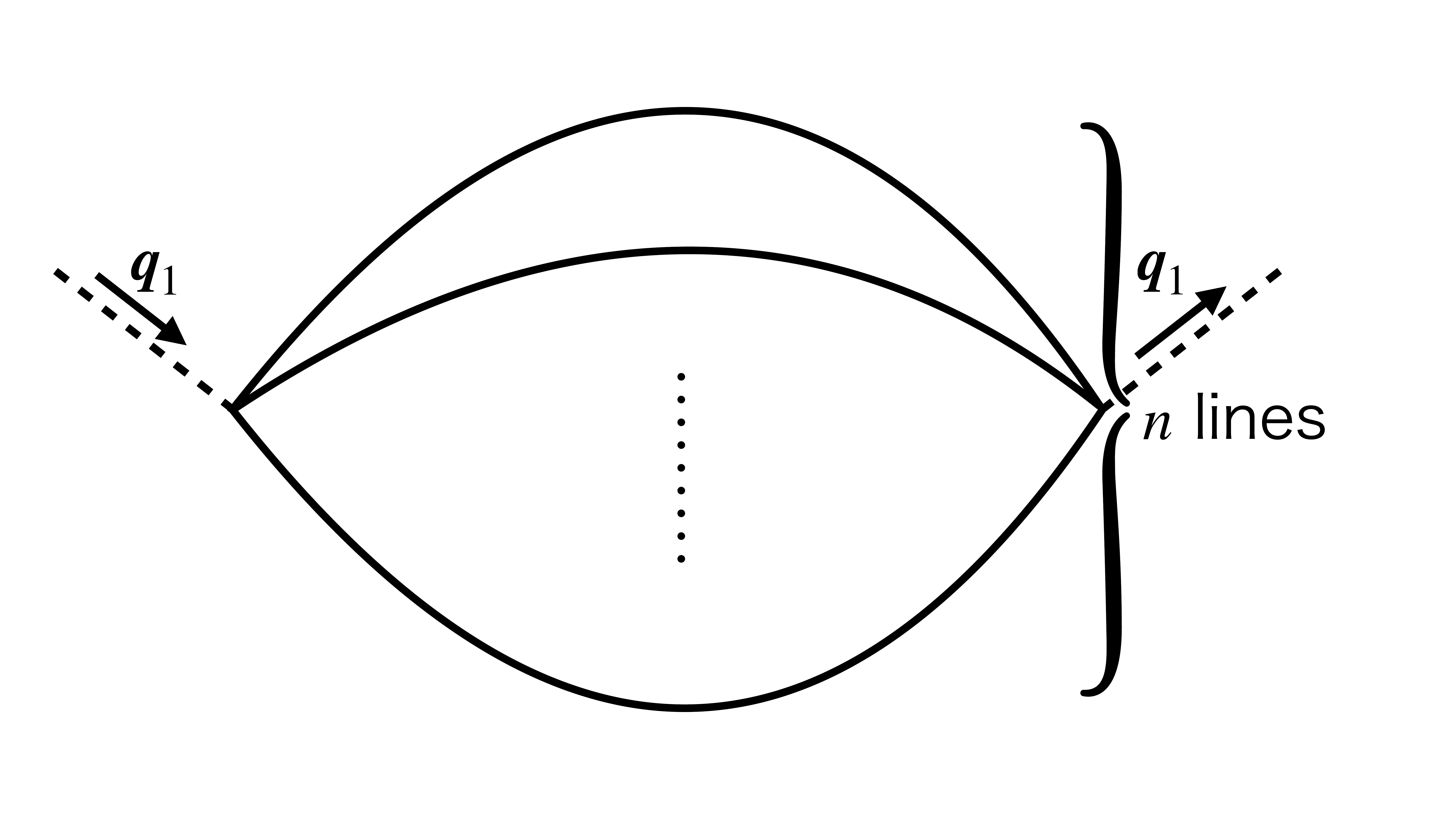}
			\subcaption{Convolved propagators.}
		\end{minipage}
		\begin{minipage}{0.5\hsize}
			\centering
			\includegraphics[width=0.7\hsize]{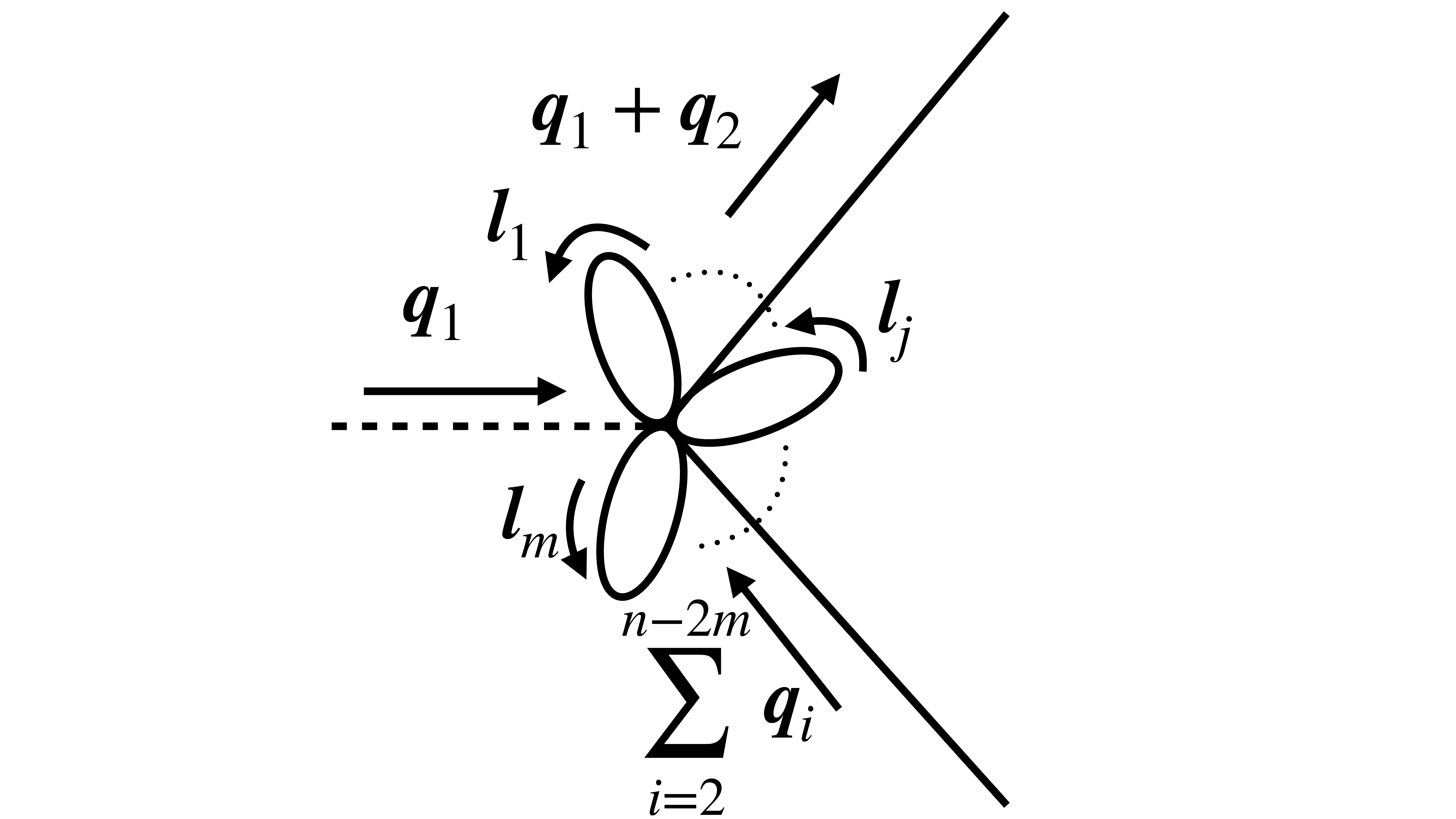}
			\subcaption{Self-closed loops.}
		\end{minipage}
		\end{tabular}
		\caption{The loop structures.}
		\label{fig:loop structs}
\end{figure}

The factor $n!$ of the coupling ii) counts up all possible connections. However, one may have some loop structures such as ``convolved propagators" and ``self-closed loops" shown in Fig.~\ref{fig:loop structs}, and in such a case, v) the diagram must be divided by the symmetric factor to avoid overcounts.  
For example, the permutation of $n$ convolved propagators yields $n!$ overcounts. Therefore, the symmetric factor is calculated as $n!$ in this case. Let us also see a $m$ self-closed loops case. The exchange of the initial and end points leads to overcount of factor 2 for each loop, and the permutation of loops themselves causes $m!$ overcounts. In total, the symmetric factor is hence $2^mm!$.

\begin{figure}
    \centering
    \begin{tabular}{c}
        \includegraphics[width=0.7\hsize]{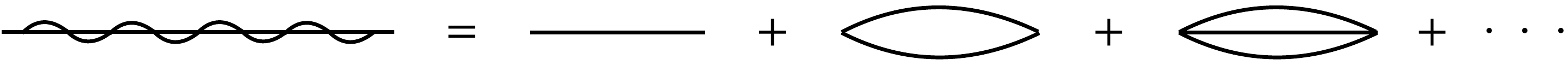} \bigskip \\
        \includegraphics[width=0.7\hsize]{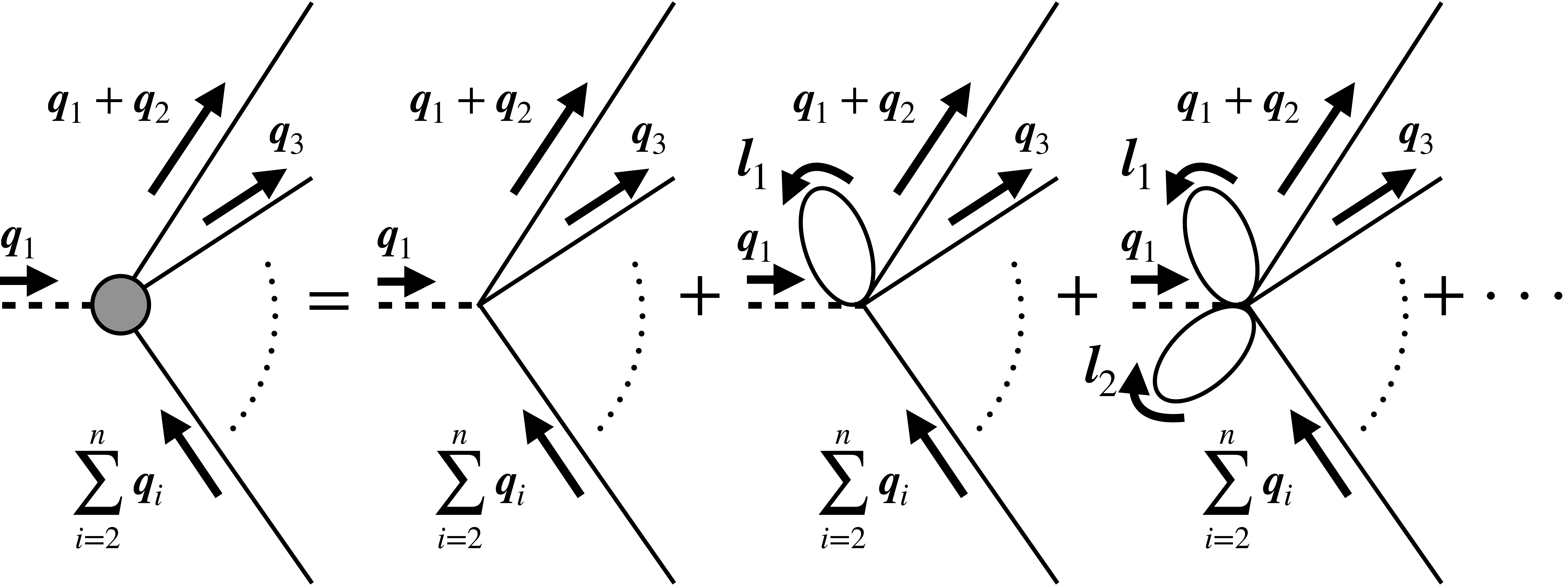}
    \end{tabular}
    \caption{Renormalized propagator of the Gaussian field (wave-plane line) and vertex (gray bubble).}
    \label{fig: renormalized P and V}
\end{figure}

For the total amplitude of \acp{GW}, all possible diagrams are summed up.
We here note that loop corrections to the propagator and vertex can be formally summarized by introducing the ``renormalized" diagrams, which are exhibited in Fig.~\ref{fig: renormalized P and V}. That is, we formally define the wave-plane line by the summation of convolved propagators of the Gaussian field, and the gray bubble by that of vertices with several closed loops. By replacing the plane propagator iii) and vertex ii) with these ``renormalized" ones, possible loop corrections are exhausted.
Note that however, the specific values of these parts depend on the number of other lines through the expansion coefficients $F_\NL^{(n)}$. Therefore, the numerical contribution must be calculated for each individual diagram.

\bfe{width=0.5\hsize}{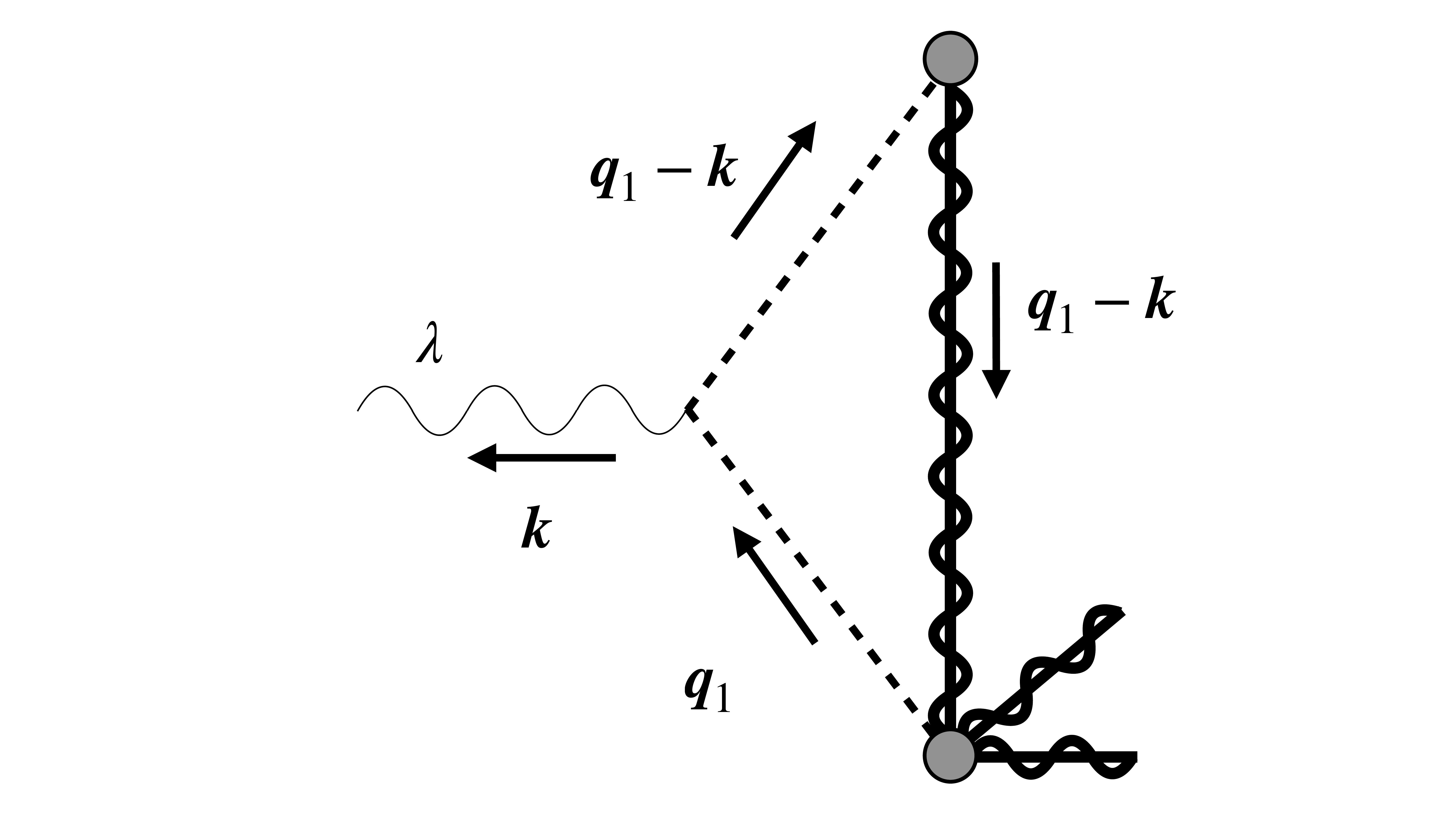}{Helically prohibited subdiagram.}{fig: helically prohibited}

One important rule is that each curvature perturbation (dotted line), which is coupled to a tensor mode (external wave line), must be connected to another dotted line coupled to the other external wave line (tensor mode) by at least a plane line (propagator of the Gaussian field),
or otherwise the diagram 
should include the subdiagram shown in Fig.~\ref{fig: helically prohibited}.
Based on the above Feynman rules,
it is found to be proportional to
\bae{
    &\int\frac{\dd[3]{\bfq_1}}{(2\pi)^3}Q_\lambda(\bfk,\bfq_1)f(\abs{\bfk-\bfq_1},q_1,\tilde{\tau})\tilde{P}_g(q_1) \nonumber \\
    &\quad=\bmte{\int_0^{2\pi}\dd{\phi}\bce{
        \cos(2\phi) & \lambda=+ \\
        \sin(2\phi) & \lambda=\times
    } \\
    \times\int\frac{q_1^2\dd{q_1}\dd{\cos\theta}}{(2\pi)^3}\frac{q_1^2}{\sqrt{2}}\sin^2(\theta) f\qty(\sqrt{k^2+q_1^2-2kq_1\cos\theta},q_1,\tilde{\tau})\tilde{P}_g(q_1)
    } \nonumber \\
    &\quad=0,
}
where $\tilde{P}_g(q_1)$ can include the loop corrections.
Thus, any diagram containing this subdiagram should vanish, which is due to the helicity conservation. 

\begin{figure}
    \centering
    \begin{tabular}{c}
        \begin{minipage}{0.5\hsize}
            \centering
            \includegraphics[width=0.8\hsize]{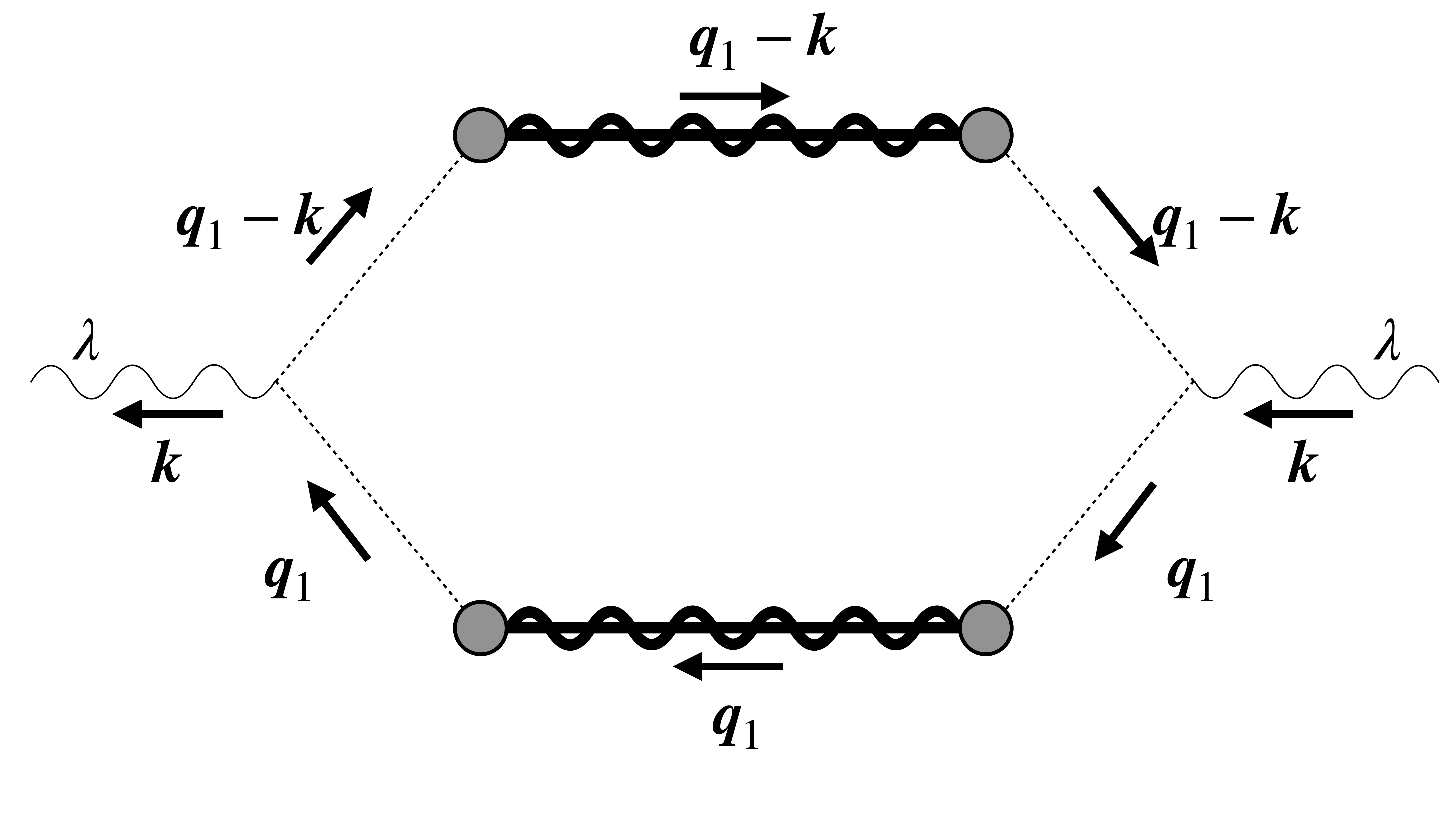}
        \end{minipage}
        \begin{minipage}{0.5\hsize}
            \centering
            \includegraphics[width=0.8\hsize]{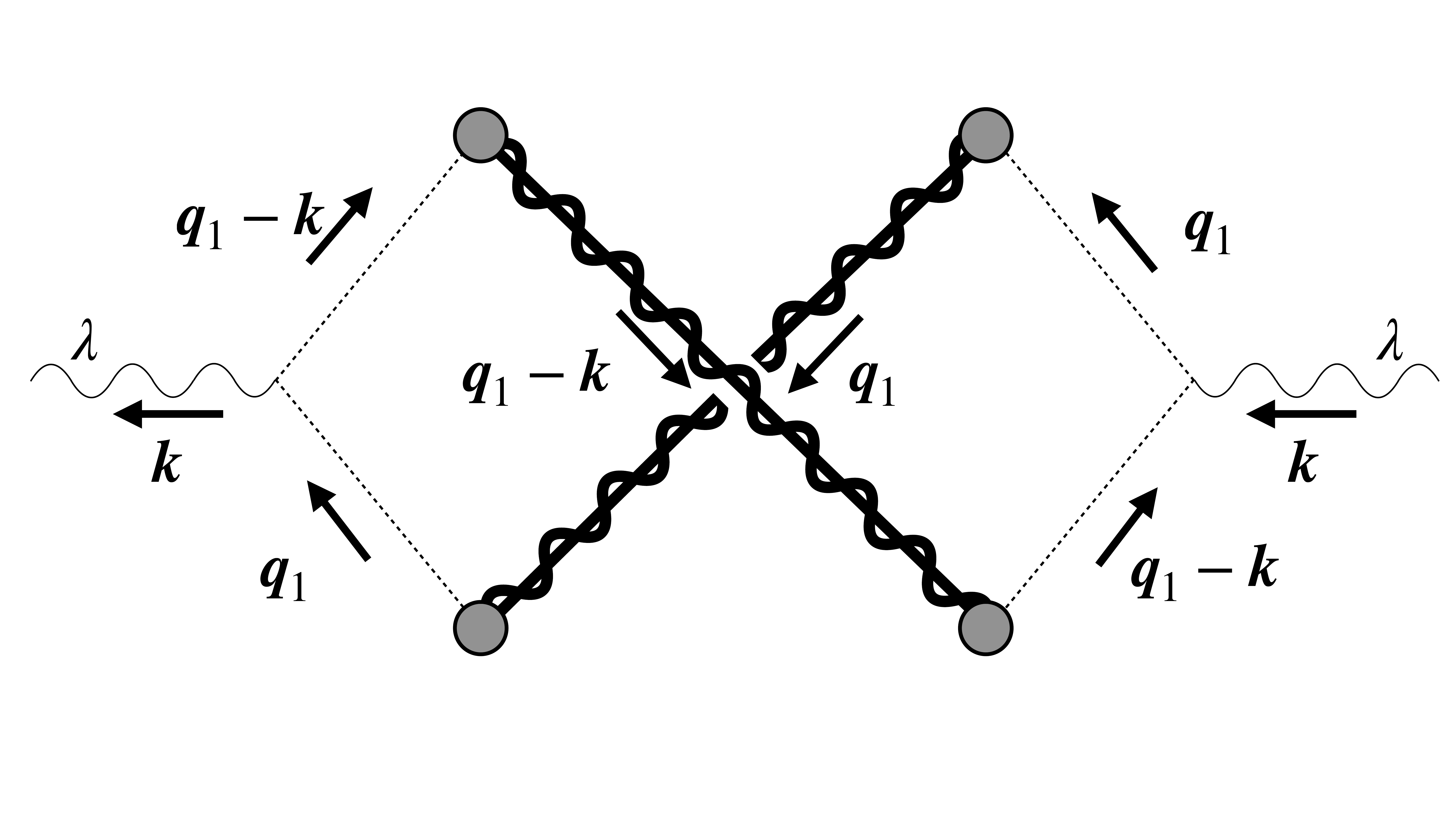}
        \end{minipage}
    \end{tabular}
    \caption{\emph{Left}: 
    ``$\mathbb{V}$anilla" diagram
    as a minimal one. \emph{Right}: the other minimal diagram, which can be taken into account just by doubling the left diagram.}
    \label{fig: tree}
\end{figure}

\begin{figure}
    \centering
    \begin{tabular}{c}
        \begin{minipage}{0.33\hsize}
            \centering
            \includegraphics[width=0.95\hsize]{Figures_paper/tree.pdf}
            \subcaption{
            $\mathbb{V}$anilla-type
            }
        \end{minipage} 
        \begin{minipage}{0.33\hsize}
            \centering
            \includegraphics[width=0.95\hsize]{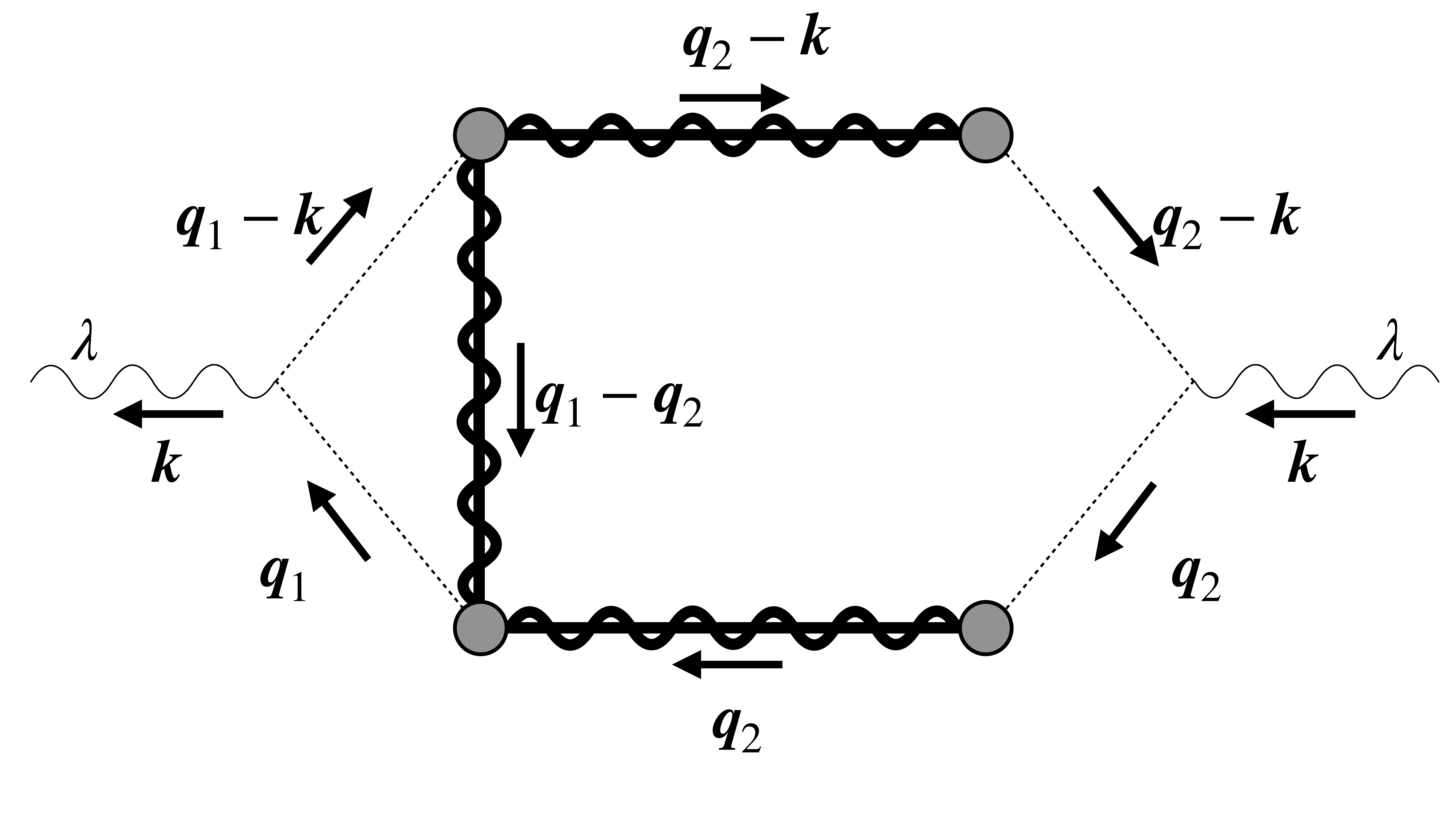}
            \subcaption{$\mathbb{C}$-type}
        \end{minipage} 
        \begin{minipage}{0.33\hsize}
            \centering
            \includegraphics[width=0.95\hsize]{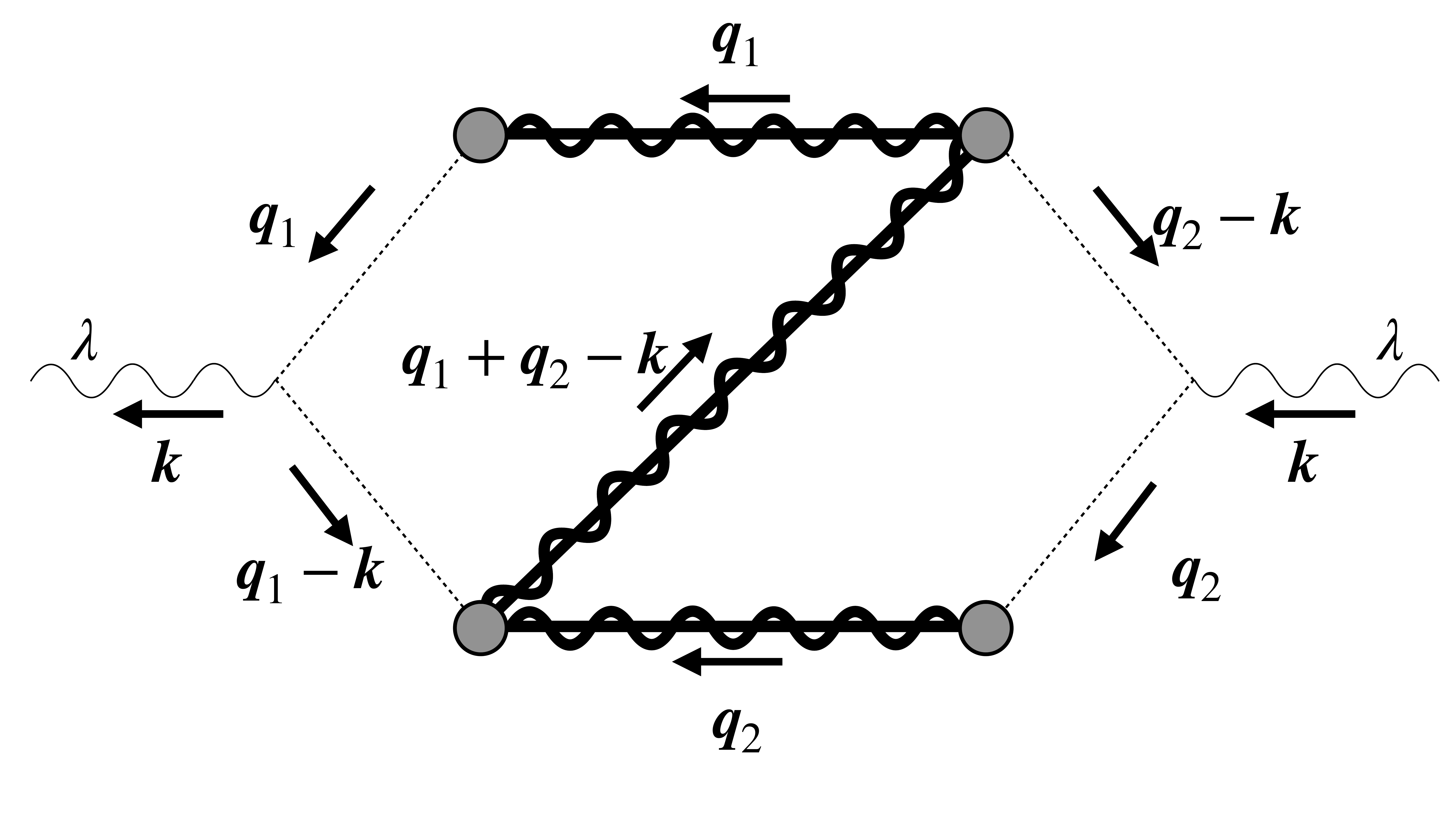}
            \subcaption{$\mathbb{Z}$-type}
        \end{minipage} 
    \end{tabular}
   
    \label{fig: 3rd}
     \centering
    \begin{tabular}{c}
        \begin{minipage}{0.33\hsize}
            \centering
            \includegraphics[width=0.95\hsize]{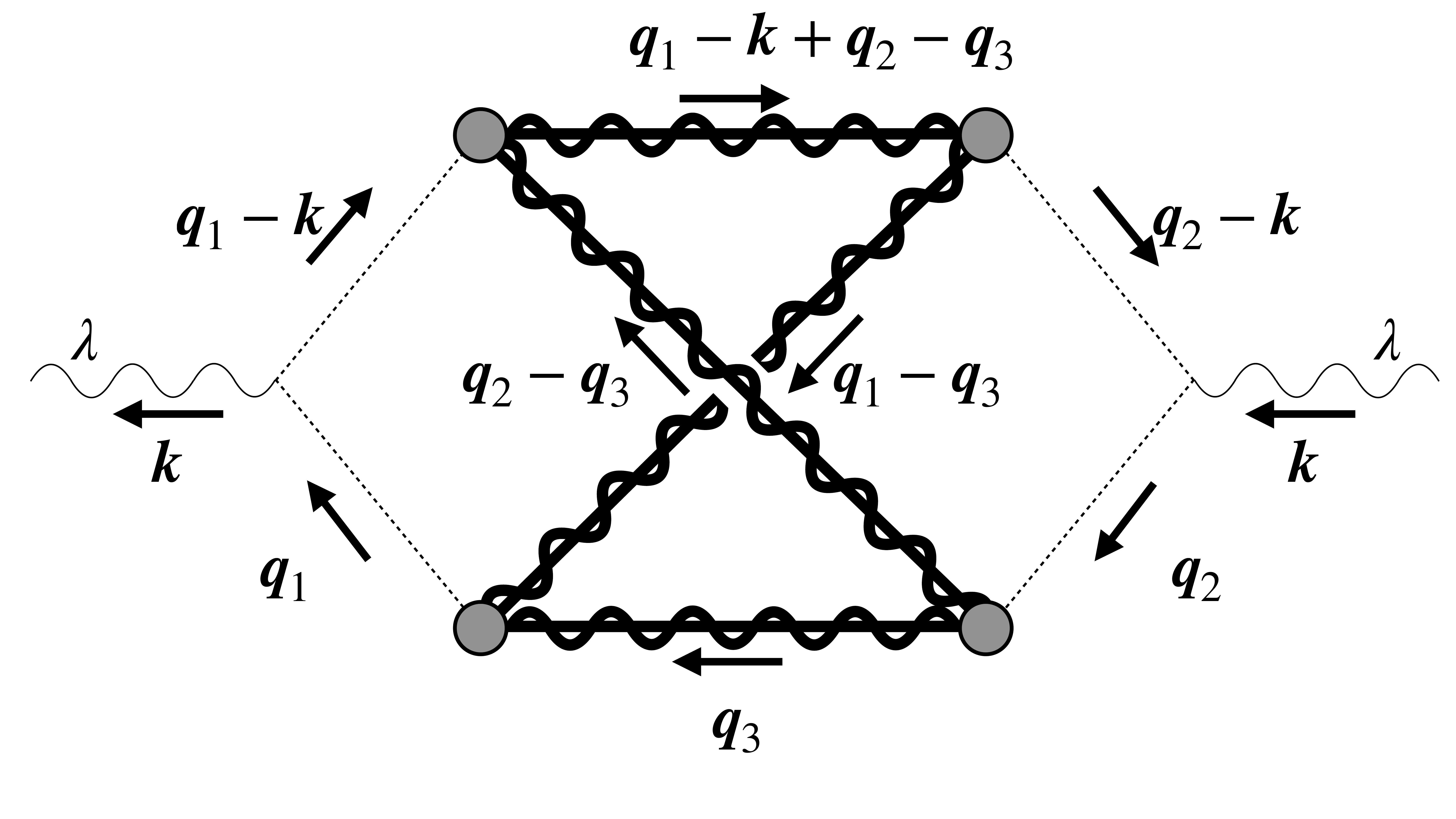}
            \subcaption{$\mathbb{X}$-type}
        \end{minipage}
        \begin{minipage}{0.33\hsize}
            \centering
            \includegraphics[width=0.95\hsize]{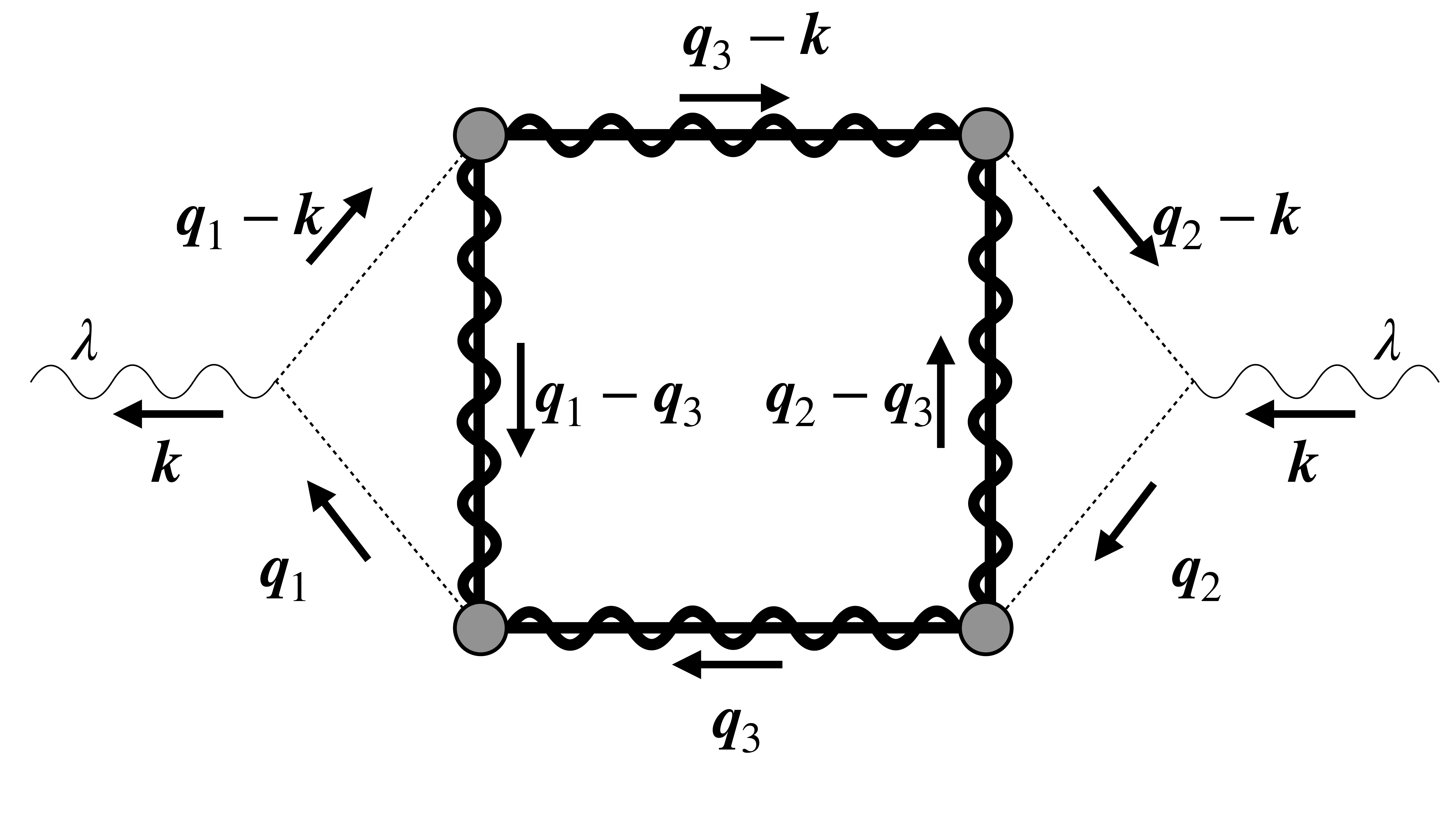}
            \subcaption{$\mathbb{B}$ox-type}
        \end{minipage}
        \begin{minipage}{0.33\hsize}
            \centering
            \includegraphics[width=0.95\hsize]{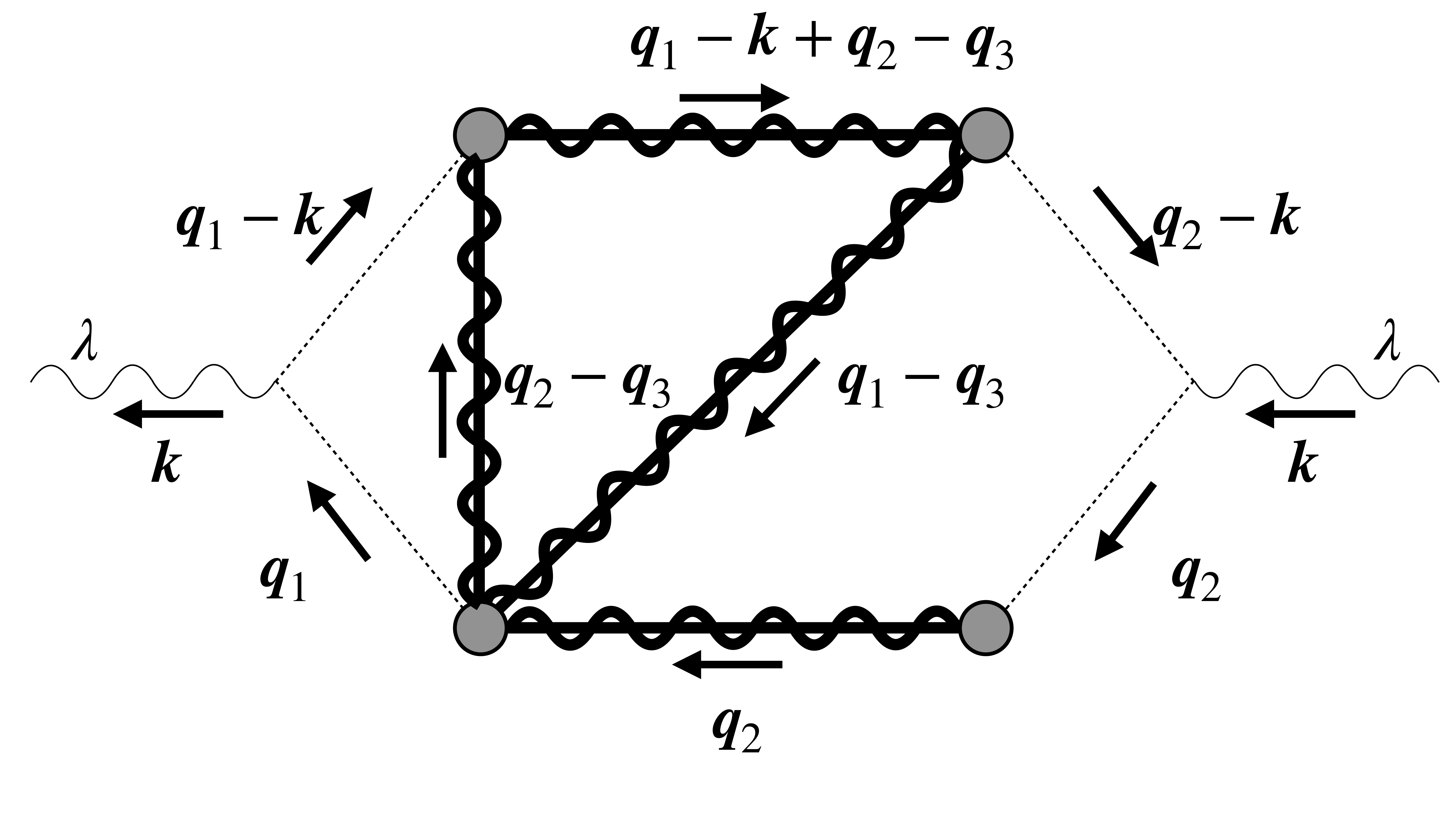}
            \subcaption{$\mathbb{CZ}$-type}
        \end{minipage} 
    \end{tabular}
   
    \label{fig: 4th}
    \centering
    \begin{tabular}{c}
        \begin{minipage}{0.33\hsize}
            \centering
            \includegraphics[width=0.95\hsize]{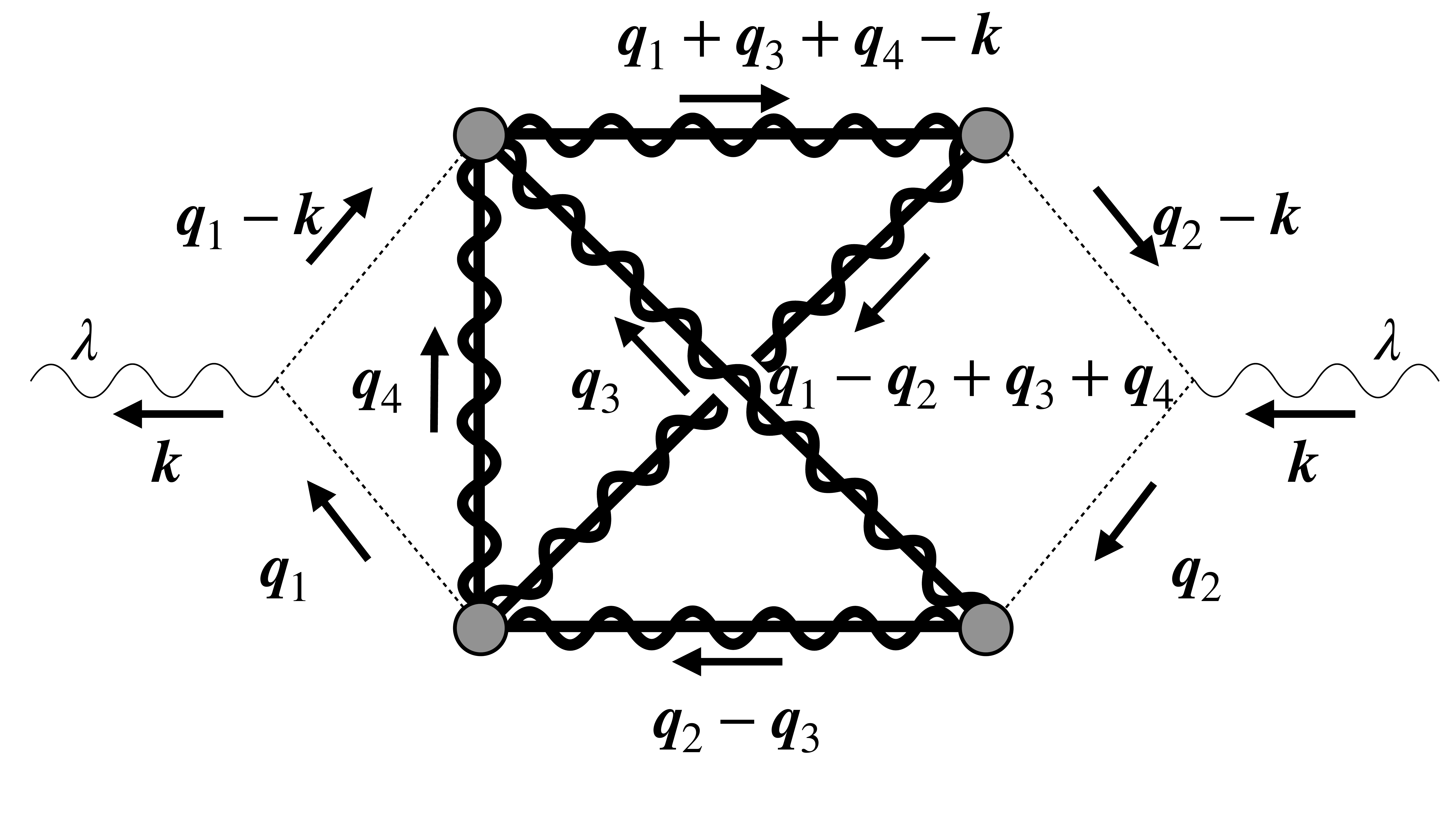}
            \subcaption{$\mathbb{CX}$-type}
        \end{minipage}
        \begin{minipage}{0.33\hsize}
            \centering
            \includegraphics[width=0.95\hsize]{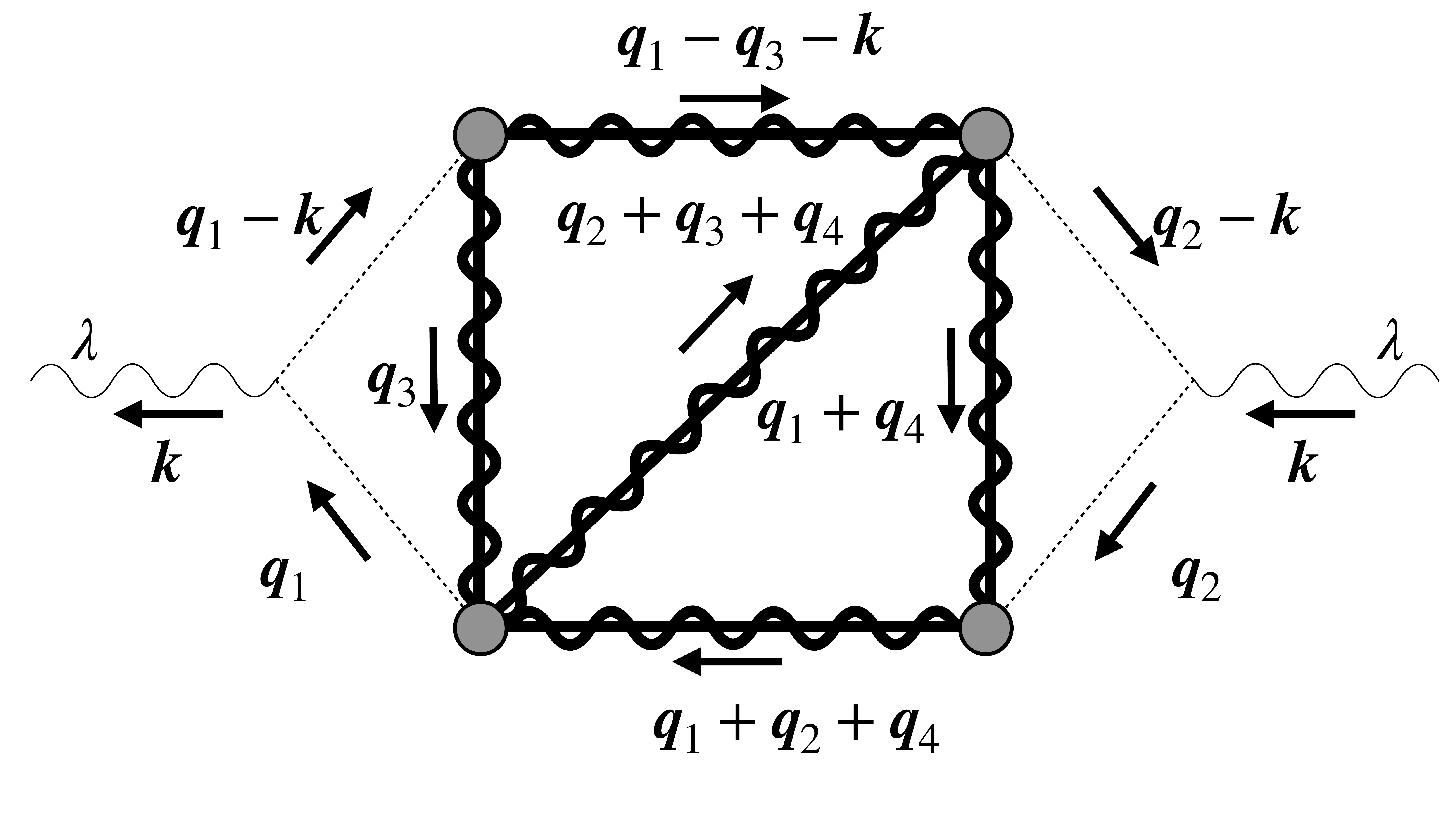}
            \subcaption{$\mathbb{ZB}$ox-type}
        \end{minipage} 
        \begin{minipage}{0.33\hsize}
            \centering
            \includegraphics[width=0.95\hsize]{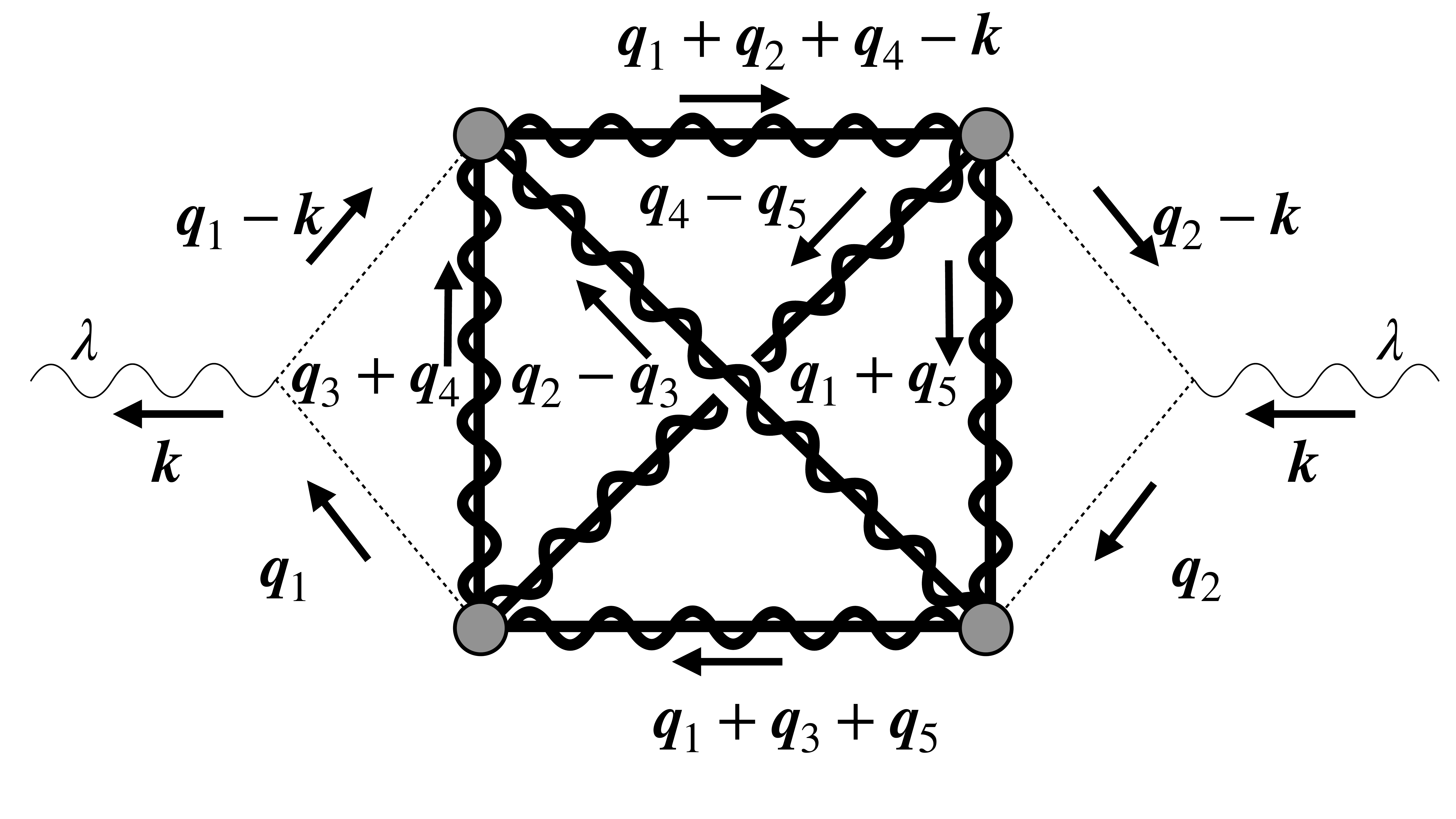}
            \subcaption{$\mathbb{XB}$ox-type}
        \end{minipage} 
    \end{tabular}
    \caption{All  
    possible contributions topologically independent.}
    \label{fig: all non-G cont}
\end{figure}

The minimal diagram is hence given by the left one shown in Fig.~\ref{fig: tree}, which we call
``$\mathbb{V}$anilla" diagram (we use the blackboard bold typeface for renormalized diagrams).
Another helpful rule is that some independent diagrams in a ``deformed" 
relation (such as ``twist" and up/down or left/right ``flip") with one diagram should give the same contribution as that diagram and hence can be taken into account just by the ``deformation factor"
$2^s$ ($s=1$, $2$, or $3$).
For example, the right diagram in Fig.~\ref{fig: tree} is independent of the left one and should be counted in. It however gives the same numerical contribution and thus we just double the left one instead of independently computing the right one.
Hereafter we hence take only the left one
as a minimal configuration 
and any non-Gaussian contribution can be expressed by adding lines to this. Then
all contributions can be summarized into nine topologically-independent diagrams shown in Fig.~\ref{fig: all non-G cont} because there are only four vertices in the diagram.

Finally, let us introduce a shorthand notation for the relevant integrals before moving on to the detailed calculation. We define $\calI_{\lambda\lambda^\prime}$ by
\bme{
    \calI_{\lambda\lambda^\prime}(\tau,\bfk\mid\bfq_1,\bfq_2\mid\bfk_1,\bfk_2,\cdots) \\
    \coloneqq\int\frac{\dd[3]{q_1}}{(2\pi)^3}\frac{\dd[3]{q_2}}{(2\pi)^3}\cdots Q_\lambda(\bfk,\bfq_1)Q_{\lambda^\prime}(\bfk,\bfq_2)I_k(\abs{\bfk-\bfq_1},q_1,\tau)I_k(\abs{\bfk-\bfq_2},q_2,\tau) \\
    \times P_g(k_1)P_g(k_2)\cdots.
}
Here $\bfk_1$, $\bfk_2$, $\cdots$ are supposed to be combinations of $\bfk$, $\bfq_1$, $\bfq_2$, $\cdots$, and the integration should be taken over all undetermined momenta $\bfq_i$ ($i=1,2,...$) other than $\bfk$.
All diagrams shown below can be summarized in this integral.

\subsubsection{Second-order contribution}

\begin{figure}
    \centering
    \begin{tabular}{c}
        \begin{minipage}{0.4\hsize}
            \centering
           \includegraphics[width=\hsize]{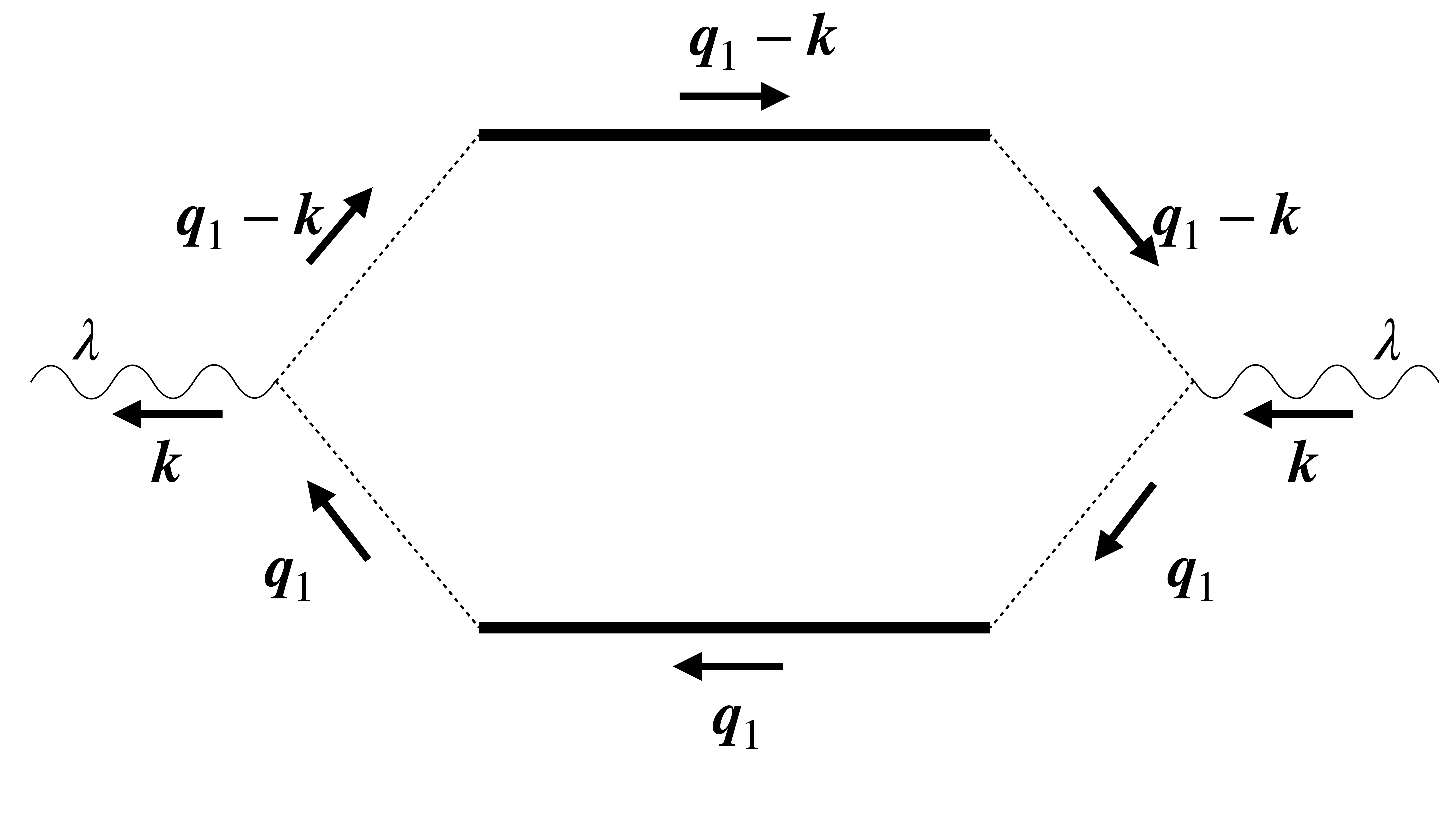}
        \end{minipage}
        \begin{minipage}{0.6\hsize}
            \centering
            \includegraphics[width=0.9\hsize]{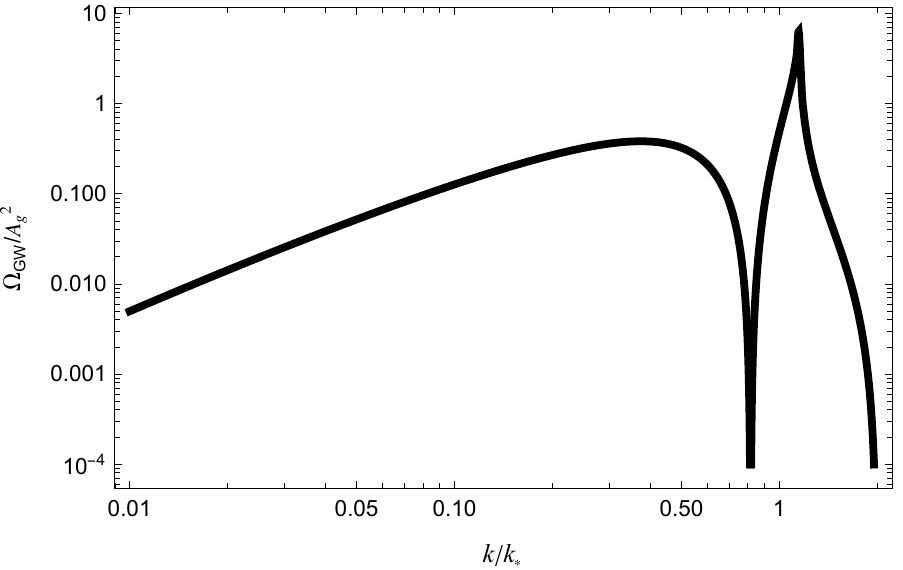}
        \end{minipage} 
    \end{tabular}
    \caption{
    The second-order (Vanilla) contribution in a diagram (left) and the resultant normalized \ac{GW} amplitude $\Omega_\GW^{(2)}/A_g^2$ (right).}
    \label{fig: GW_Gauss}
\end{figure}

Let us see specific examples  
order by order in our monochromatic power spectrum of the curvature perturbation~\eqref{eq: monochromatic P}.
There is only one topologically-independent diagram for the leading order contribution ($\propto A_g^2$), shown in the left panel of Fig.~\ref{fig: GW_Gauss}. 
Either for $++$ 
or $\times\times$ mode, this diagram 
reads 
\bae{
    P_{\lambda\lambda}^{\text{Vanilla}}(\tau,k)=
    \calI_{\lambda\lambda}(\tau,\bfk\mid\bfq,\bfq\mid\bfq,\bfk-\bfq).
}
The symmetry factor is unity because it has no loop structure, and the deformation factor is two as it can be only ``twisted" (any ``flip" does not yield an independent diagram). Taking account of the two polarization patterns, 
the amplitude of \ac{GW} spectrum~\eqref{eq: OmegaGW} then reads
\bae{\label{eq: OmegaGauss}
    \Omega^{(2)}_{\text{GW}}(k)&=2^2\times\frac{1}{48}\lr{\frac{k}{aH}}^2\overline{\mathcal{P}_{++}^{\text{Vanilla}}(\tau\to\infty,k)} \nonumber\\
    &=\frac{3A_g^2}{1024}\tilde{k}^2\Theta(2-\tilde{k})\left(\tilde{k}^2-4\right)^2\left(3\tilde{k}^2-2\right)^2 \nonumber \\
    &\qquad\times\left(\pi^2\left(3\tilde{k}^2-2\right)^2\Theta(2\sqrt{3}-3\tilde{k})+\left[4+\left(3\tilde{k}^2-2\right)\ln{\abs{\frac{4}{3\tilde{k}^2}-1}}\right]\right),
}
where $\tilde{k}=k/k_*$, and we used the asymptotic formula~\eqref{eq: I2 average} of the kernel function $I_k$. 
The \ac{GW} spectrum has a sharp peak as one can see in the right panel of Fig.~\ref{fig: GW_Gauss}. 
This is because we assume a monochromatic power spectrum.

\subsubsection{Third-order contributions}

\begin{figure}
    \centering
    \begin{tabular}{c}
        \begin{minipage}{0.5\hsize}
            \centering
            \includegraphics[width=0.8\hsize]{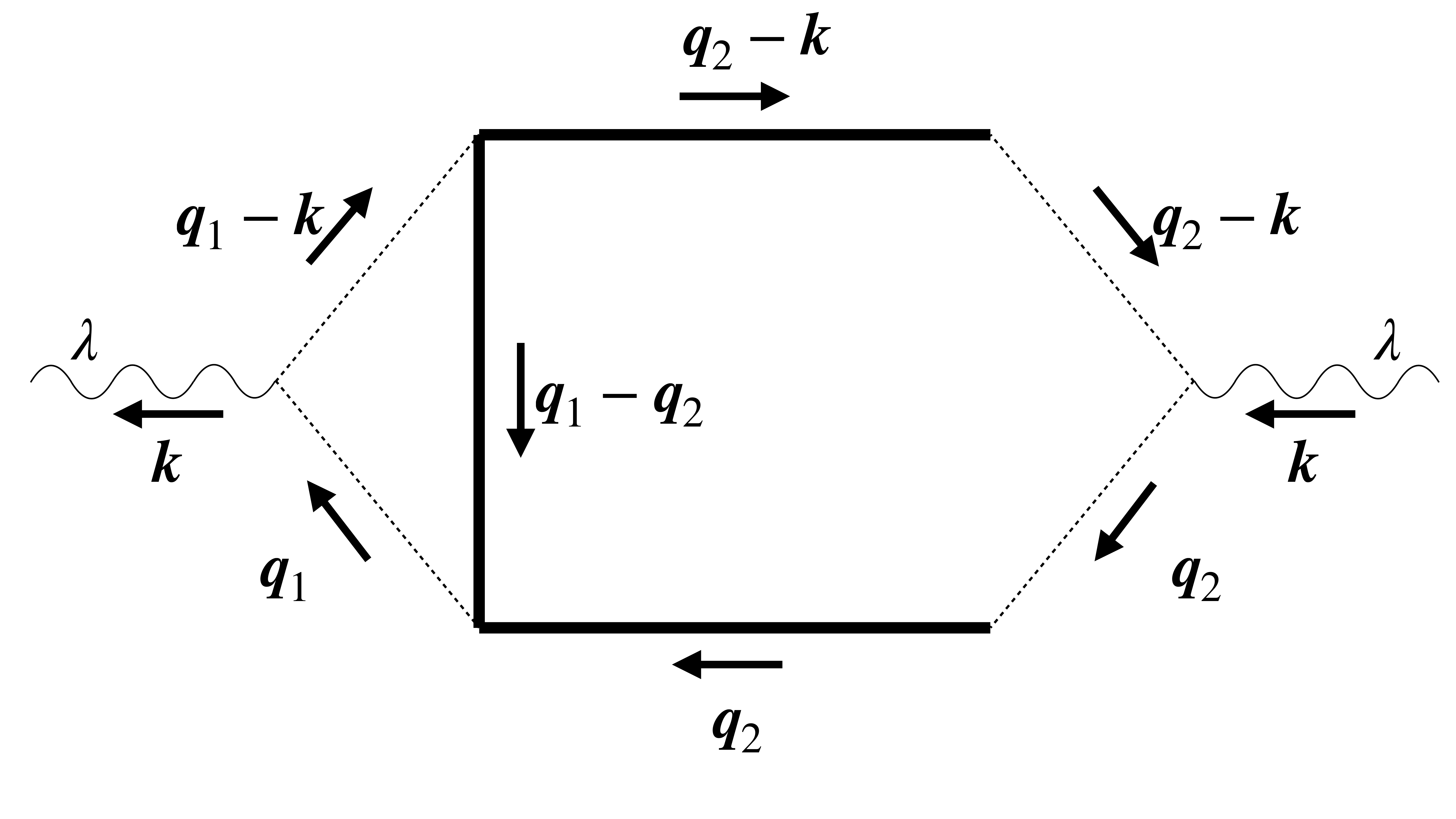}
            \subcaption{C : $P^{\text{C}}_{\lambda\lambda}$ 
            }
        \end{minipage} 
        \begin{minipage}{0.5\hsize}
            \centering
            \includegraphics[width=0.8\hsize]{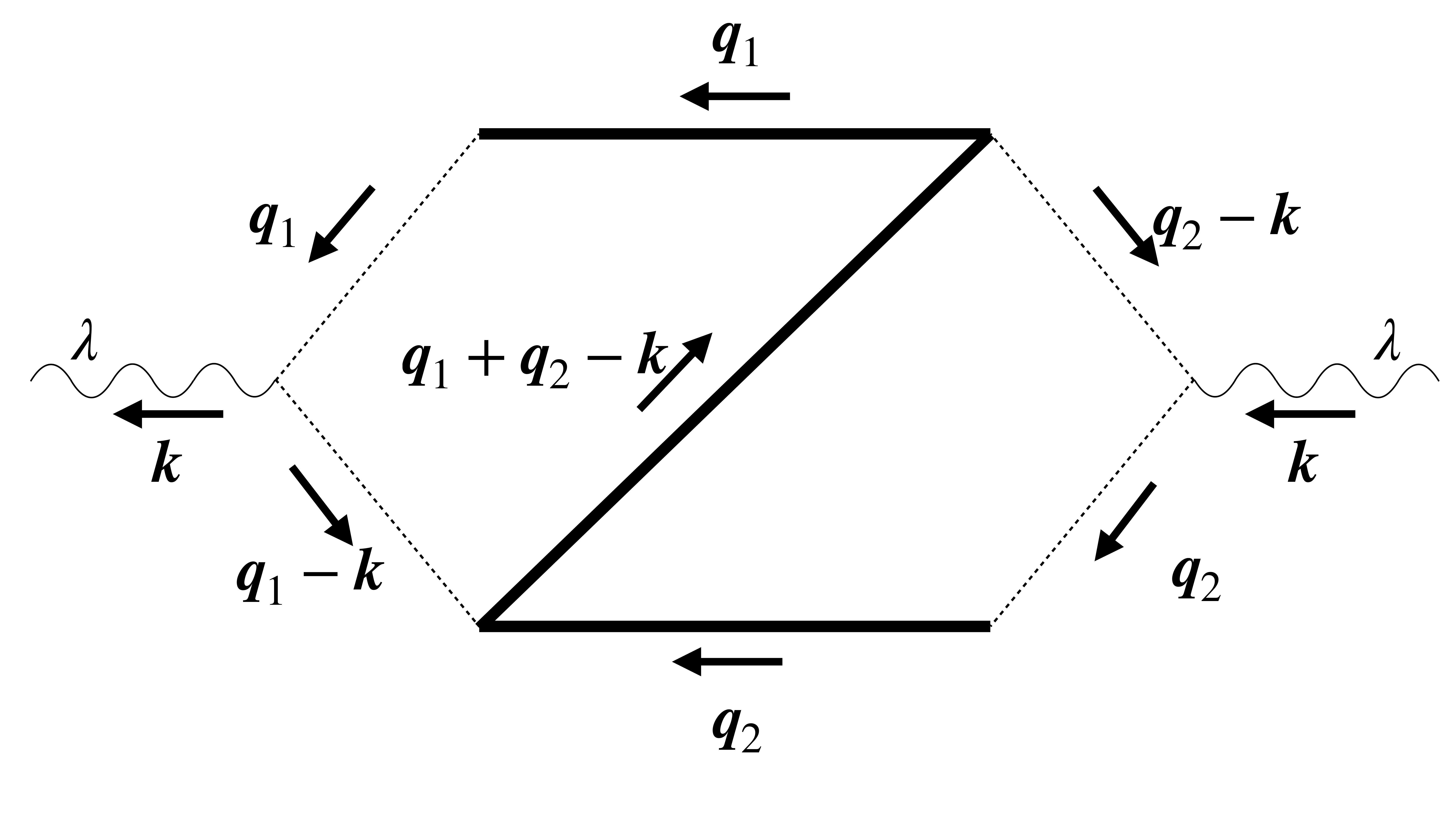}
            \subcaption{Z : $P^{\text{Z}}_{\lambda\lambda}$}
        \end{minipage}\cr\cr
        \begin{minipage}{0.5\hsize}
            \centering
            \includegraphics[width=0.8\hsize]{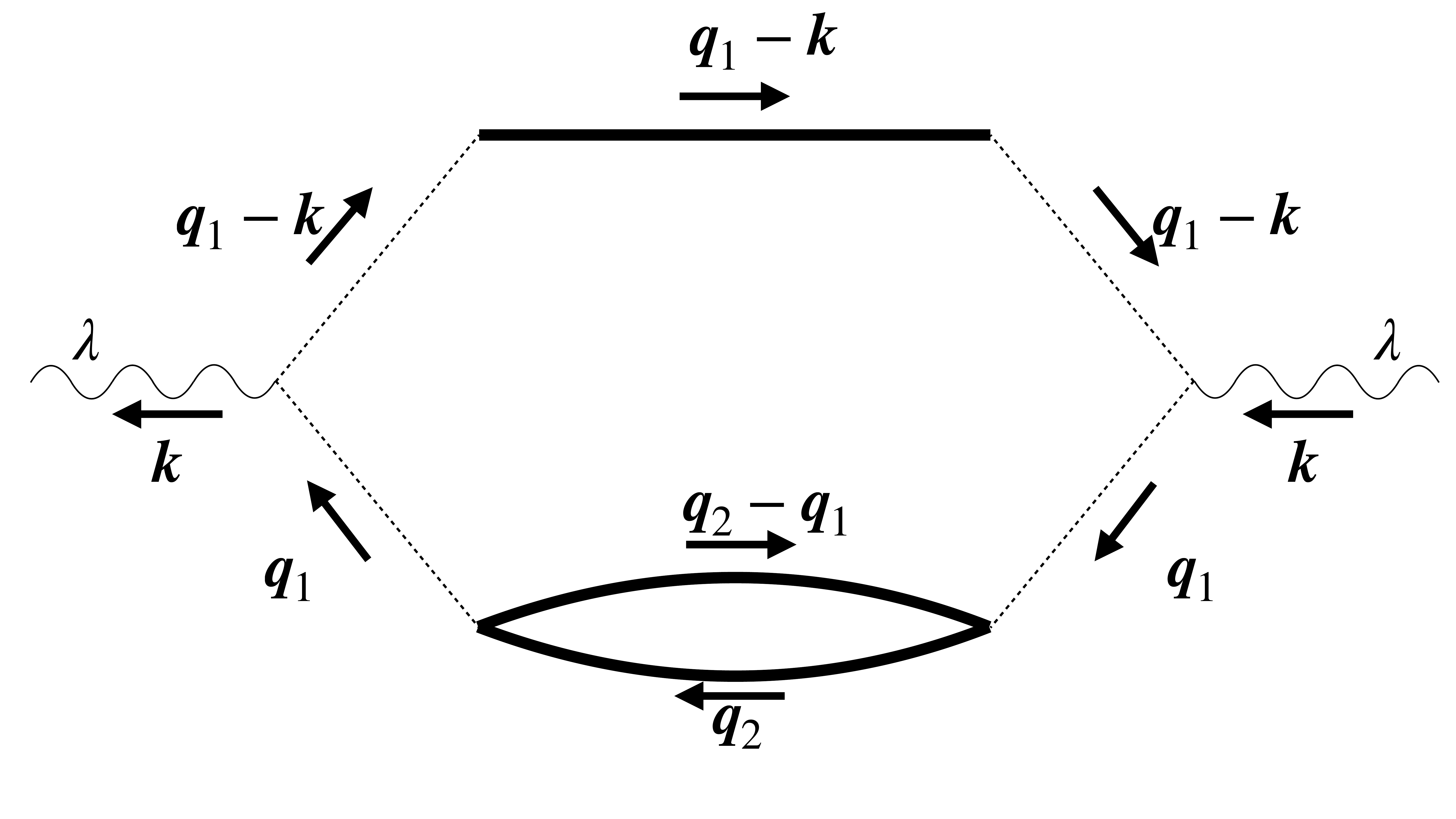}
            \subcaption{1-convolution : $P^{\text{1c}}_{\lambda\lambda}$} 
        \end{minipage} 
        \begin{minipage}{0.5\hsize}
            \centering
            \includegraphics[width=0.8\hsize]{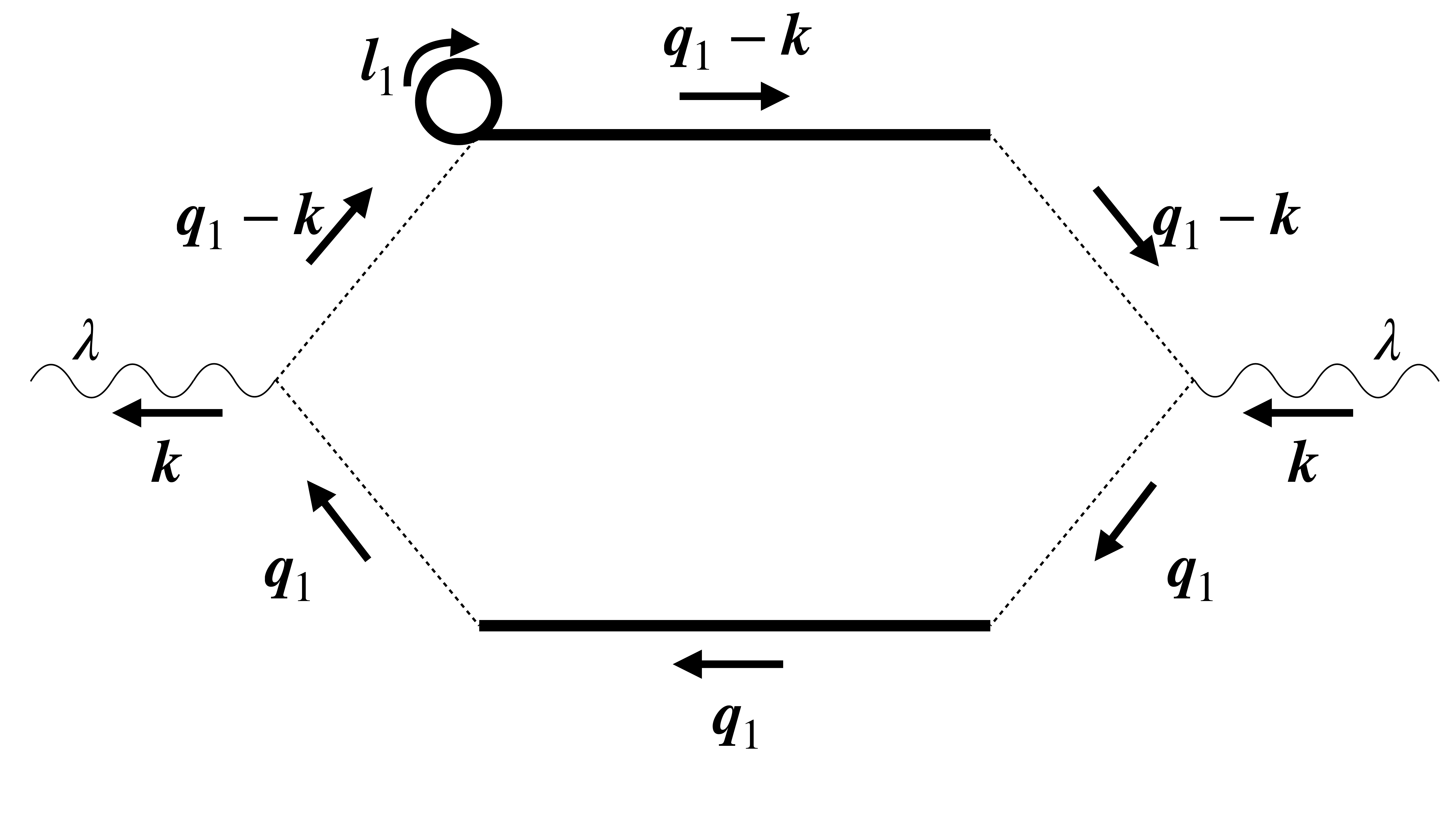}
            \subcaption{
            1-loop : $P^{\text{1$\ell$}}_{\lambda\lambda}$ }
        \end{minipage}
    \end{tabular}
    \caption{Third-order contributions.}
    \label{fig: diag_O(P^3)}
\end{figure}

The third-order contributions ($\propto A_g^3$) are summarized in Fig.~\ref{fig: diag_O(P^3)}. The symmetric factor is unity for the C and Z terms, while it is two for the 1-convolution term and 1-loop term. Hence they are summarized as
\beae{\label{eq: 3rd order P}
    P^{\text{C}}_{\lambda\lambda}(\tau,k)&=(2!F_\NL)^2\calI_{\lambda\lambda}(\tau,\bfk\mid\bfq_1,\bfq_2\mid\bfq_2,\bfk-\bfq_2,\bfq_1-\bfq_2), \\
    P^{\text{Z}}_{\lambda\lambda}(\tau,k)&=(2!F_\NL)^2\calI_{\lambda\lambda}(\tau,\bfk\mid\bfq_1,\bfq_2\mid\bfq_1,\bfq_2,\bfk-\bfq_1-\bfq_2), \\
    P_{\lambda\lambda}^{\text{1c}}(\tau,k)&=
    \frac{(2!F_\NL)^2}{2!}\calI_{\lambda\lambda}(\tau,\bfk\mid\bfq_1,\bfq_1\mid\bfk-\bfq_1,\bfq_2,\bfq_1-\bfq_2), \\
    P^{\text{1$\ell$}}_{\lambda\lambda}(\tau,k)&=\pqty{\frac{3!G_\NL}{2!}\int\frac{\dd[3]l_1}{(2\pi)^3} P_g(l_1)} P_{\lambda\lambda}^{\text{Vanilla}}(k)=3G_\NL A_g P_{\lambda\lambda}^{\text{Vanilla}}.
}
One finds in the 1-loop term that adding self-closed loops to some diagram can be practically realized by multiplying the original diagram by the expansion coefficients and the perturbation amplitude $A_g$.

The deformation factors read $2^2$ for the C, Z, and 1-convolution terms and $2^3$ for the 1-loop term. Therefore, the third-order \ac{GW} spectrum is given by
\bae{
    \Omega^{(3)}_\GW(k)=2\times\frac{1}{48}\pqty{\frac{k}{aH}}^2\bqty{2^2\times\overline{\calP^\text{C}_{++}(k)}+2^2\times\overline{\calP^\text{Z}_{++}(k)}+2^2\times\overline{\calP^\text{1c}_{++}(k)}+2^3\times\overline{\calP^\text{1$\ell$}_{++}(k)}}.
}
Though the integrations cannot be done analytically in contrast to the Vanilla case~\eqref{eq: OmegaGauss}, we show the numerical results in Fig.~\ref{fig: GW_O(P^3)}, which include two polarizations and the deformation factors.
We do not explicitly show the 1-loop term because it is just a constant multiplication of the Vanilla term shown in Fig.~\ref{fig: GW_Gauss}.

\bfe{width=0.6\hsize}{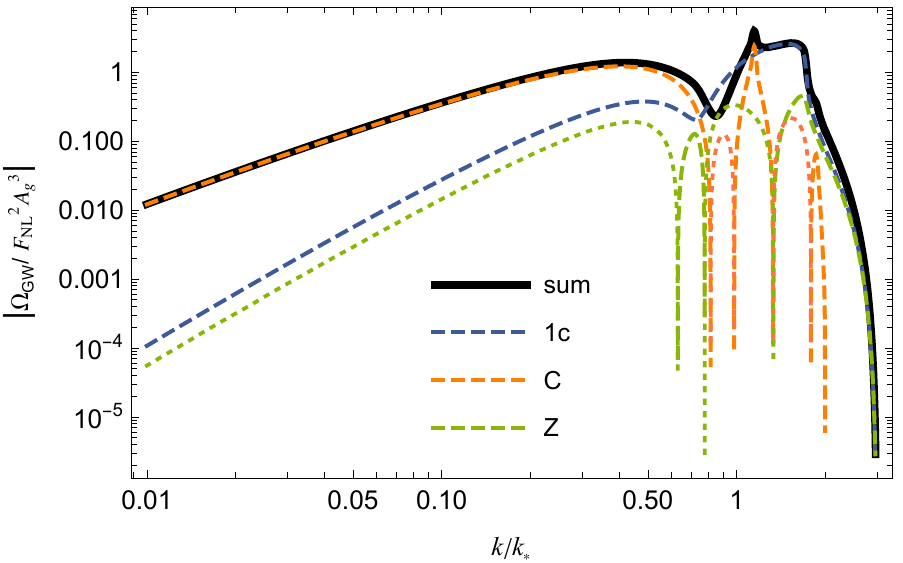}{The normalized GW amplitude of third-order contributions except for the  
1-loop term, which is just a constant multiplication of the Vanilla term shown in Fig.~\ref{fig: GW_Gauss}.
We represent the third-order total amplitude as the black solid line.
Dashed and dotted lines respectively show where the sign of $\Omega_\GW$ is positive and negative.
We note that this plot is including two polarizations and the deformation factors.}{fig: GW_O(P^3)}

\subsubsection{Fourth-order contributions}

\begin{figure}
    \centering
    \begin{tabular}{c}
        \begin{minipage}[t]{0.33\hsize}
            \centering
            \includegraphics[width=0.95\hsize]{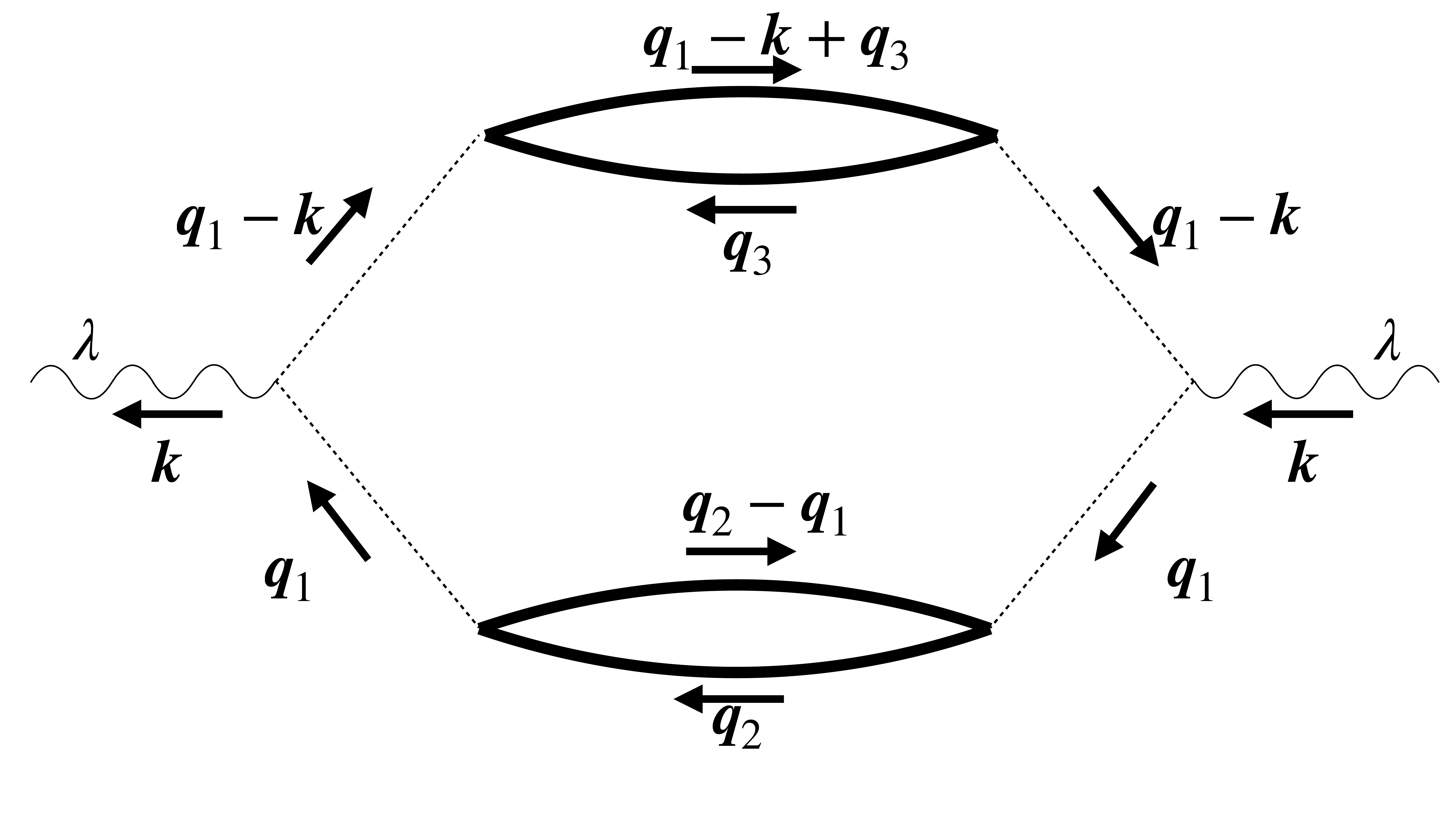}
            \subcaption{
            (1,1)-conv. : $P^{\text{(1,1)c}}_{\lambda\lambda}$}
        \end{minipage} 
        \begin{minipage}[t]{0.33\hsize}
            \centering
            \includegraphics[width=0.95\hsize]{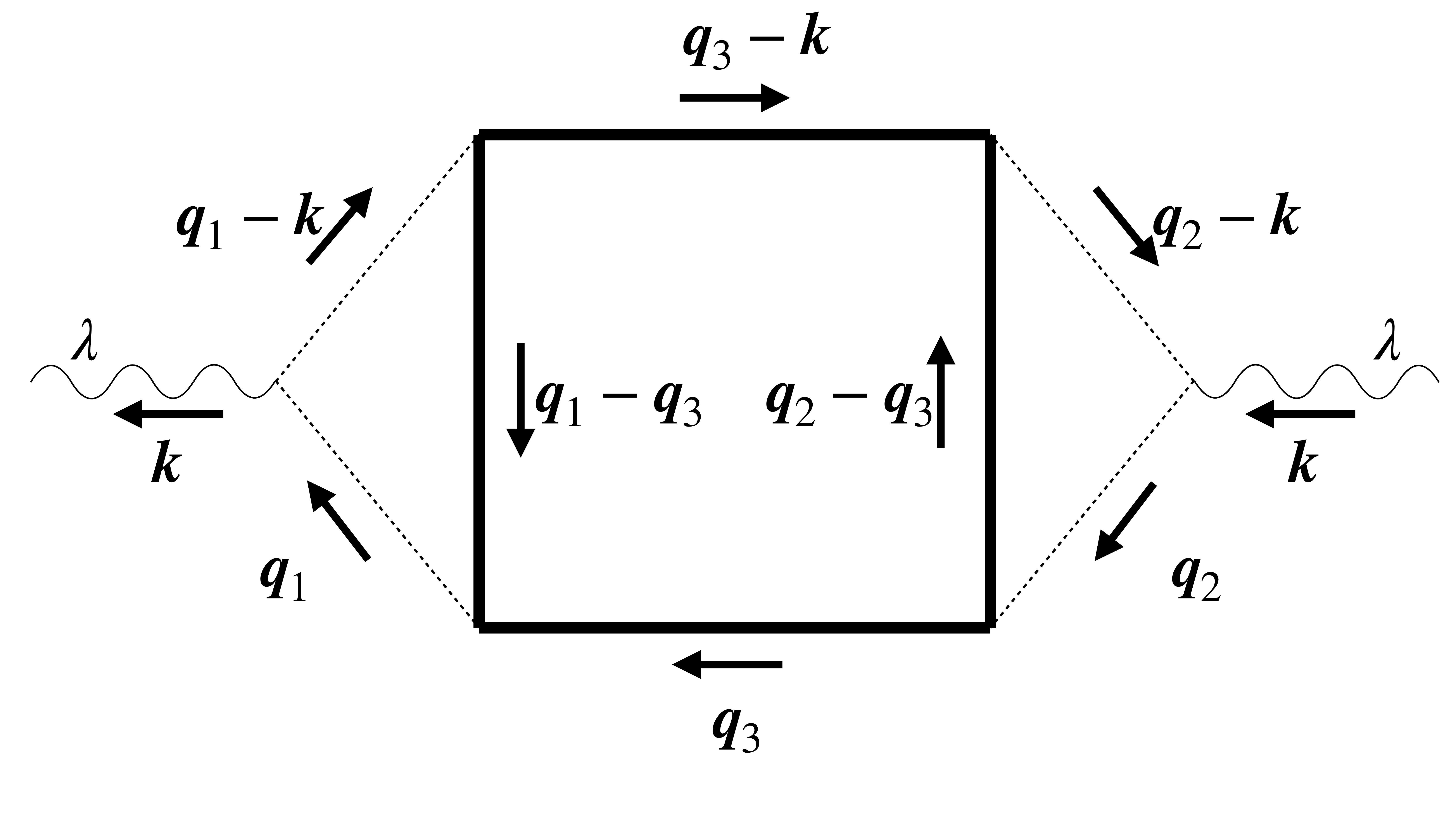}
            \subcaption{Box : $P^{\text{Box}}_{\lambda\lambda}$}
        \end{minipage} 
        \begin{minipage}[t]{0.33\hsize}
            \centering
            \includegraphics[width=0.95\hsize]{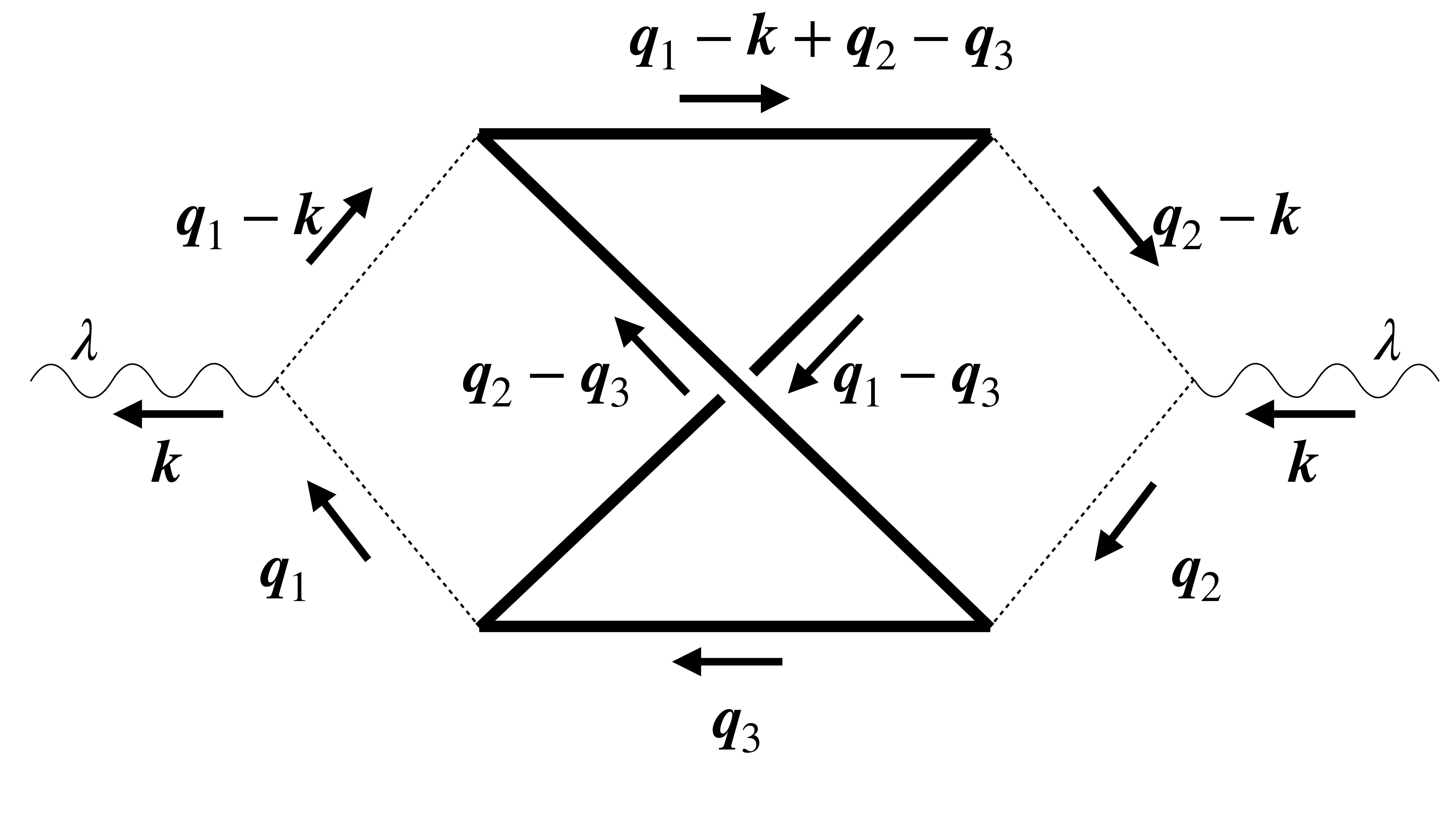}
            \subcaption{X  : $P^{\text{X}}_{\lambda\lambda}$}
        \end{minipage} \cr\cr
        \begin{minipage}[t]{0.33\hsize}
            \centering
            \includegraphics[width=0.95\hsize]{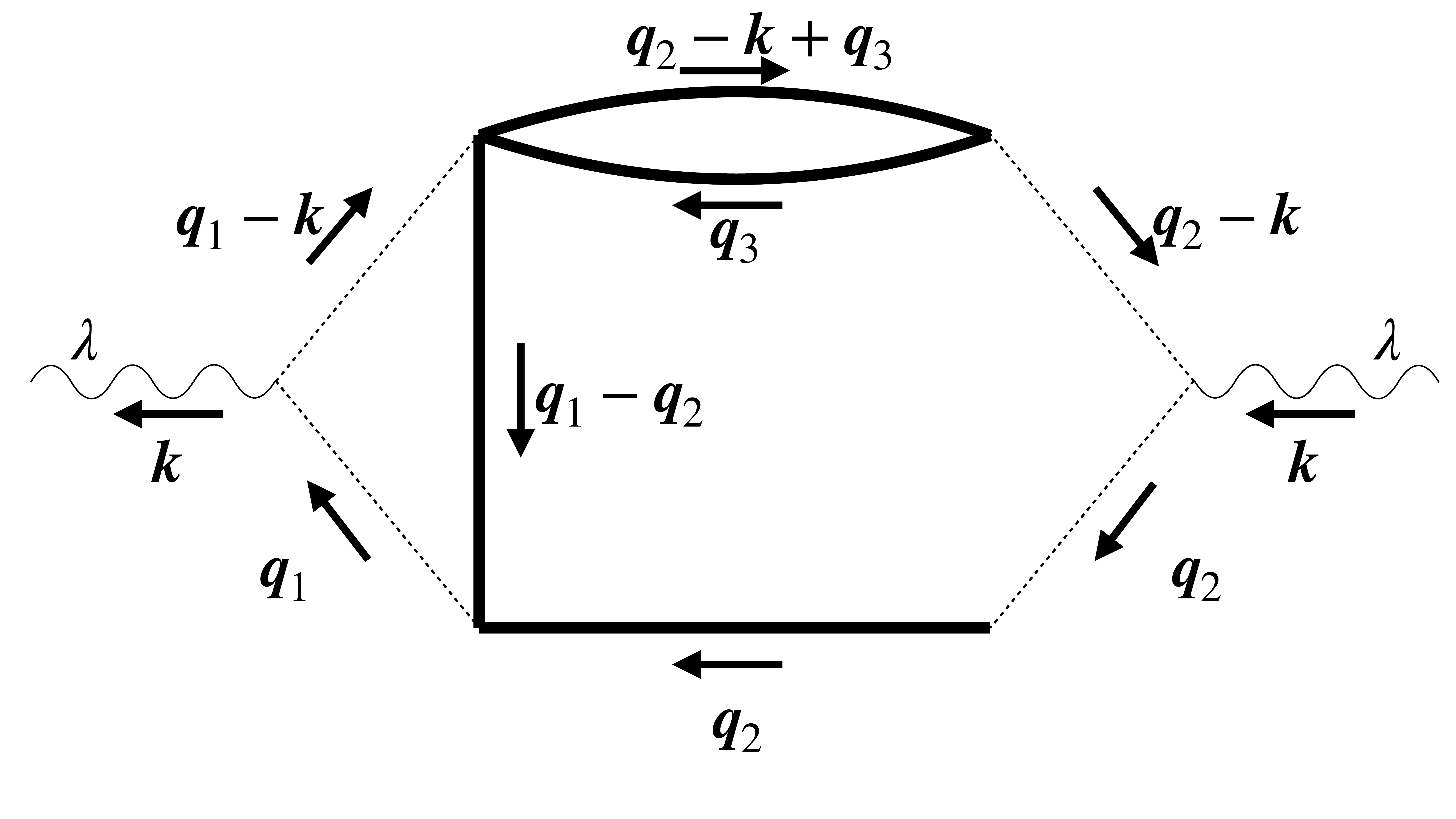}
            \subcaption{1-conv. C type-1 : $P^{\text{1c-C1}}_{\lambda\lambda}$}
        \end{minipage} 
        \begin{minipage}[t]{0.33\hsize}
            \centering
            \includegraphics[width=0.95\hsize]{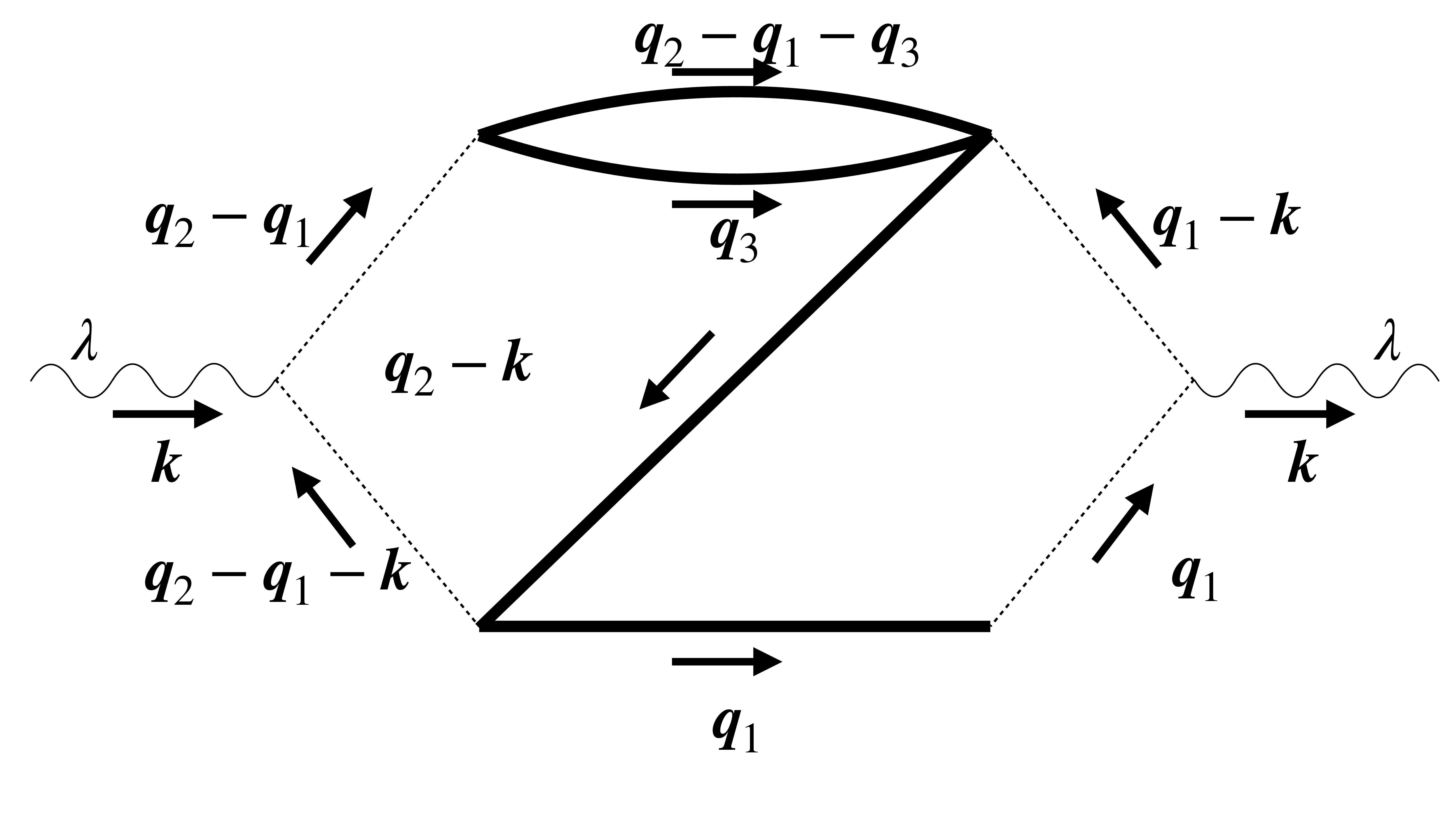}
            \subcaption{1-conv. Z type-1 : $P^{\text{1c-Z1}}_{\lambda\lambda}$}
        \end{minipage} 
        \begin{minipage}[t]{0.33\hsize}
            \centering
            \includegraphics[width=0.95\hsize]{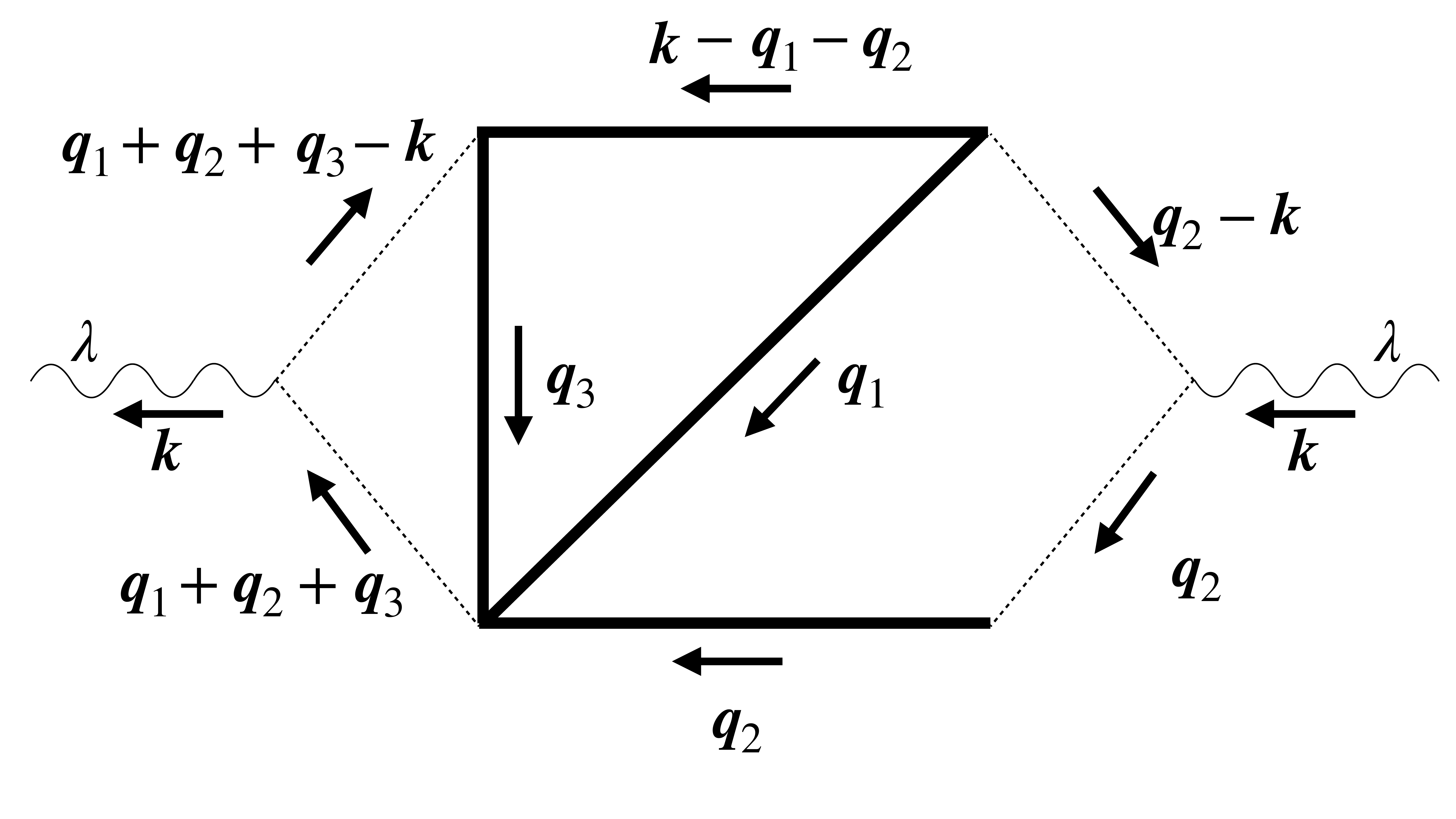}
            \subcaption{CZ : $P^{\text{CZ}}_{\lambda\lambda}$}
        \end{minipage} \cr\cr
        \begin{minipage}[t]{0.33\hsize}
            \centering
            \includegraphics[width=0.95\hsize]{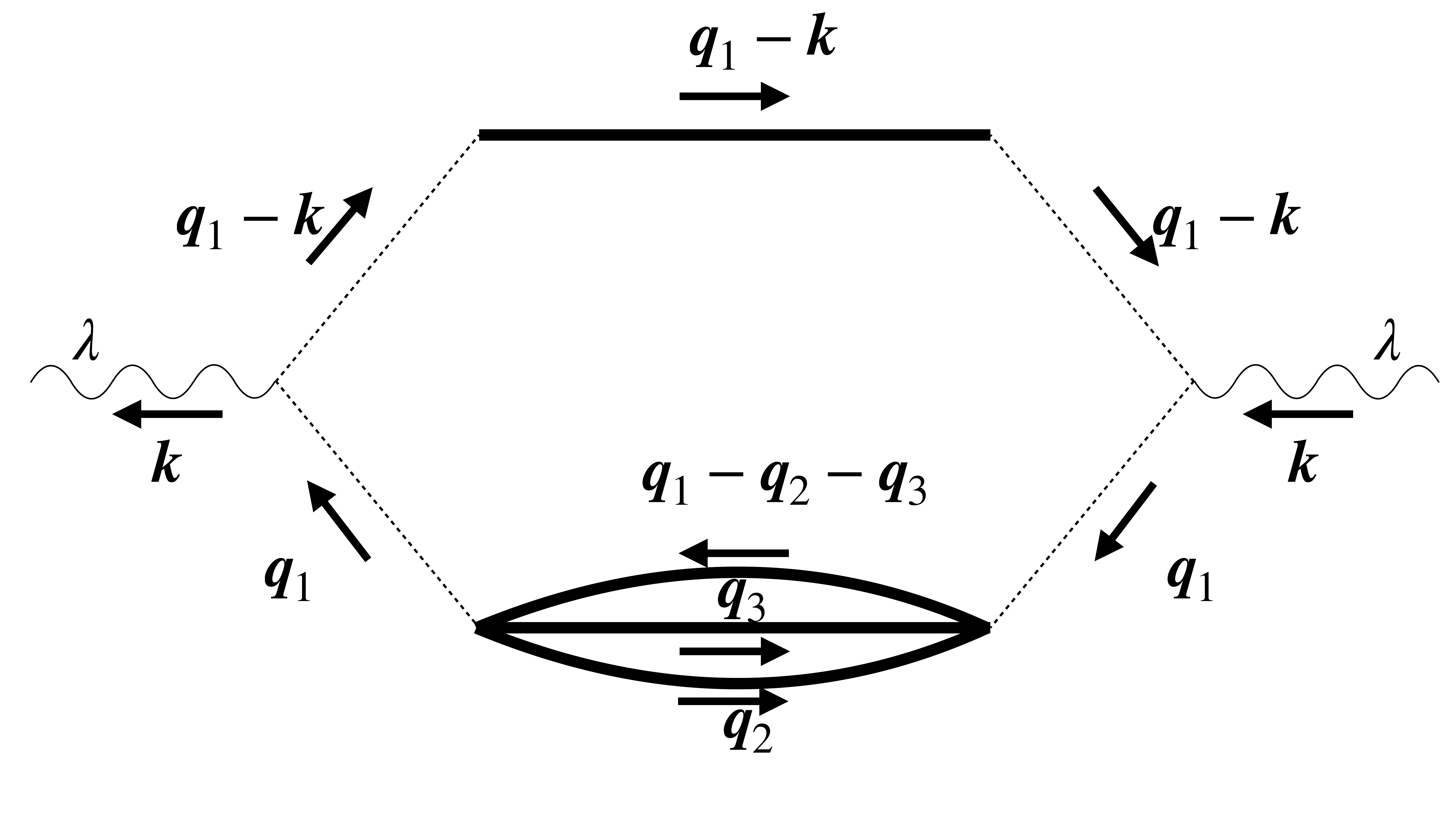}
            \subcaption{
            2-conv. : $P^{\text{2c}}_{\lambda\lambda}$
            }
        \end{minipage} 
        \begin{minipage}[t]{0.33\hsize}
            \centering
            \includegraphics[width=0.95\hsize]{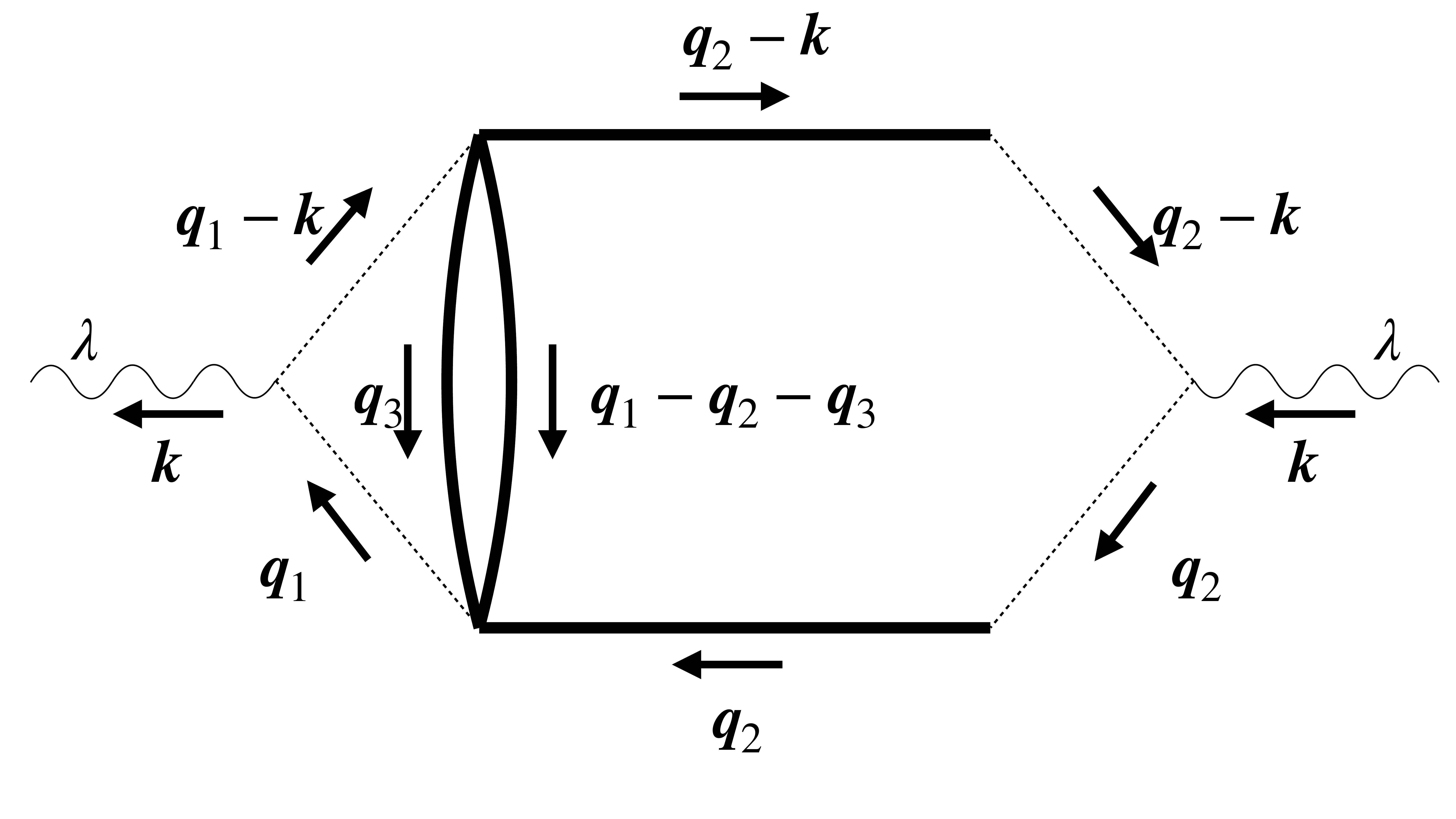}
            \subcaption{1-conv. C type-2 : $P^{\text{1c-C2}}_{\lambda\lambda}$}
        \end{minipage} 
        \begin{minipage}[t]{0.33\hsize}
            \centering
            \includegraphics[width=0.95\hsize]{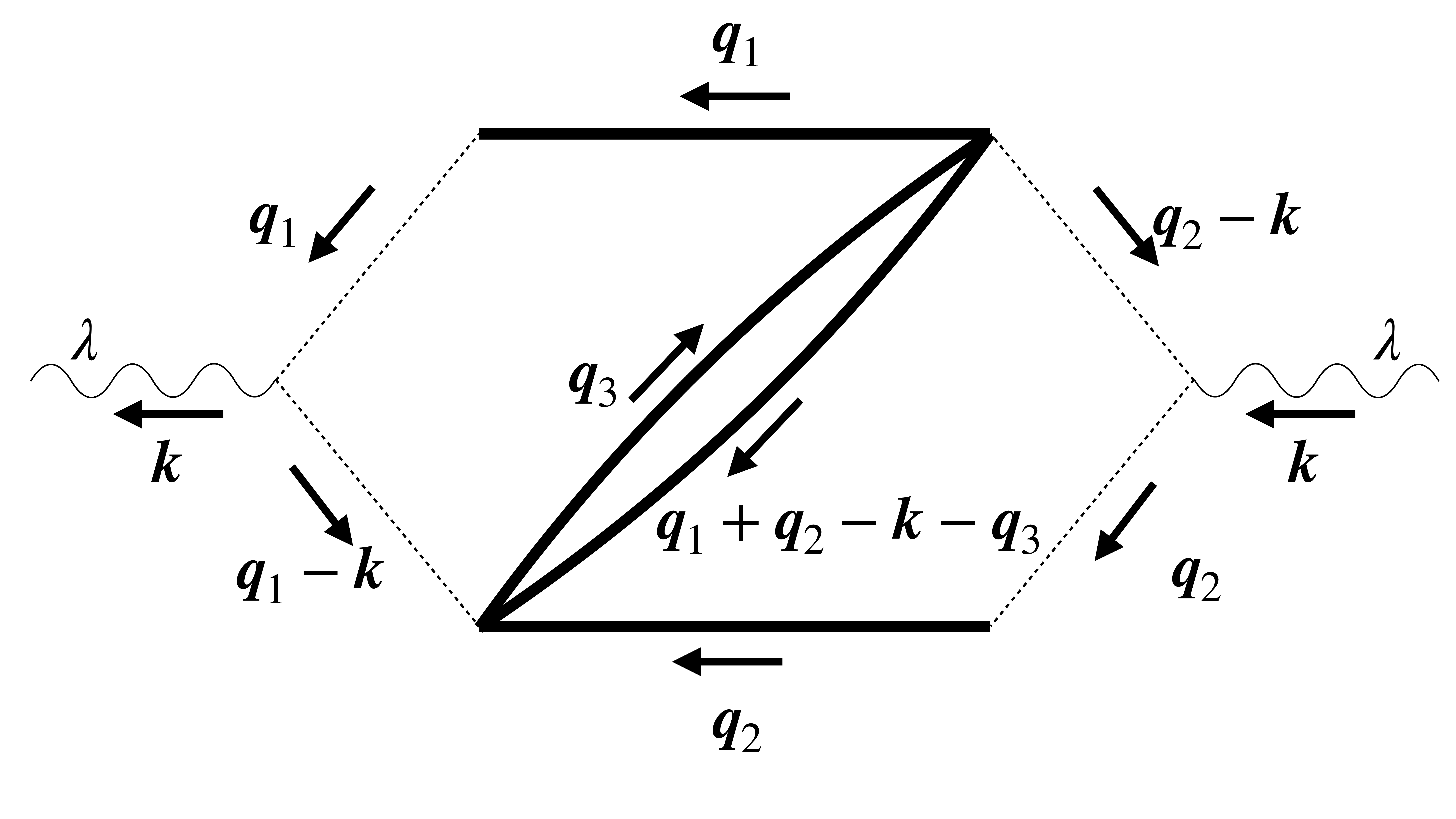}
            \subcaption{1-conv. Z type-2 : $P^{\text{1c-Z2}}_{\lambda\lambda}$}
        \end{minipage}
    \end{tabular}
    \caption{Fourth-order contributions without self-closed loops.}
    \label{fig: diag_fnl^4}
\end{figure}

\begin{figure}
    \centering
    \begin{tabular}{c}
        \begin{minipage}[t]{0.33\hsize}
            \centering
            \includegraphics[width=0.95\hsize]{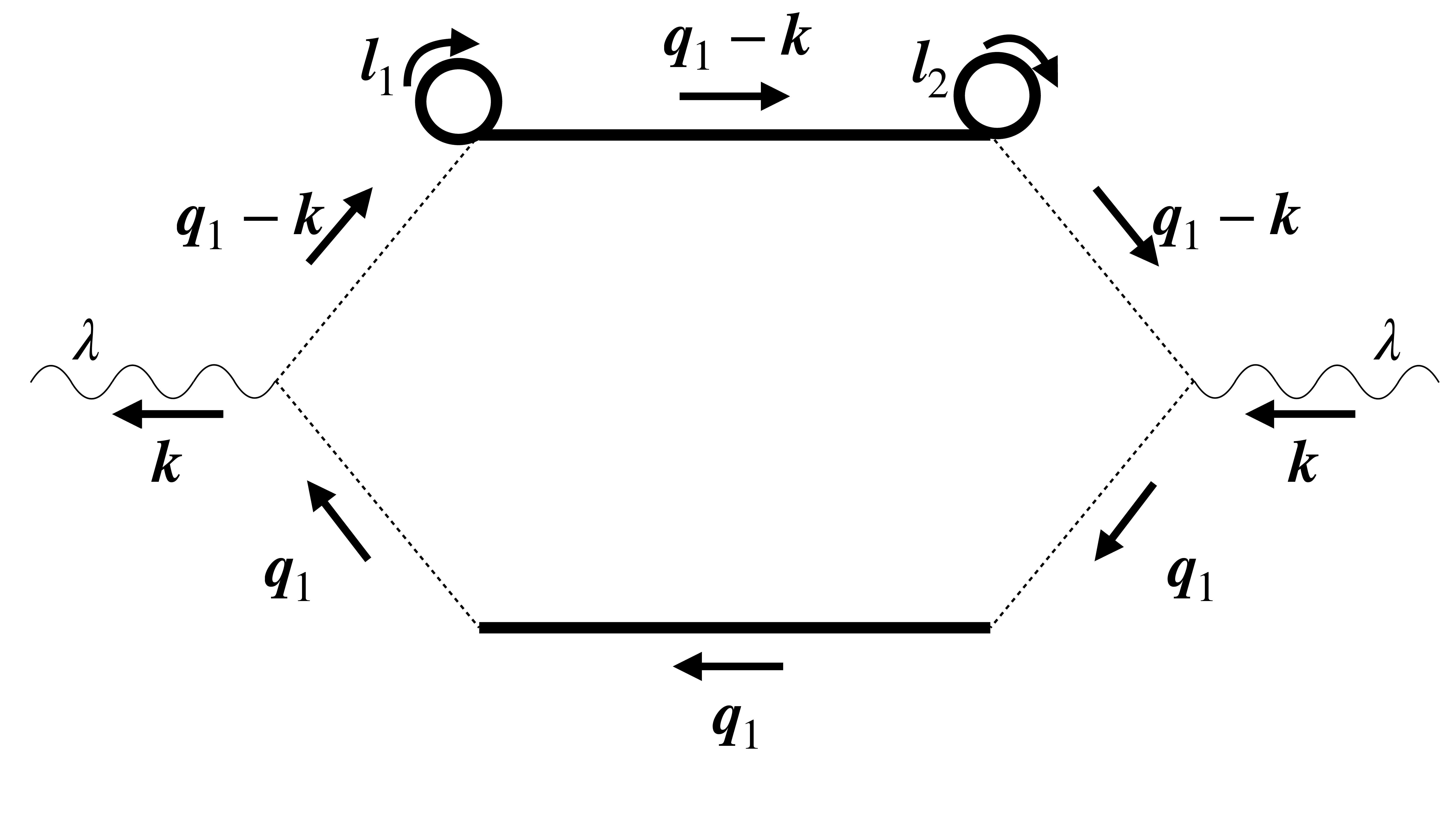}
            \subcaption{\text{(1,1)-loop type-1}~:~$P^{\text{(1,1)$\ell$-1}}_{\lambda\lambda}$}
        \end{minipage} 
        \begin{minipage}[t]{0.33\hsize}
            \centering
            \includegraphics[width=0.95\hsize]{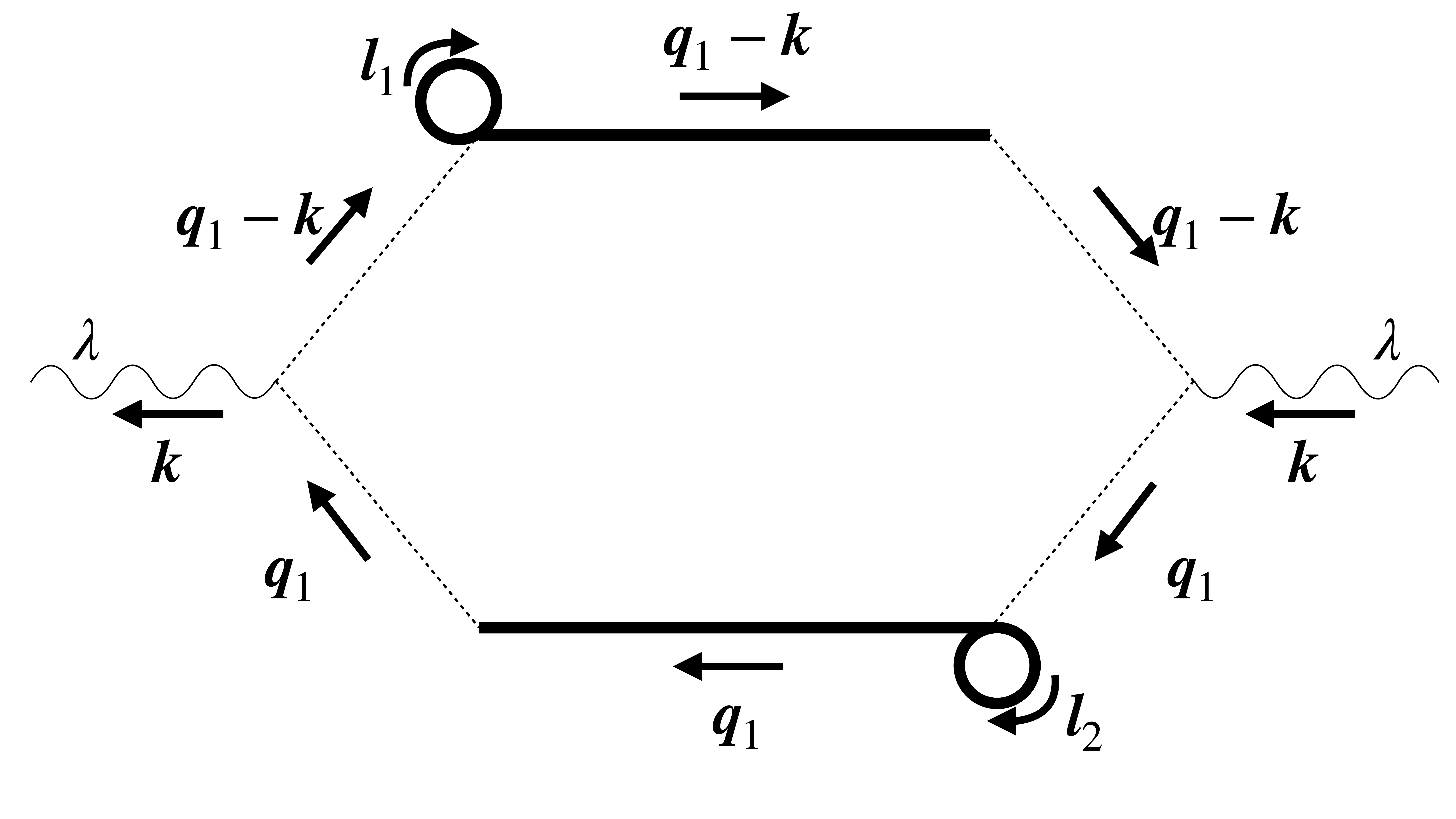}
            \subcaption{
            \text{(1,1)-loop type-2}~:~$P^{\text{(1,1)$\ell$-2}}_{\lambda\lambda}$
            }
        \end{minipage} 
        \begin{minipage}[t]{0.33\hsize}
            \centering
            \includegraphics[width=0.95\hsize]{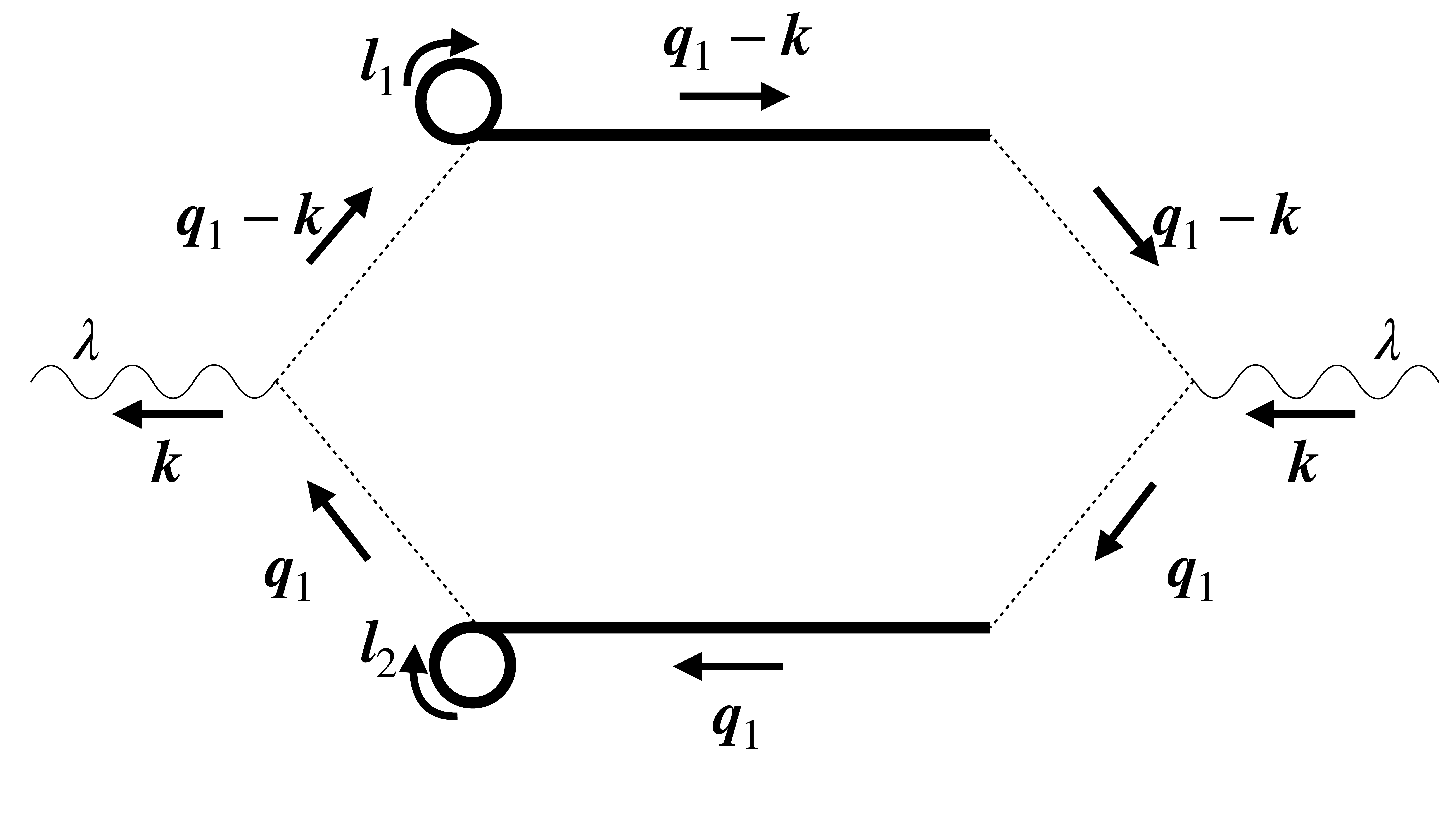}
            \subcaption{
            \text{(1,1)-loop type-3}~:~$P^{\text{(1,1)$\ell$-3}}_{\lambda\lambda}$
            }
        \end{minipage}
        \cr\cr
        \begin{minipage}[t]{0.33\hsize}
            \centering
            \includegraphics[width=0.95\hsize]{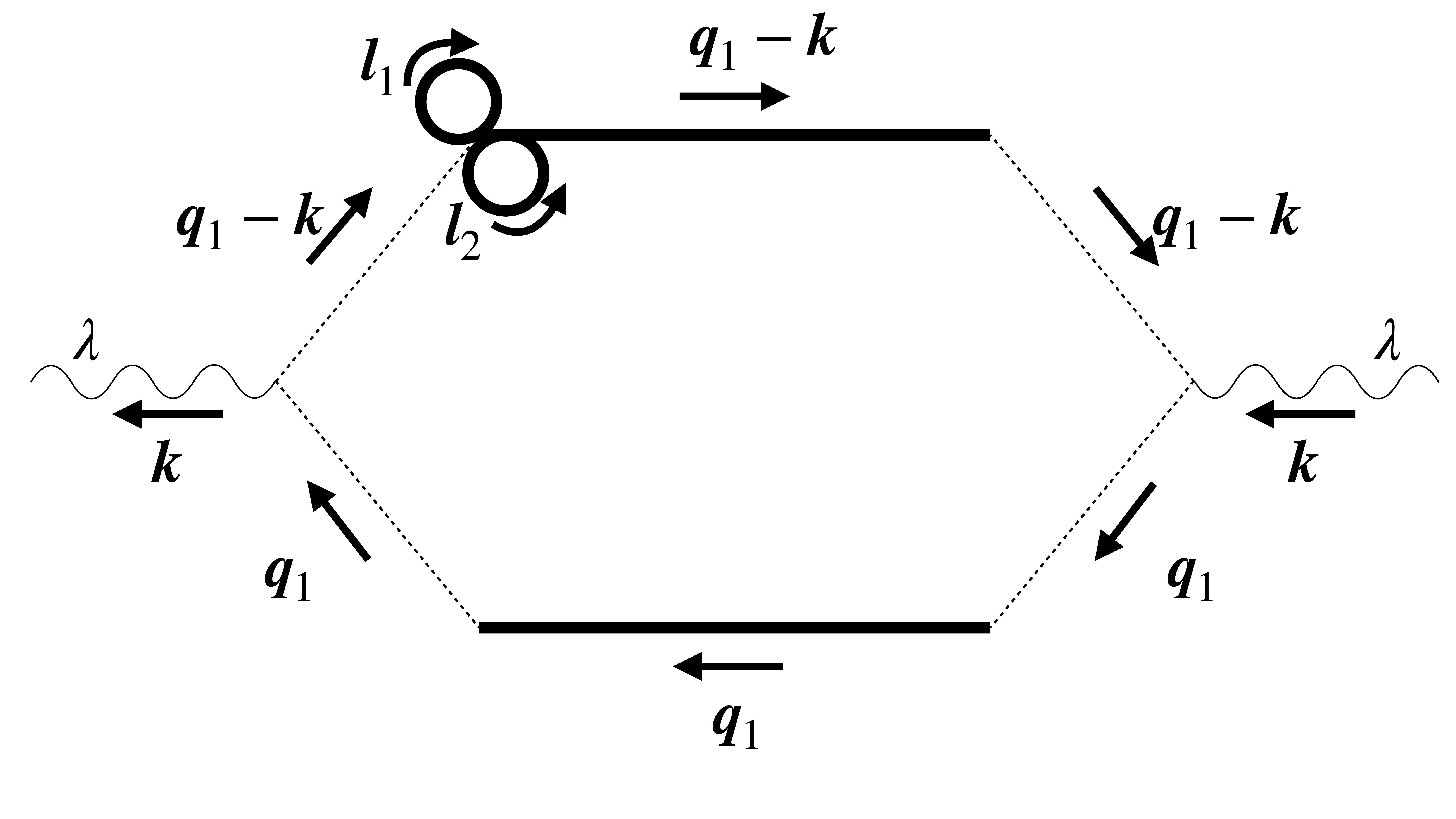}
            \subcaption{
            \text{2-loop}~:~$P^{\text{2$\ell$}}_{\lambda\lambda}$
            }
        \end{minipage} 
        \begin{minipage}[t]{0.33\hsize}
            \centering
            \includegraphics[width=0.95\hsize]{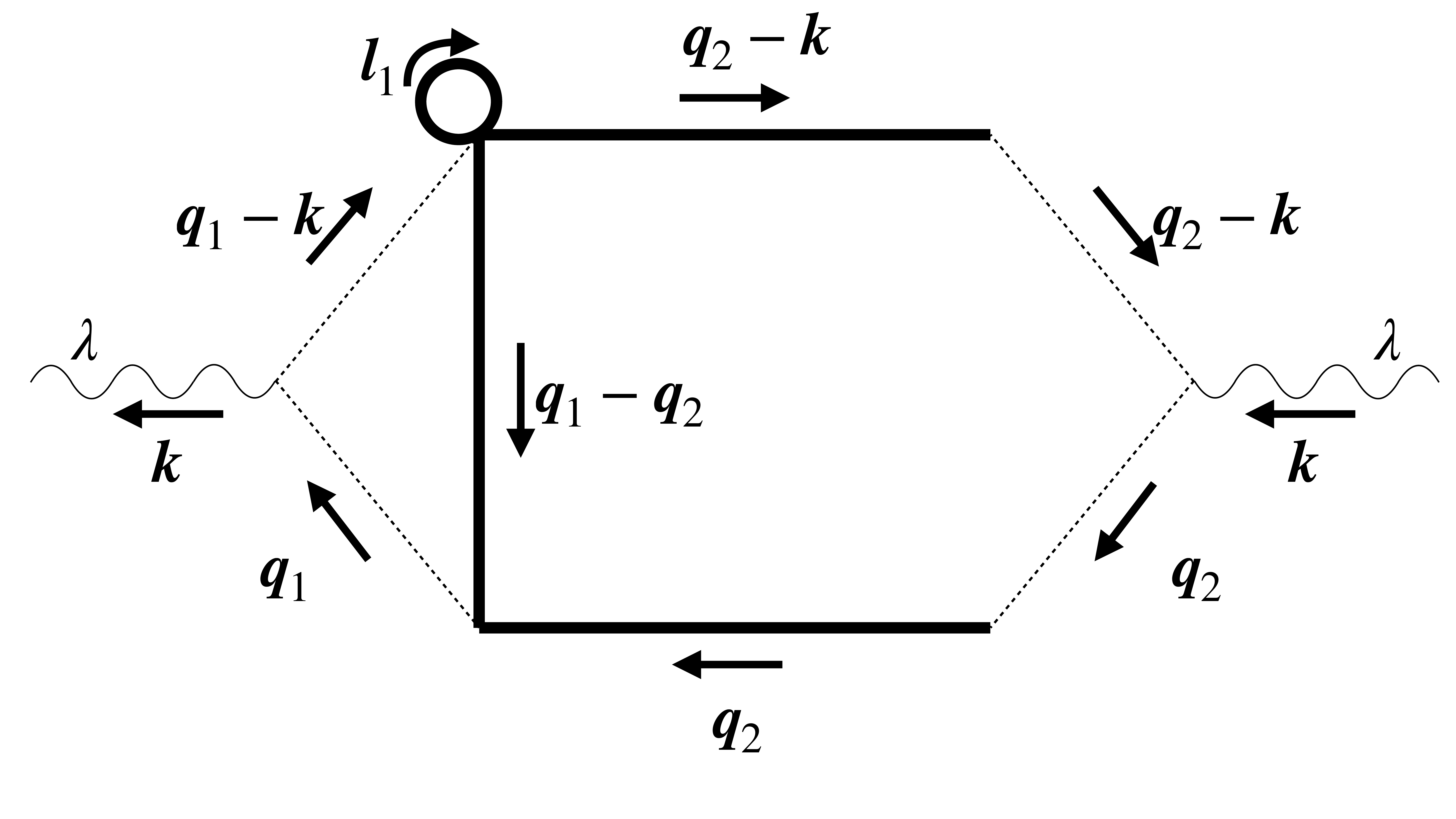}
            \subcaption{\text{1-loop C type-1}~:~$P^{\text{1$\ell$-C1}}_{\lambda\lambda}$  
            }
        \end{minipage} 
        \begin{minipage}[t]{0.33\hsize}
            \centering
            \includegraphics[width=0.95\hsize]{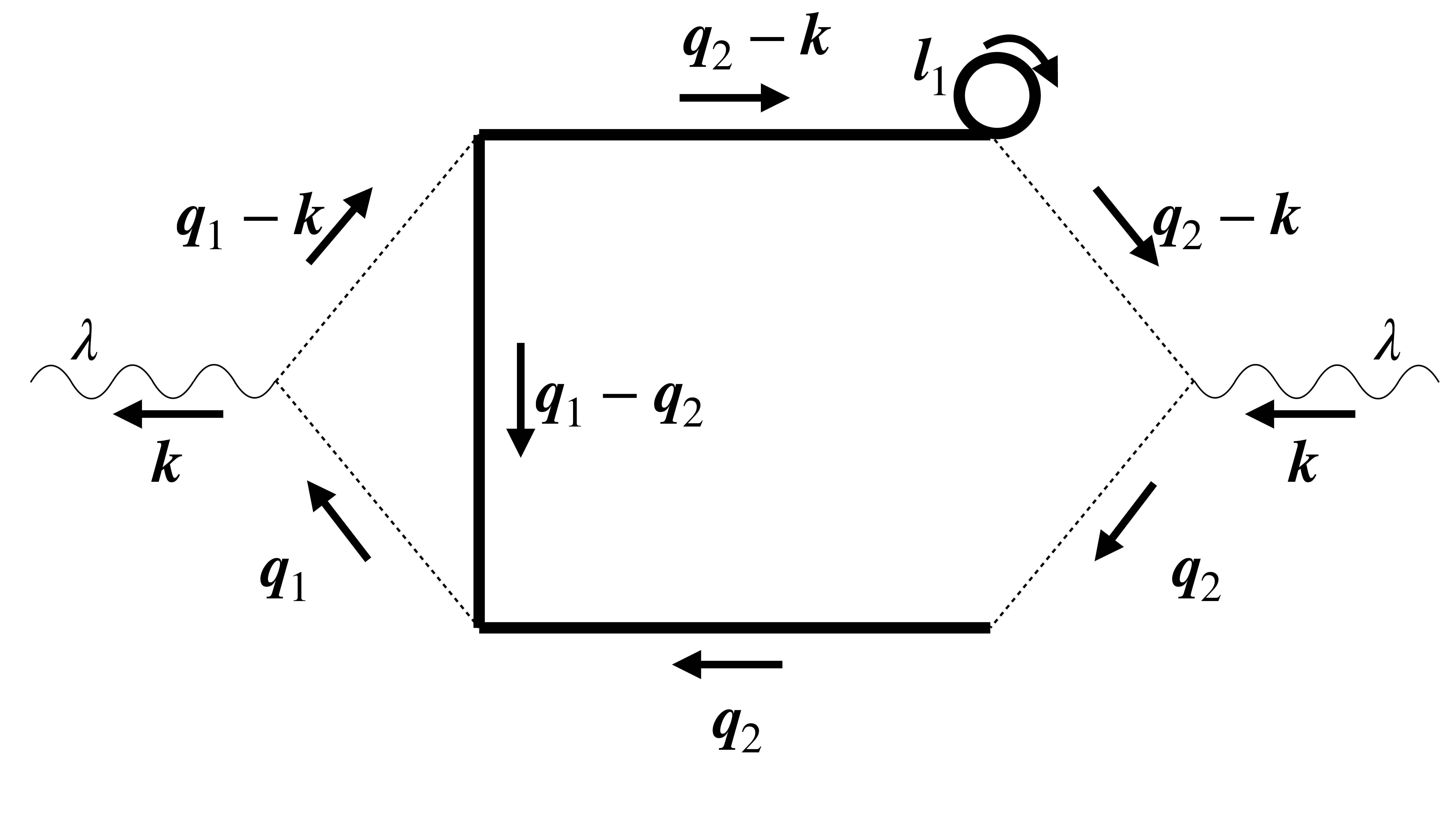}
            \subcaption{\text{1-loop C type-2}~:~$P^{\text{1$\ell$-C2}}_{\lambda\lambda}$
            }
        \end{minipage} \cr\cr
        \begin{minipage}[t]{0.33\hsize}
            \centering
            \includegraphics[width=0.95\hsize]{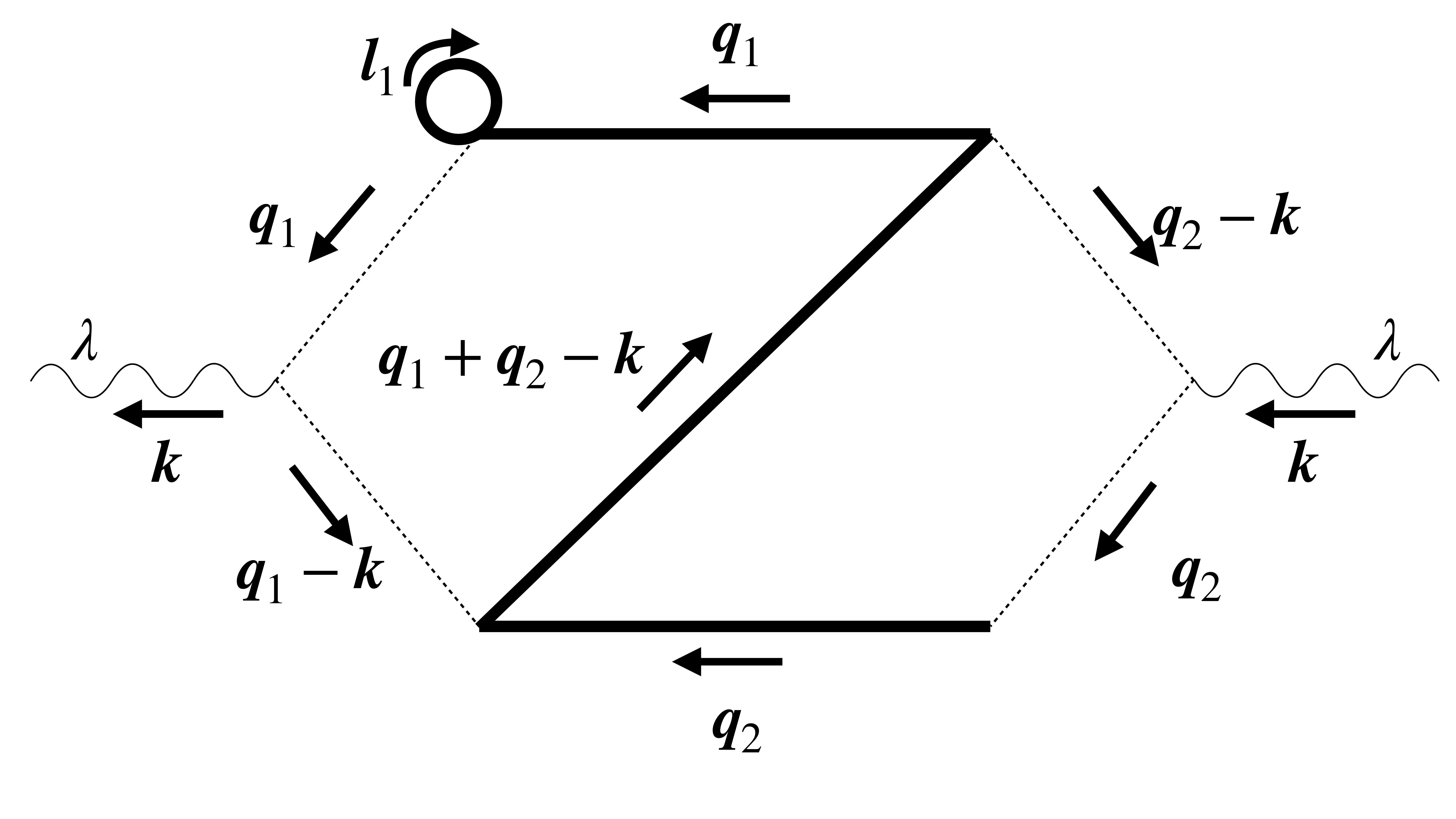}
            \subcaption{\text{1-loop Z type-1}~:~$P^{\text{1$\ell$-Z1}}_{\lambda\lambda}$
            }
        \end{minipage} 
        \begin{minipage}[t]{0.33\hsize}
            \centering
            \includegraphics[width=0.95\hsize]{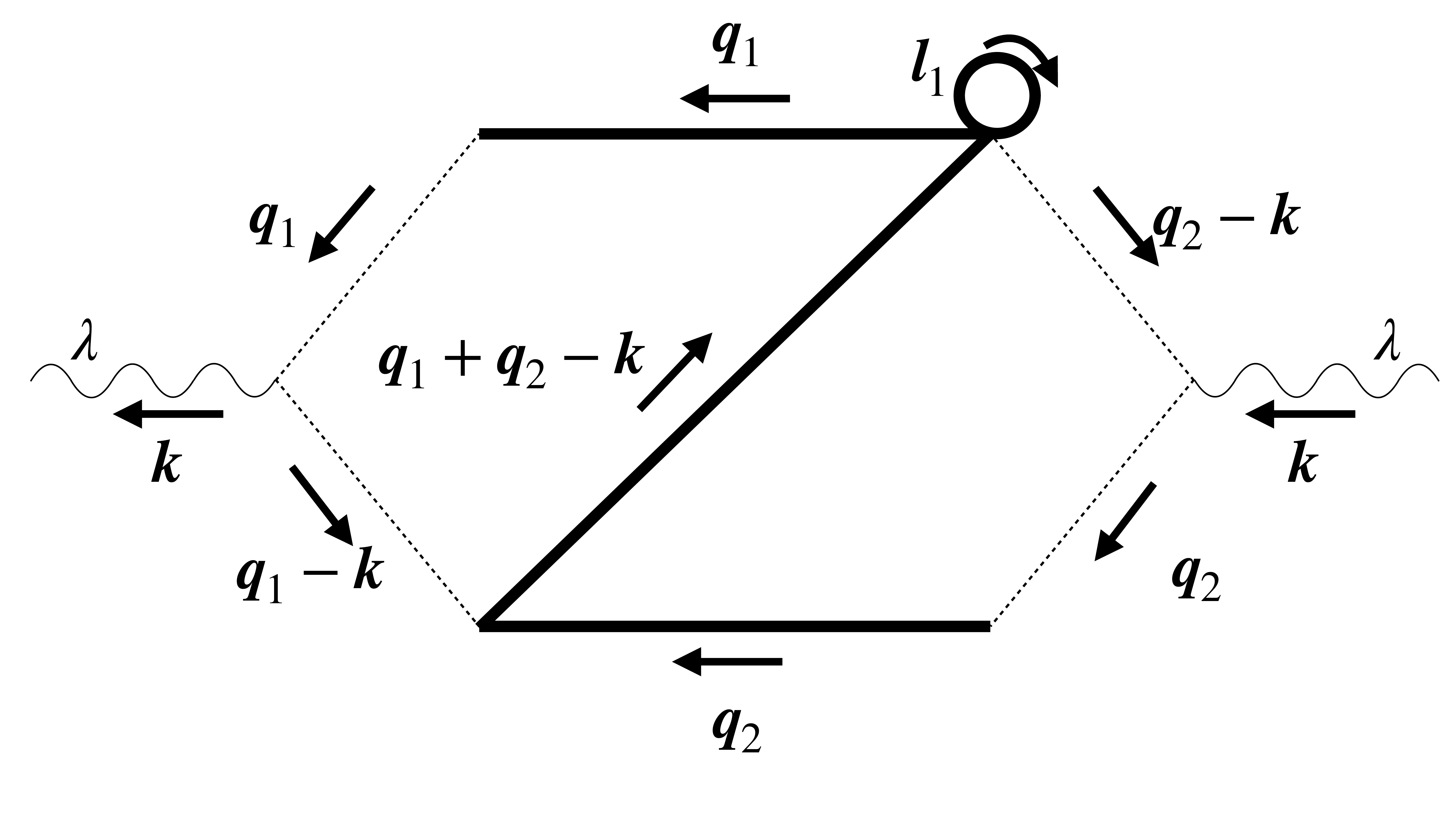}
            \subcaption{\text{1-loop Z type-2}~:~$P^{\text{1$\ell$-Z2}}_{\lambda\lambda}$
            }
        \end{minipage} 
        \begin{minipage}[t]{0.33\hsize}
            \centering
            \includegraphics[width=0.95\hsize]{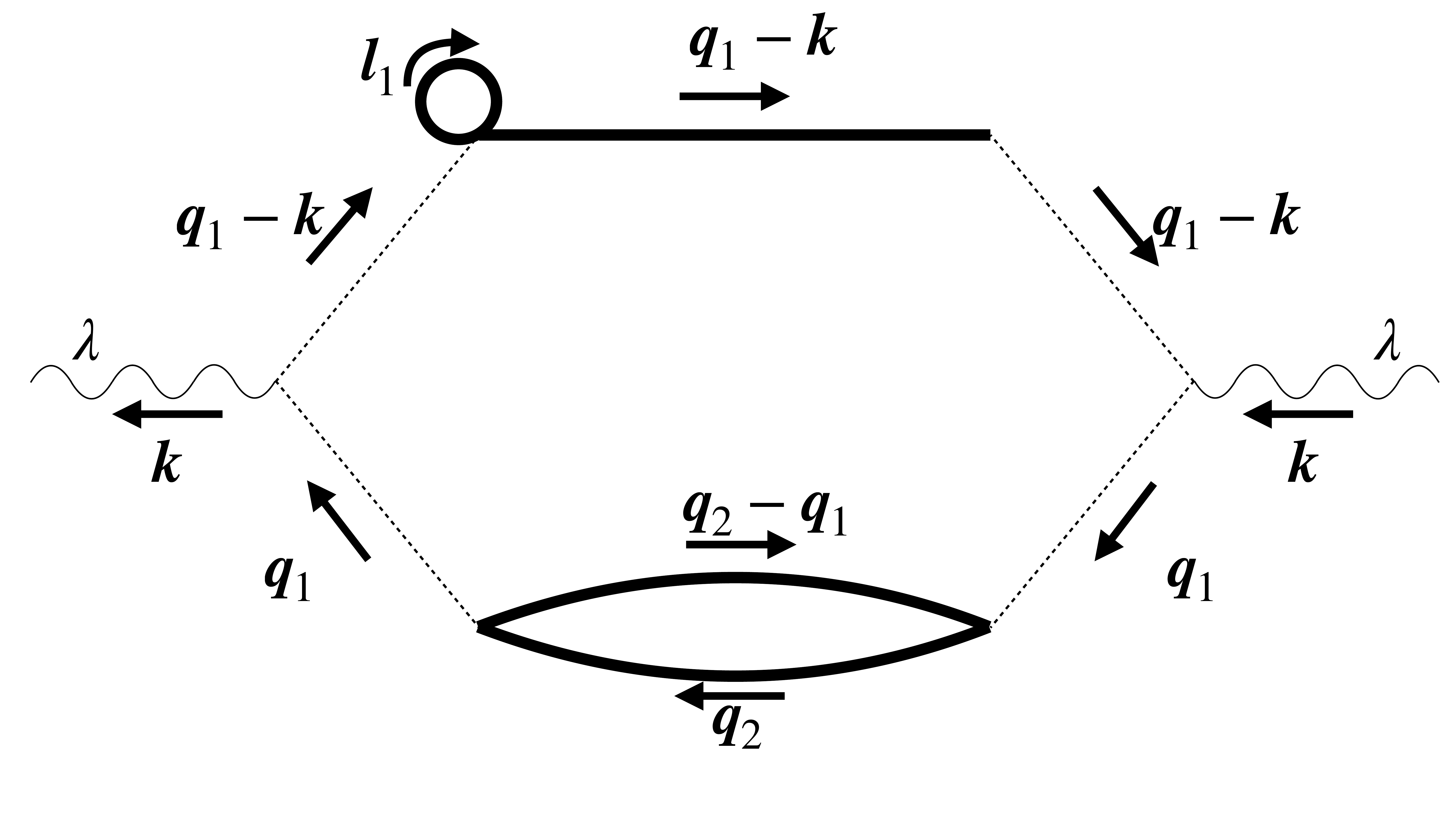}
            \subcaption{
            \text{1-loop 1-conv. type-1}~:~$P^{\text{1$\ell$1c-1}}_{\lambda\lambda}$
            }
        \end{minipage}\cr\cr 
        \begin{minipage}[t]{0.33\hsize}
            \centering
            \includegraphics[width=0.95\hsize]{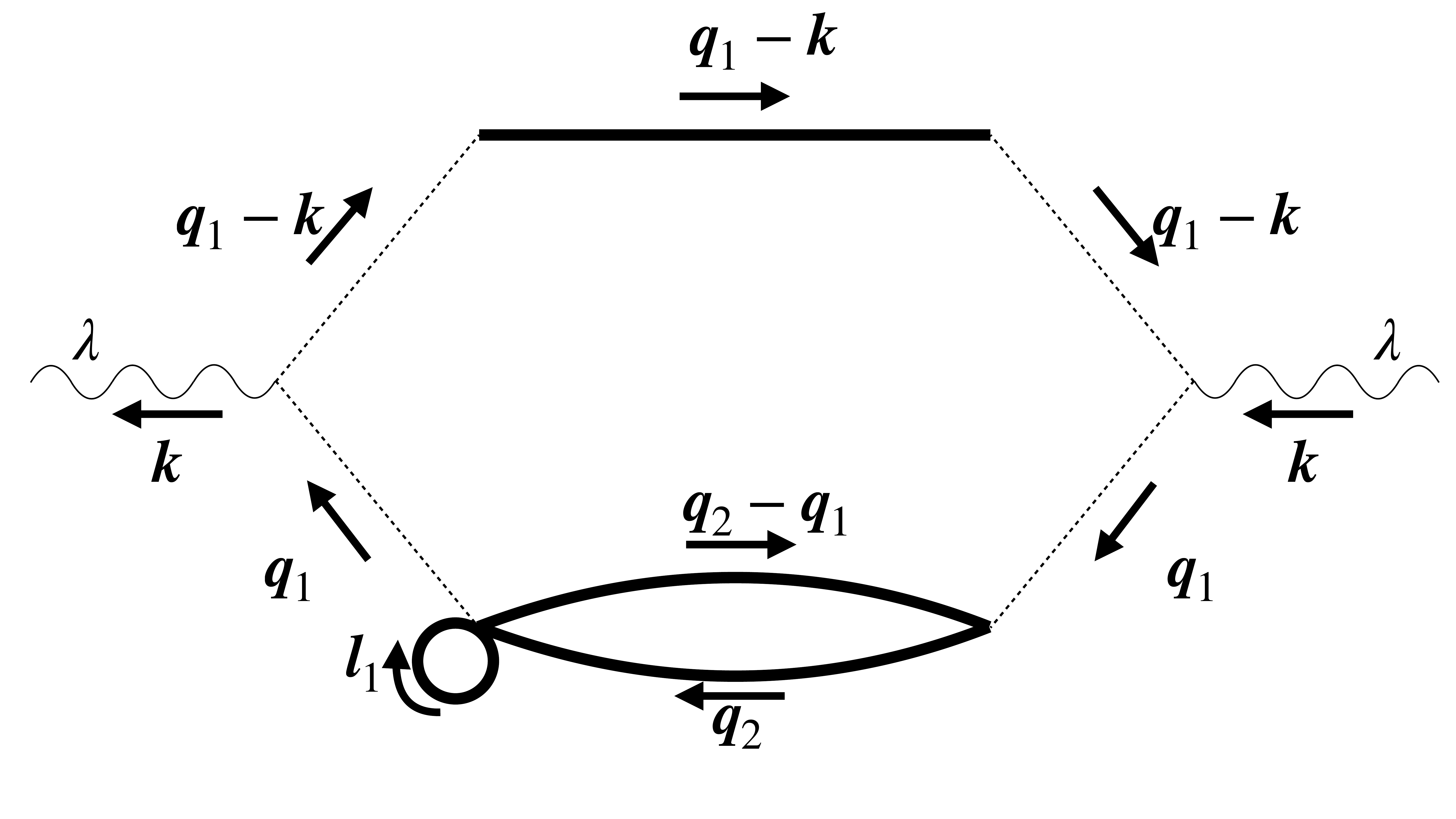}
            \subcaption{
            \text{1-loop 1-conv. type-2}~:~$P^{\text{1$\ell$1c-2}}_{\lambda\lambda}$
            }
        \end{minipage} 
    \end{tabular}
    \caption{Fourth-order contributions with self-closed loops.}
    \label{fig: diag_loop}
\end{figure}

Fourth-order contributions ($\propto A_g^4$) are summarized in Figs.~\ref{fig: diag_fnl^4} and \ref{fig: diag_loop}.
(1,1)-conv, Box, and X terms have been provided in ~\cite{Adshead:2021hnm} and 2-conv term has been introduced by ~\cite{Yuan:2020iwf}, while other contributions in Fig.~\ref{fig: diag_fnl^4} and all contributions in Fig.~\ref{fig: diag_loop} are our new findings.  They read 
\beae{
    P^{\text{(1,1)c}}_{\lambda\lambda}(\tau,k)&=\frac{(2!F_\NL)^4}{(2!)^2}\calI_{\lambda\lambda}(\tau,\bfk\mid\bfq_1,\bfq_1\mid\bfq_1-\bfk+\bfq_3,\bfq_3,\bfq_2,\bfq_2-\bfq_1), \\
    P^\text{Box}_{\lambda\lambda}(\tau,k)&=(2!F_\NL)^4\calI_{\lambda\lambda}(\tau,\bfk\mid\bfq_1,\bfq_2\mid\bfq_1-\bfq_3,\bfq_2-\bfq_3,\bfq_3,\bfq_3-\bfk), \\
    P^\text{X}_{\lambda\lambda}(\tau,k)&=(2!F_\NL)^4\calI_{\lambda\lambda}(\tau,\bfk\mid\bfq_1,\bfq_2\mid\bfq_1-\bfk+\bfq_2-\bfq_3,\bfq_1-\bfq_3,\bfq_2-\bfq_3,\bfq_3),
}
for ones proportional to $F_\NL^4$,
\beae{
    P^{\text{1c-C1}}_{\lambda\lambda}(\tau,k)&=\frac{(2!F_\NL)^23!G_\NL}{2!}\calI_{\lambda\lambda}(\tau,\bfk\mid\bfq_1,\bfq_2\mid\bfq_1-\bfq_2,\bfq_2,\bfq_3,\bfq_2-\bfk+\bfq_3), \\
    P^{\text{1c-Z1}}_{\lambda\lambda}(\tau,k)&=\frac{(2!F_\NL)^23!G_\NL}{2!}\calI_{\lambda\lambda}(\tau,\bfk\mid\bfq_1,\bfq_2-\bfq_1\mid \bfq_1,\bfq_2-\bfk,\bfq_3,\bfq_2-\bfq_1-\bfq_3), \\
    P^\text{CZ}_{\lambda\lambda}(\tau,k)&=(2!F_\NL)^23!G_\NL\calI_{\lambda\lambda}(\tau,\bfk\mid\bfq_1+\bfq_2+\bfq_3,\bfq_2\mid\bfq_1,\bfq_2,\bfq_3,\bfk-\bfq_1-\bfq_2),
}
for ones proportional to $F_\NL^2G_\NL$,\footnote{Here we note that for the 1c-Z1 and CZ terms the assignment of $\bfq_1$ and $\bfq_2$ in the diagram are different from those for the other terms, which is just for the computational reason (see Appendix \ref{sec: appendix}).}
\beae{
    P^{\text{2c}}_{\lambda\lambda}(\tau,k)&=\frac{(3!G_\NL)^2}{3!}\calI_{\lambda\lambda}(\tau,\bfk\mid\bfq_1,\bfq_2\mid\bfk-\bfq_1,\bfq_2,\bfq_3,\bfq_1-\bfq_2-\bfq_3), \\
    P^{\text{1c-C2}}_{\lambda\lambda}(\tau,k)&=\frac{(3!G_\NL)^2}{2!}\calI_{\lambda\lambda}(\tau,\bfk\mid\bfq_1,\bfq_2\mid\bfq_1-\bfq_2-\bfq_3,\bfq_2,\bfq_3,\bfq_2-\bfk), \\
    P^{\text{1c-Z2}}_{\lambda\lambda}(\tau,k)&=\frac{(3!G_\NL)^2}{2!}\calI_{\lambda\lambda}(\tau,\bfk\mid\bfq_1,\bfq_2\mid\bfq_1,\bfq_2,\bfq_3,\bfq_1+\bfq_2-\bfk-\bfq_3),
}
for ones proportional to $G_\NL^2$ (above nine diagrams are shown in Fig.~\ref{fig: diag_fnl^4}), 
and
\bege{
    P^{\text{(1,1)$\ell$-1}}_{\lambda\lambda}(\tau,k)=P^{
    \text{(1,1)$\ell$-2}}_{\lambda\lambda}(\tau,k)=P^{
    \text{(1,1)$\ell$-3}}_{\lambda\lambda}(\tau,k)=\frac{(3!G_\NL)^2}{(2!)^2}A_g^2P^{
    \text{Vanilla}}_{\lambda\lambda}(\tau,k), \\ 
    \begin{aligned}
        P^{\text{2$\ell$}}_{\lambda\lambda}(\tau,k)&=\frac{5!I_\NL}{2^22!}A_g^2P^{\text{Vanilla}}_{\lambda\lambda}(\tau,k), &
        P^\text{1$\ell$-C1}_{\lambda\lambda}(\tau,k)&=\frac{4!H_\NL}{2!}\frac{A_gP^\text{C}_{\lambda\lambda}(\tau,k)}{2!F_\NL}, \\
        P^\text{1$\ell$-C2}_{\lambda\lambda}(\tau,k)&=\frac{3!G_\NL}{2!}A_gP^\text{C}_{\lambda\lambda}(\tau,k), &
        P^\text{1$\ell$-Z1}_{\lambda\lambda}(\tau,k)&=\frac{3!G_\NL}{2!}A_gP^\text{Z}_{\lambda\lambda}(\tau,k), \\
        P^\text{1$\ell$-Z2}_{\lambda\lambda}(\tau,k)&=\frac{4!H_\NL}{2!}\frac{A_gP^\text{Z}_{\lambda\lambda}(\tau,k)}{2!F_\NL}, &
        P^{\text{1$\ell$1c-1}}_{\lambda\lambda}(\tau,k)&=\frac{3!G_\NL}{2!}A_gP^{\text{1c}}_{\lambda\lambda}(\tau,k), \\
        P^{\text{1$\ell$1c-2}}_{\lambda\lambda}(\tau,k)&=\frac{4!H_\NL}{2!}\frac{A_gP^{\text{1c}}_{\lambda\lambda}(\tau,k)}{2!F_\NL},
    \end{aligned}
}
for ones including self-closed loops shown in Fig.~\ref{fig: diag_loop}. 
Fourth-order diagrams basically include highly multi-dimensional integrals and their specific computations require several techniques. Particularly for the 1-convolution C (1c-C1 and 1c-C2), 1-convolution Z (1c-Z1 and 1c-Z2), and CZ terms, we describe the detailed calculations in Appendix~\ref{sec: appendix}.

\begin{figure}
    \centering
    \begin{tabular}{c}
        \begin{minipage}{0.5\hsize}
            \centering
            \includegraphics[width=0.95\hsize]{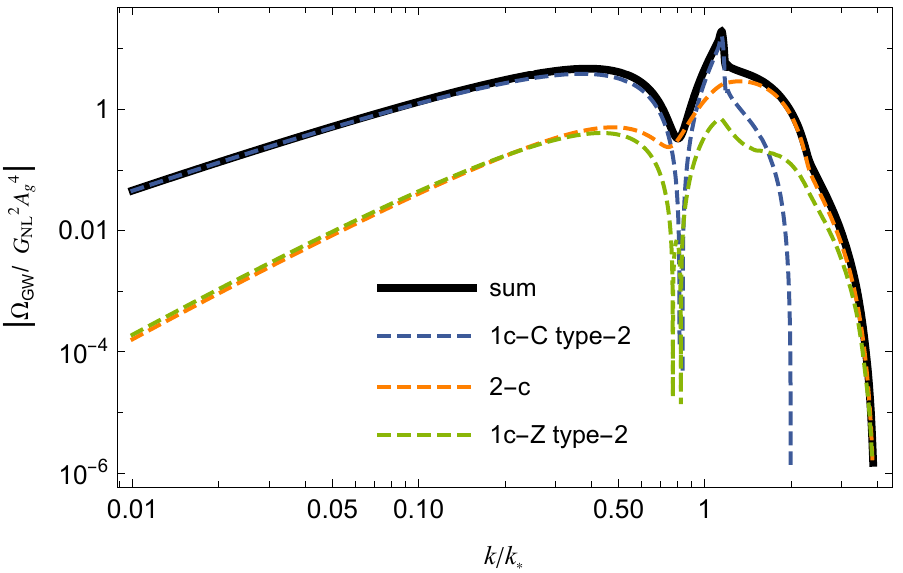}
        \end{minipage} 
        \begin{minipage}{0.5\hsize}
            \centering
            \includegraphics[width=0.95\hsize]{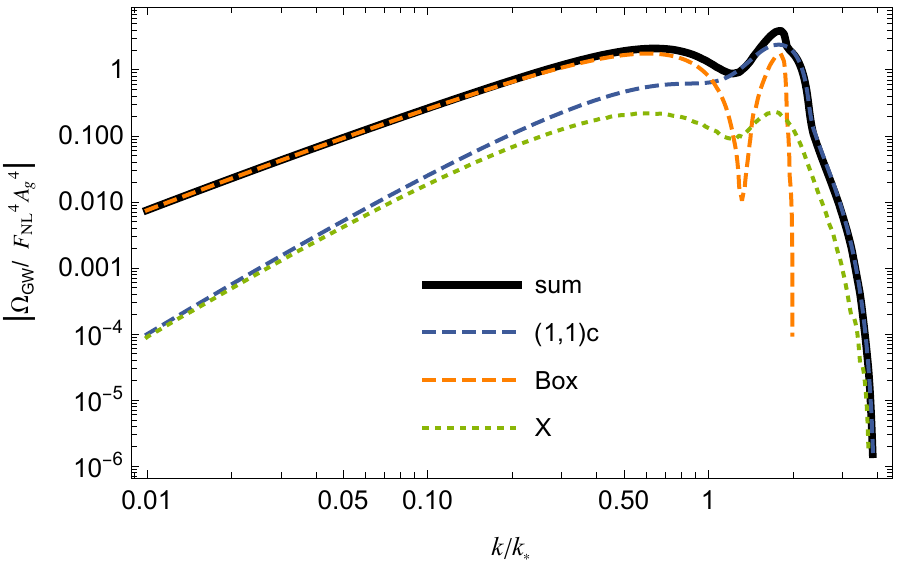}
        \end{minipage}\\
        \begin{minipage}{0.5\hsize}
            \centering
            \includegraphics[width=0.95\hsize]{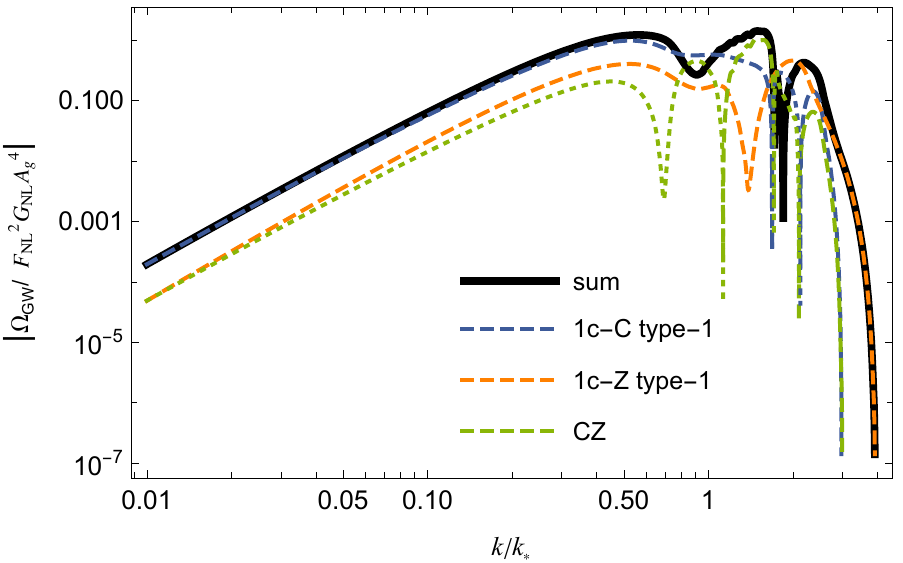}
        \end{minipage} 
    \end{tabular}
    \caption{The normalized GW amplitude of fourth-order contributions 
    except for ones with self-closed loops.
    In each figure, we represent the fourth-order total amplitude as the black solid line. 
    The dashed and dotted lines show the spectrum of $\Omega_\GW$ where the sign is positive and negative respectively.
    The high-frequency side of the X term is wavering, which is due to numerical error. 
    We note that these plots are including two polarizations and the deformation factors.}
    \label{fig: GW_O(P^4)}
\end{figure}

Including the deformation factors, the \ac{GW} spectrum is summarized as
\bme{
    \Omega_\GW^{(4)}(k)=\frac{2}{48}\pqty{\frac{k}{aH}}^2\left[2\overline{\calP^\text{(1,1)c}_{++}(k)}+2\overline{\calP^\text{Box}_{++}(k)}+\overline{\calP^\text{X}_{++}(k)}+2^3\overline{\calP^{\text{1c-C1}}_{++}(k)}+2^3\overline{\calP^{\text{1c-Z1}}(k)} \right. \\
    +2^3\overline{\calP^\text{CZ}(k)}
    +2^2\overline{\calP^{\text{2c}}_{++}(k)}+2^2\overline{\calP^{\text{1c-C2}}_{++}(k)}+2\overline{\calP^{\text{1c-Z2}}_{++}(k)}+3\times2^2\overline{\calP^{
    \text{(1,1)$\ell$}}_{++}(k)}+2^3\overline{\calP^{\text{2$\ell$}}_{++}(k)} \\
    \left.+2^3\overline{\calP^\text{1$\ell$-C1}_{++}(k)}+2^3\overline{\calP^\text{1$\ell$-C2}_{++}(k)}+2^3\overline{\calP^\text{1$\ell$-Z1}_{++}(k)}+2^3\overline{\calP^\text{1$\ell$-Z2}_{++}(k)}+2^3\overline{\calP^{\text{1$\ell$1c-1}}_{++}(k)}+2^3\overline{\calP^{\text{1$\ell$1c-2}}_{++}(k)} \right].
}
The numerical results are shown in Fig.~\ref{fig: GW_O(P^4)}. Again we do not show the contributions with self-closed loops because they are constant multiplications of lower-order diagrams.

\section{Application to the exponential tail case} \label{sec: result&diss}

\begin{figure}
    \centering
    \begin{tabular}{c}
        \begin{minipage}{0.8\hsize}
            \centering
            \includegraphics[width=0.95\hsize]{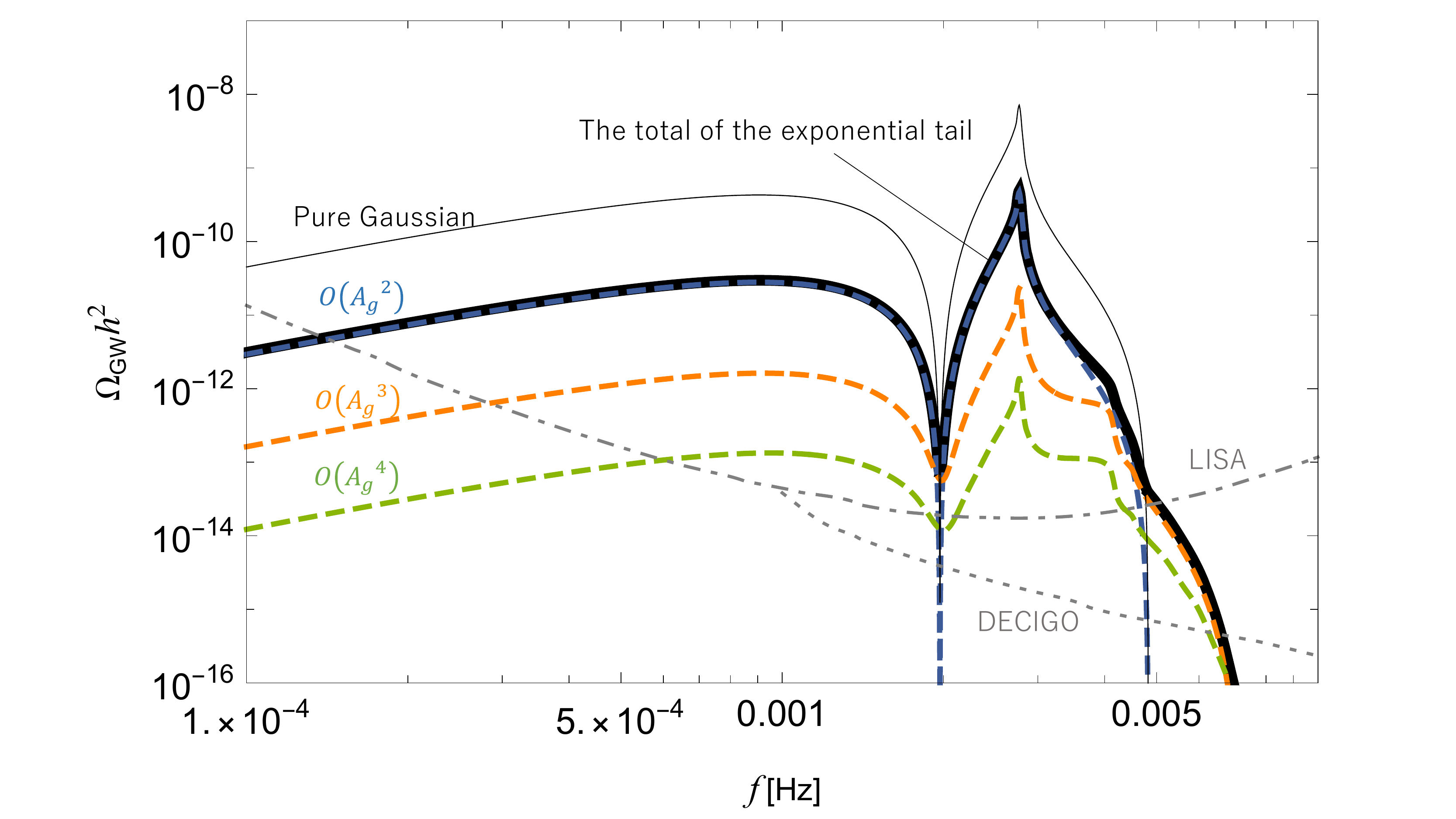}
        \end{minipage} 
    \end{tabular}
    \caption{The prediction of the current induced GW spectrum associated with the \ac{PBH} \ac{DM} scenario shown in 
    Fig.~\ref{fig: mass function} in the exponential tail case (black thick line). Contributions of $\calO(A_g^2)$, $\calO(A_g^3)$, and $\calO(A_g^4)$ are shown by blue-dashed, orange-dashed, and green-dashed lines, respectively. The leading contribution $\sim\calO(A_g^2)$ is dominant even in the non-perturbative exponential tail case because of the smallness of $A_g$ ($=1.32\times10^{-3}$). As a comparison, the black thin line shows the prediction for $f_\PBH=1$ if the primordial curvature perturbation is purely Gaussian with the same peak scale $k_*=1.56\times10^{-12}\,\si{Mpc^{-1}}$. The expected \ac{GW} amplitude is reduced by $\sim\pqty{\frac{1.32\times10^{-3}}{5.17\times10^{-3}}}^2\simeq0.065$ in the exponential tail case because the required primordial amplitude $A_g$ is reduced from $5.17\times10^{-3}$ to $1.32\times10^{-3}$. Nevertheless, it is still detectable by LISA, whose sensitivity is illustrated by the gray dot-dashed line.
    Deeper observations such as DECIGO shown by the gray-dotted line may distinguish the non-Gaussian signature in the high-frequency tail, though it requires more thorough investigations (see the text). Both sensitivities are taken from Ref.~\cite{Schmitz:2020syl}. 
    }
    \label{fig: totGW}
\end{figure}

Having been armed with all weapons, we here show the scalar-induced \acp{GW} spectrum associated with \ac{PBH} \ac{DM} in the exponential tail case.
The explicit expansion of the exponential tail mapping~\eqref{eq: expansion of zeta} first specifies the expansion coefficients~\eqref{eq: perturbexpand} as $F_\NL=3/2$, $G_{\NL}=3$, $H_\NL=27/4$, $I_\NL=81/5$, etc.
We fix the perturbation amplitude $A_g$ and the peak scale $k_*$ as
$A_g=1.32\times10^{-3}$ and $k_*=1.56\times10^{-12}\;\si{Mpc}^{-1}$ similarly to Sec.~\ref{sec: exptail}.
\Acp{PBH} account for the full \ac{DM} abundance around $M\sim10^{22}\,\si{g}$ in this case as shown in 
Fig.~\ref{fig: mass function}.
The corresponding induced \ac{GW} spectrum is then shown in Fig.~\ref{fig: totGW} in terms of its frequency $f=k/(2\pi)$.
We first confirm that the leading order \text{Vanilla} contribution $\sim\calO(A_g^2)$ is dominant and the series expansion of the \ac{GW} spectrum soon converges due to the smallness of $A_g$ even in the non-perturbative exponential-tail non-Gaussianity case as expected in the previous section.
The leading order contribution is enough for the LISA's sensitivity and it would hold true if one includes the higher order corrections in the gravitational potential $\Phi$ which we mentioned in footnote~\ref{footnote: Phi3} because the nonlinearity parameters due to gravity are expected to be order-unity.
The leading order one is simply proportional to $A_g^2$ and thus the \ac{GW} amplitude is reduced by $\pqty{\frac{1.32\times10^{-3}}{5.17\times10^{-3}}}^2\simeq0.065$ compared with the case where $\zeta$ is purely Gaussian (recall that $A_g=5.17\times10^{-3}$ is required for $f_\PBH=1$ in the Gaussian case as shown in 
Fig.~\ref{fig: fPBHtot}).
The induced \ac{GW} can still be detected by LISA thanks to its high sensitivity.
Note that the perturbation amplitude $A_g$ and hence the \ac{GW} amplitude are really insensitive to the small change of $f_\PBH$ as shown in 
Fig.~\ref{fig: fPBHtot}. 
Fixing the amplitude $A_g$ by the requirement of
$f^{\tot}_{\PBH}= 1$ then it could be said that the typical amplitude of induced \acp{GW} is determined by the non-Gaussian nature of the primordial perturbation. We also note that the relation between the \ac{PBH} mass and the \ac{GW} frequency does not change so much due to the non-Gaussianity (see Ref.~\cite{Kitajima:2021fpq}) but is almost determined through the mass-scale relation~\eqref{eq: Mk}.

\section{Conclusions}
\label{sec:conclusion}

The scalar-induced stochastic \acp{GW} accompanying the enhanced primordial fluctuations to form the \acp{PBH} is one of the probes to test the \ac{PBH} \ac{DM} model.
In this work, we have investigated the induced \ac{GW} spectrum associated with the model where the primordial curvature perturbations have the exponential-tail non-Gaussian distribution.

We first review the \ac{PBH} abundance prediction with the exponential-tail non-Gaussianity in Sec.~\ref{sec: exptail}, following Ref.~\cite{Kitajima:2021fpq}.
In Sec.~\ref{sec:inducedGWs}, we then extend the formulation of the two-point function of the \acp{GW} induced by the second-order scalar perturbations to the case where the primordial curvature perturbations show the general local-type non-Gaussianity.
To take account of the non-Gaussian corrections into the trispectrum of the curvature perturbation, we have employed the diagrammatic approach developed in Refs.~\cite{Unal:2018yaa,Adshead:2021hnm}.
The minimal configuration is represented by the ``$\mathbb{V}$anilla" diagram shown in Fig.~\ref{fig: tree}, and any non-Gaussian contribution can be represented by adding lines to this ``$\mathbb{V}$anilla" diagram. 
We found that all non-Gaussian contributions can be summarized into nine topologically-independent diagrams shown in Fig.~\ref{fig: all non-G cont}. 

In Sec.~\ref{sec: result&diss}, we have adopted this general formalism to the exponential-tail-type curvature perturbations in the model where \acp{PBH} with the mass of $10^{22}~\si{g}$ are whole \ac{DM}.
We calculated the scalar-induced \ac{GW} spectrum with the non-Gaussian contributions up to the fourth order in terms of the amplitude parameter for the Gaussian field of the curvature perturbations, $A_g$.
Even though the non-perturbative nature of the exponential tail is crucial for the \ac{PBH} abundance, the \ac{GW} amplitude is well given by the leading order contribution $\sim\calO(A_g^2)$ thanks to the smallness of $A_g=1.32\times10^{-3}$.
The expected \ac{GW} amplitude is reduced by $\pqty{\frac{1.32\times10^{-3}}{5.17\times10^{-3}}}^2\simeq0.065$ compared with the pure Gaussian case, but it is still large enough to be detected by LISA.

It is worth mentioning that the non-Gaussian contributions can appear on the high-frequency side. This is because, while the leading term can produce \acp{GW} only up to $k=2k_*$ due to the momentum conservation, higher order terms can go beyond that through more complicated momentum configurations.
Though they cannot be distinguished in the LISA's sensitivity, it might be possible to obtain information about the primordial non-Gaussianity from \ac{GW} observation with deeper sensitivity such as DECIGO, although the precise spectrum would depend on the UV behavior of the power spectrum of the curvature perturbation~\cite{Atal:2021jyo}. The enhancement of the primordial power spectrum induced by the ultra slow-roll models requires a typical width being broad~\cite{Byrnes:2018txb}. Although we simply analyzed the Dirac delta function, in practice, it needs to take an effect of some finite width into account for the evaluation of the GWs spectrum.
One also has to include gravitational higher-order corrections such as $h\sim\Phi^3$, $\Phi^4$, $\cdots$ which we have neglected in this work~\cite{Yuan:2019udt,Zhou:2021vcw,Chang:2022nzu}.
Gauge issues~\cite{Matarrese:1997ay,Boubekeur:2008kn,Arroja:2009sh,Hwang:2017oxa,Domenech:2017ems,Gong:2019mui,Tomikawa:2019tvi,Inomata:2019yww,Yuan:2019fwv,Chang:2020iji,Chang:2020mky,Domenech:2020xin} would be relevant, too.
We leave this possibility for future works.
\acknowledgments

This work is supported by JST FOREST Program JPMJFR20352935 (R.I.) and JSPS KAKENHI Grants
No.~JP20J22260 (K.T.A.),
No.~JP21K13918 (Y.T.),
No.~JP20H01932 (S.Y.), and No.~JP20K03968 (S.Y.).

\appendix

\section{Detailed computations for fourth-order diagrams} \label{sec: appendix}

\subsection{1-convolution C term}

The explicit expressions of 1c-C1 and 1c-C2 terms are given by 
\beae{
    &\bmte{P^{\text{1c-C1}}_{\lambda\lambda'}(\tau,k)=
    \frac{(2!F_{\NL})^2 3!G_\NL }{2!}\int\frac{\dd[3]{q_1}}{(2\pi)^3}\frac{\dd[3]{q_2}}{(2\pi)^3}\frac{\dd[3]{q_3}}{(2\pi)^3}Q_\lambda(\bfk,\bfq_1)Q_{\lambda'}(\bfk,\bfq_2) \\
    \times I_k(\abs{\bfk-\bfq_1},q_1,\tau)I_k(\abs{\bfk-\bfq_2},q_2,\tau)P_g(q_2)P_g(\abs{\bfk-\bfq_2-\bfq_3})P_g(q_3)P_g(\abs{\bfq_1-\bfq_2}),} \\
    &\bmte{P^{\text{1c-C2}}_{\lambda\lambda'}(\tau,k)=
    \frac{(3!G_\NL)^2}{2!} \int\frac{\dd[3]{q_1}}{(2\pi)^3}\frac{\dd[3]{q_2}}{(2\pi)^3}\frac{\dd[3]{q_3}}{(2\pi)^3}Q_\lambda(\bfk,\bfq_1)Q_{\lambda'}(\bfk,\bfq_2) \\
    \times I_k(\abs{\bfk-\bfq_1},q_1,\tau)I_k(\abs{\bfk-\bfq_2},q_2,\tau)P_g(q_2)P_g(\abs{\bfk-\bfq_2})P_g(q_3)P_g(\abs{\bfq_1-\bfq_2-\bfq_3}).}
}
We hereafter allow different polarizations for $\lambda$ and $\lambda^\prime$. Under the monochromatic assumption~\eqref{eq: monochromatic P},
\bae{
    P_g=\frac{2\pi^2}{k^3}A_g\delta(\ln k-\ln k_*),
}
these multiple integrations are simplified to some extent.

\begin{figure}
    \centering
    \begin{tabular}{c}
        \begin{minipage}{0.5\hsize}
            \centering
            \includegraphics[width=0.95\hsize]{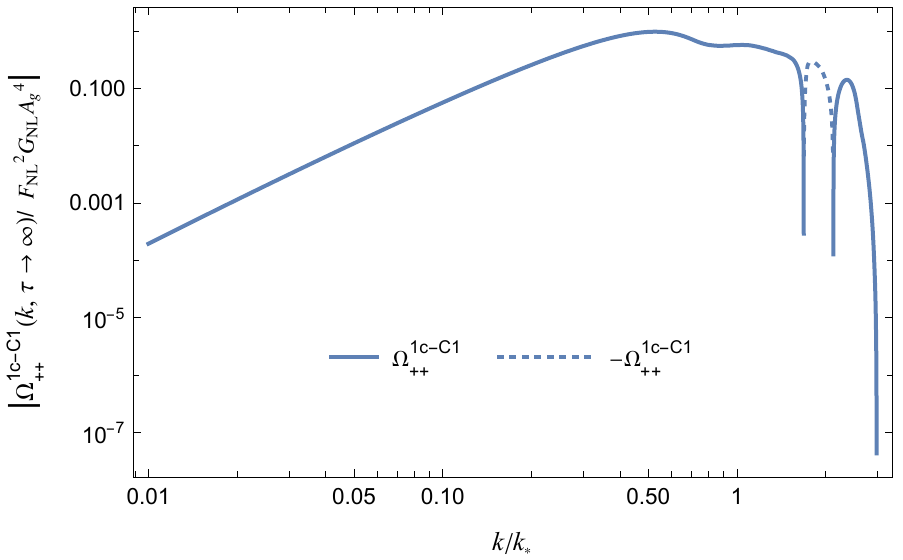}
        \end{minipage}
        \begin{minipage}{0.5\hsize}
            \centering
            \includegraphics[width=0.95\hsize]{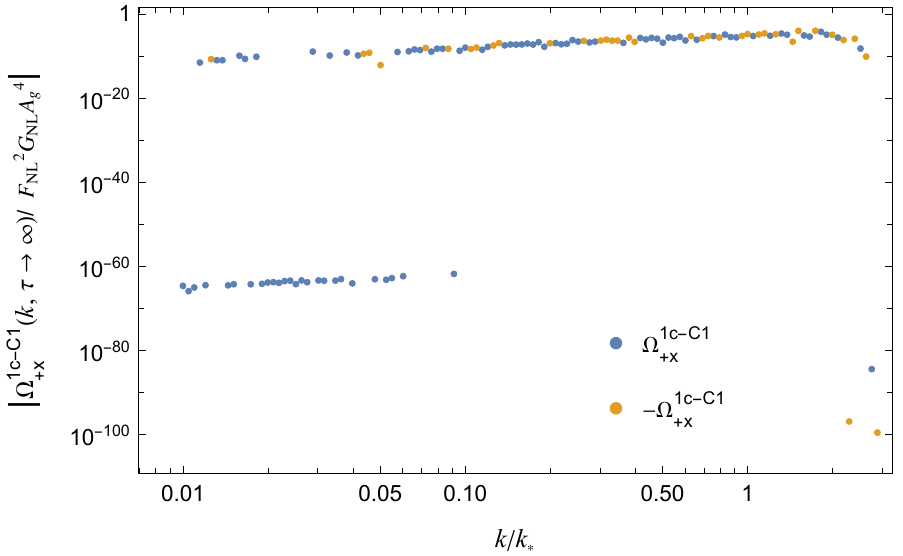}
        \end{minipage}
    \end{tabular}
    \caption{The normalized GW amplitudes $\abs{\Omega^{\text{1c-C1}}_{++}/F_\NL^2G_\NL A_g^4}$ (left) and $\abs{\Omega^{\text{1c-C1}}_{+\times}/F_\NL^2G_\NL A_g^4}$ (right). The plane and dotted lines are for positive and negative values respectively in the left panel, while the blue and orange points are for positive and negative in the right panel.}
    \label{fig: 1c-C1}
\end{figure}

First of all, the convolved propagator part ($\bfq_3$ integral) can be calculated as the following formula,
\bae{\label{eq: convolved propagator}
    \int\frac{\dd[3]{q}}{(2\pi)^3}P_g(q)P_g(\abs{\bfk-\bfq})=\frac{\pi^2A_g^2}{kk_*^2}\Theta(2k_*-k).
}
For the 1c-C1 term, the remaining momentum constraints are rewritten as 
\bae{
    \delta(\ln\abs{\bfq_1-\bfq_2}-\ln k_*)\delta(\ln q_2-\ln k_*)=\frac{2k_*^2}{q_{1*}^2}\delta(\ln q_1-\ln q_{1*})\delta(\ln q_2-\ln k_*),
}
where
\bae{
    q_{1*}=2k_*(\sin\theta_1\sin\theta_2\cos(\phi_1-\phi_2)+\cos\theta_1\cos\theta_2),
}
in the polar coordinate expression $\bfq_i=q_i(\sin\theta_i\cos\phi_i,\sin\theta_i\sin\phi_i,\cos\theta_i)$.
Accordingly, the (dimensionless) power spectrum for the 1c-C1 term reduces to 
\bme{
    \calP^\text{1c-C1}_{\lambda\lambda^\prime}(\tau,k)=\frac{3F_\NL^2G_\NL A_g^4}{4\pi^2}\frac{k^3}{k_*^3}\int\dd{\cos\theta_1}\dd{\phi_1}\dd{\cos\theta_2}\dd{\phi_2}Q_\lambda(\bfk,\bfq_1)Q_{\lambda^\prime}(\bfk,\bfq_2) \\
    \times \eval{I_k(\abs{\bfk-\bfq_1},q_1,\tau)I_k(\abs{\bfk-\bfq_2},q_2,\tau)\frac{q_1}{\abs{\bfk-\bfq_2}}\Theta(2k_*-\abs{\bfk-\bfq_2})}_{q_1=q_{1*}, \, q_2=k_*}.
}
Changing the integration variables from $\phi_1$ and $\phi_2$ to $\varphi=\phi_1-\phi_2$ and $\phi_2$, one finds that the second line does not depend on $\phi_2$. 
Therefore, the $\phi_2$ integration can be summarized in the polarization part as
\bae{\label{eq: Q^2}
    Q_{\lambda\lambda^\prime}^2(\bfk,\bfq_1,\bfq_2)\coloneqq{}&\int\dd{\phi_2}Q_\lambda(\bfk,\bfq_1)Q_{\lambda^\prime}(\bfk,\bfq_2) \nonumber \\
    ={}&\frac{q_1^2q_2^2\pi}{2}\sin^2\theta_1\sin^2\theta_2\times
    \bce{
        \cos(2\varphi) & \text{for $\lambda=\lambda^\prime$}, \\
        \sin(2\varphi) & \text{for $\lambda=\times$ and $\lambda^\prime=+$}, \\
        -\sin(2\varphi) & \text{for $\lambda=+$ and $\lambda^\prime=\times$}.
    }
}
Including the deformation factor $2^3$ for the 1c-C1 term, the corresponding \ac{GW} density parameter reads 
\bae{
    \Omega_{\lambda\lambda^\prime}^\text{1c-C1}(k)&=
    \lim_{\tau\to\infty}\frac{2^3}{48}(k\tau)^2\overline{\calP_{\lambda\lambda^\prime}^\text{1c-C1}(\tau,k)} \nonumber \\
    &\bmbe{=\frac{F_\NL^2G_\NL A_g^4}{8\pi^2}\frac{k^3}{k_*^3}\int\dd{\cos\theta_1}\dd{\cos\theta_2}\dd{\varphi}Q_{\lambda\lambda^\prime}^2(\bfk,\bfq_1,\bfq_2) \\
    \times \eval{\overline{J_k^2(\abs{\bfk-\bfq_1},q_1;\abs{\bfk-\bfq_2},q_2)}\frac{q_1}{\abs{\bfk-\bfq_2}}\Theta(2k_*-\abs{\bfk-\bfq_2})}_{q_1=q_{1*},\,q_2=k_*}.}
}
The numerical result of this integral is shown in Fig.~\ref{fig: 1c-C1}.
One finds that the \ac{GW} amplitude indeed vanishes for $\lambda\neq\lambda^\prime$ within the numerical error.

\begin{figure}
    \centering
    \begin{tabular}{c}
        \begin{minipage}{0.5\hsize}
            \centering
            \includegraphics[width=0.95\hsize]{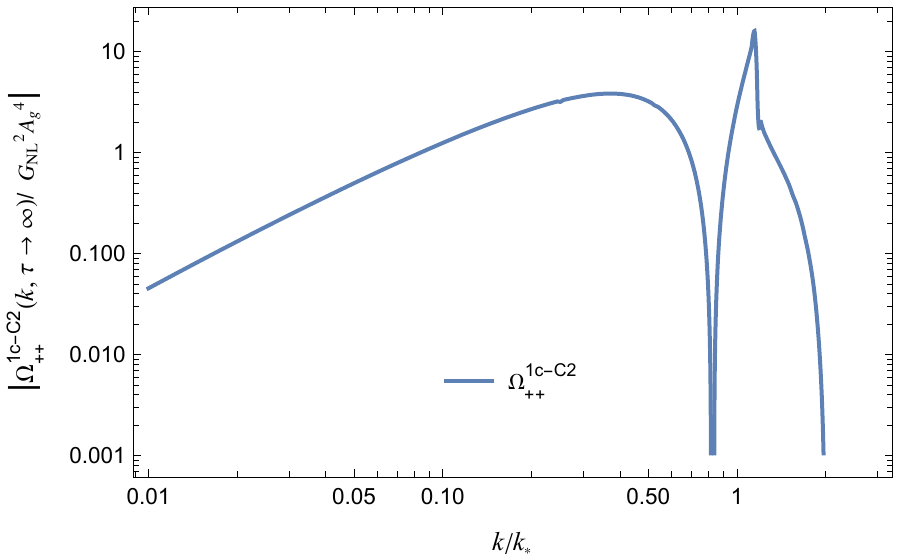}
        \end{minipage}
        \begin{minipage}{0.5\hsize}
            \centering
            \includegraphics[width=0.95\hsize]{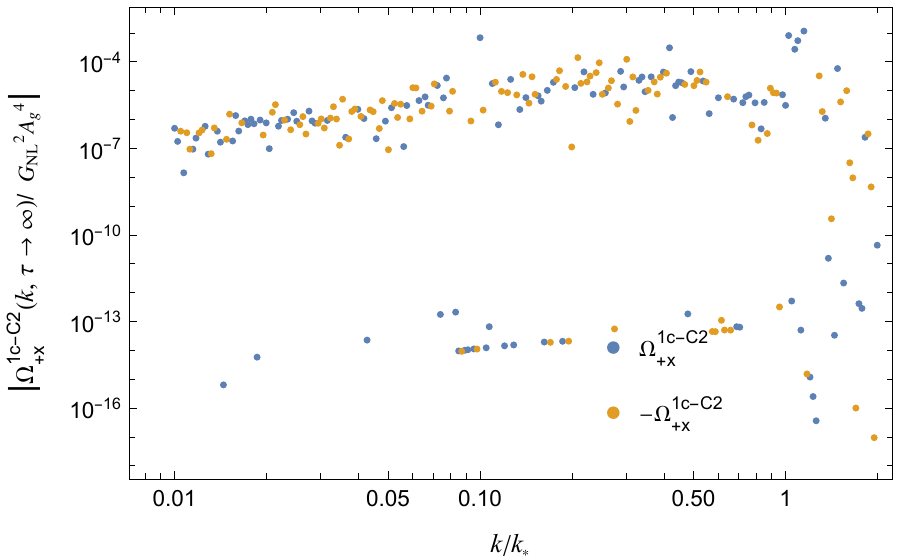}
        \end{minipage}
    \end{tabular}
    \caption{The normalized GW amplitudes $\abs{\Omega^{\text{1c-C2}}_{++}/G_\NL^2 A_g^4}$ (left) and $\abs{\Omega^{\text{1c-C2}}_{+\times}/G_\NL^2  A_g^4}$ (right) in the similar style to Fig.~\ref{fig: 1c-C1}.}
    \label{fig: 1c-C2}
\end{figure}

For the 1c-C2 term, the remaining constraints after the $\bfq_3$ integral are rewritten as 
\bae{
	\delta(\ln q_2-\ln k_*)\delta(\ln\abs{\bfk-\bfq_2}-\ln k_*)=\frac{k_*}{k}\delta(\ln q_2-\ln k_*)\delta(\cos\theta_2-\mu_{2*}),
}
where $\mu_{2*}=k/(2k_*)$. The corresponding power spectrum then reads 
\bme{
	\calP^\text{1c-C2}_{\lambda\lambda^\prime}(\tau,k)=\frac{9G_\NL^2A_g^4}{16\pi^2}\frac{k^2}{k_*^4}\int\dd{q_1}\dd{\cos\theta_1}\dd{\varphi}Q_{\lambda\lambda^\prime}^2(\bfk,\bfq_1,\bfq_2) \\
	\times \eval{I_k(\abs{\bfk-\bfq_1},q_1,\tau)I_k(\abs{\bfk-\bfq_2},q_2,\tau)\frac{q_1^2}{\abs{\bfq_1-\bfq_2}}\Theta(2k_*-\abs{\bfq_1-\bfq_2})}_{q_2=k_*,\,\cos\theta_2=\mu_{2*}}.
}
Including the deformation factor $2^2$, the corresponding \ac{GW} density parameter reads 
\bae{
    \Omega_{\lambda\lambda^\prime}^\text{1c-C2}(k)&=\lim_{\tau\to\infty}\frac{2^2}{48}(k\tau)^2\overline{\calP_{\lambda\lambda^\prime}^\text{1c-C2}(\tau,k)} \nonumber \\
    &\bmbe{=\frac{3G_\NL^2A_g^4}{64\pi^2}\frac{k^2}{k_*^4}\int\dd{q_1}\dd{\cos\theta_1}\dd{\varphi}Q_{\lambda\lambda^\prime}^2(\bfk,\bfq_1,\bfq_2) \\
    \times \eval{\overline{J_k^2(\abs{\bfk-\bfq_1},q_1;\abs{\bfk-\bfq_2},q_2)}\frac{q_1^2}{\abs{\bfq_1-\bfq_2}}\Theta(2k_*-\abs{\bfq_1-\bfq_2})}_{q_2=k_*,\,\cos\theta_2=\mu_{2*}}.}
}
Its numerical result is shown in Fig.~\ref{fig: 1c-C2}.

\subsection{1-convolution Z term}

The 1c-Z1 and 1c-Z2 terms are given by
\beae{
    &\bmte{P^{\text{1c-Z1}}_{\lambda\lambda'}(\tau,k)=\frac{(2!F_\NL)^2 3!G_\NL}{2!} \int\frac{\dd[3]{q_1}}{(2\pi)^3}\frac{\dd[3]{q_2}}{(2\pi)^3}\frac{\dd[3]{q_3}}{(2\pi)^3}Q_\lambda(\bfk,\bfq_1)Q_{\lambda'}(\bfk,\bfq_2-\bfq_1) \\ 
    \times I_k(\abs{\bfk-\bfq_1},q_1,\tau)I_k(\abs{\bfk-\bfq_2+\bfq_1},\abs{\bfq_2-\bfq_1},\tau) \\ 
    \times P_g(q_1)P_g(
    \abs{\bfq_2-\bfk})P_g(q_3)P_g(
    \abs{\bfq_2-\bfq_1-\bfq_3}),
    } \\
    &\bmte{P^{\text{1c-Z2}}_{\lambda\lambda'}(\tau,k)=\frac{(3!G_\NL)^2}{2!}\int\frac{\dd[3]{q_1}}{(2\pi)^3}\frac{\dd[3]{q_2}}{(2\pi)^3}\frac{\dd[3]{q_3}}{(2\pi)^3}Q_\lambda(\bfk,\bfq_1)Q_{\lambda'}(\bfk,\bfq_2) \\
    \times I_k(\abs{\bfk-\bfq_1},q_1,\tau)I_k(\abs{\bfk-\bfq_2},q_2,\tau)P_g(q_1)P_g(q_2)P_g(q_3)P_g(\abs{\bfq_1+\bfq_2-\bfk-\bfq_3}).
    }
}
$\bfq_3$ integrations can be again done by Eq.~\eqref{eq: convolved propagator}.
For the 1c-Z1 term, the remaining constraints read
\bae{
    \delta(\ln q_1-\ln k_*)
    \delta(\ln\abs{\bfq_2-\bfk}-\ln k_*)=\frac{k_*^2}{k q_2}
    \delta(\ln q_1-\ln k_*)
    \delta(\cos{\theta_2}-\tilde{\mu}_{2*}),
}
where $\tilde{\mu}_{2*}=\frac{q_2^2+k^2-k_*^2}{2kq_2}$. Accordingly, the power spectrum for the \text{1c-Z1} term can be reduced to
\bme{\label{eq: PZh1''}
    \mathcal{P}^{\text{1c-Z1}}_{\lambda\lambda'}(\tau,k)=\frac{3 F_\NL^2G_\NL A_g^4}{8\pi^2}\frac{k^2}{k_{*}^3}\int \dd{\cos{\theta_1}} \dd{\phi_1} \dd{q_2} \dd{\phi_2}Q_\lambda(\bfk,\bfq_1)Q_{\lambda'}(\bfk,\bfq_2-\bfq_1) \\
    \times \eval{I_k(\abs{\bfk-\bfq_1},q_1,\tau)I_k(\abs{\bfk-\bfq_2+\bfq_1},
    \abs{\bfq_2-\bfq_1},\tau)\frac{q_2\Theta(2k_{*}-\abs{\bfq_2-\bfq_1})}{\abs{\bfq_2-\bfq_1}}}_{q_1=k_*,\,\cos\theta_2=\tilde{\mu}_{2*}}.
}

Changing the integration variables from $\phi_1$ and $\phi_2$ to $\varphi=\phi_1-\phi_2$ and $\phi_2$, one finds that the second line does not depend on $\phi_2$. 
Therefore, the polarization factors can be summarized again as a $\phi_2$ integration of $Q_\lambda(\bfk,\bfq_1)Q_{\lambda^{\prime}}(\bfk,\bfq_2-\bfq_1)$: 
\bae{
    \tilde{Q}_{\lambda\lambda^\prime}^2(\bfk,\bfq_1,\bfq_2-\bfq_1)\coloneqq\int\dd{\phi_2}Q_\lambda(\bfk,\bfq_1)Q_{\lambda^\prime}(\bfk,\bfq_2-\bfq_1),
}
which is obtained as
\beae{
    \tilde{Q}_{++}^2&=\tilde{Q}_{\times\times}^2=\frac{\pi q_1^2\sin^2{\theta_1} }{2}\left(2 q_1 q_2\sin{\theta_1} \sin{\theta_2} \cos {\varphi }+q_1^2\sin^2{\theta_1} +q_2^2\sin^2{\theta_2}\cos {2 \varphi}\right), \\
    \tilde{Q}_{+\times}^2&=-\tilde{Q}_{\times+}^2=- \pi q_1^2 q_2\sin{\theta_1}^2  \sin{\theta_2}\left(- q_1\sin{\theta_1}+ q_2\sin{\theta_2} \cos{\varphi}\right) \sin{\varphi}.
}

\begin{figure}
    \centering
    \begin{tabular}{c}
        \begin{minipage}{0.5\hsize}
            \centering
            \includegraphics[width=0.95\hsize]{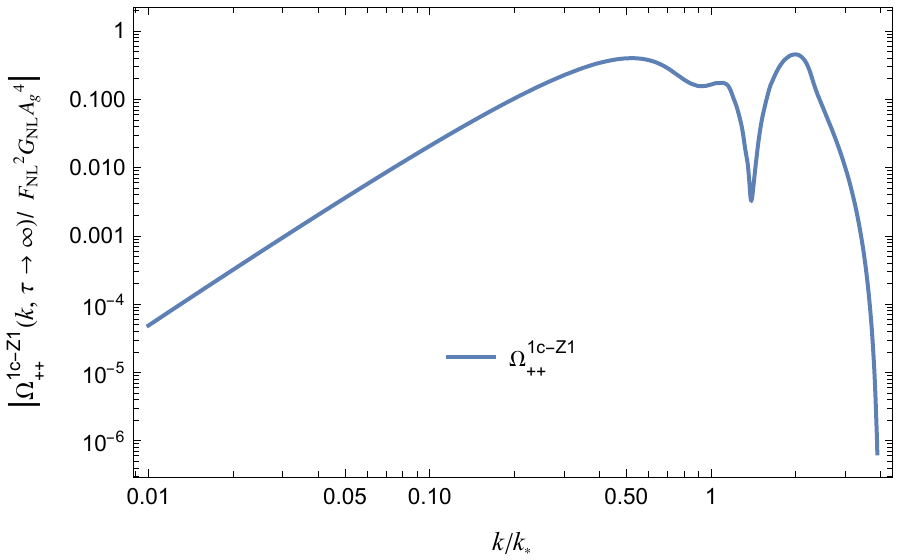}
        \end{minipage}
        \begin{minipage}{0.5\hsize}
            \centering
            \includegraphics[width=0.95\hsize]{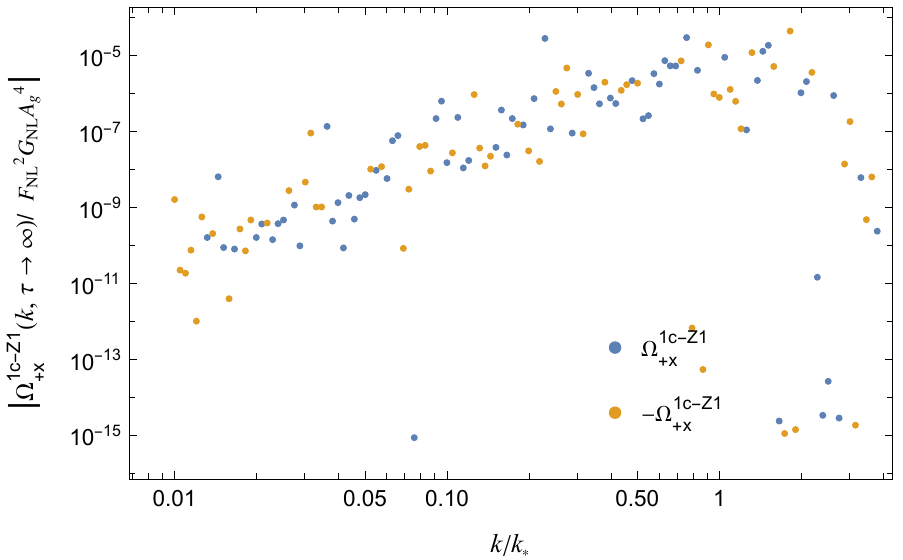}
        \end{minipage}
    \end{tabular}
    \caption{The normalized GW amplitudes $\abs{\Omega^{\text{1c-Z1}}_{++}/F_\NL^2 G_\NL A_g^4}$ (left) and $\abs{\Omega^{\text{1c-Z1}}_{+\times}/F_\NL^2 G_\NL A_g^4}$ (right) in the similar style to Fig.~\ref{fig: 1c-C1}}
    \label{fig: 1c-Z1}
\end{figure}

Including the deformation factor $2^3$ for the $\text{1c-Z1}$ term, the density parameter of the induced \ac{GW} in the \ac{RD} era can be obtained as follows.
\bae{
    \Omega^{\text{1c-Z1}}_{\lambda\lambda'}(k) &=\lim_{\tau\to\infty}\frac{2^3}{48}(k\tau)^2\overline{\mathcal{P}^{\text{1c-Z1}}_{\lambda\lambda'}(\tau,k)} \nonumber\\
    &\bmbe{=\frac{F_\NL^2G_\NL A_g^4}{16\pi^2}\frac{k^2}{k_{*}^3}\int \dd{\cos{\theta_1}} \dd{\varphi} \dd{q_2} \tilde{Q}^2_{\lambda\lambda'}(\bfk,\bfq_1,\bfq_2-\bfq_1) \\
    \times \eval{\overline{J_k^2(\abs{\bfk-\bfq_1},q_1;\abs{\bfk-\bfq_2+\bfq_1},\abs{\bfq_2-\bfq_1})}\frac{q_2\Theta(2k_{*}-\abs{\bfq_2-\bfq_1})}{\abs{\bfq_2-\bfq_1}}}_{q_1=k_*,\,\cos\theta_2=\tilde{\mu}_{2*}}.}
}
The numerical resultant power spectrum is exhibited in Fig.~\ref{fig: 1c-Z1}.

For $\text{1c-Z2}$ term, the remaining constraints after the $\bfq_3$ integral are trivial as
\bae{
   \delta(\ln q_1-\ln k_*)\delta(\ln q_2-\ln k_*).
}
The corresponding power spectrum then reads
\bme{
    \mathcal{P}^{\text{1c-Z2}}_{\lambda\lambda'}(\tau,k)=\frac{9G_\NL^2 A_g^4}{16\pi^2}\frac{k^3}{k_{*}^2}\int \dd{\cos{\theta_1}} \dd{\varphi} \dd{\cos{\theta_2}} Q^2_{\lambda\lambda'}(\bfk,\bfq_1,\bfq_2) \\
    \times \eval{I_k(\abs{\bfk-\bfq_1},q_1,\tau)I_k(\abs{\bfk-\bfq_2},q_2,\tau)\frac{\Theta(2k_{*}-\abs{\bfq_1+\bfq_2-\bfk})}{\abs{\bfq_1+\bfq_2-\bfk}}}_{q_1=q_2=k_*}.
}
where $Q_{\lambda\lambda^\prime}^2$ is given by Eq.~(\ref{eq: Q^2}).

\begin{figure}
    \centering
    \begin{tabular}{c}
        \begin{minipage}{0.5\hsize}
            \centering
            \includegraphics[width=0.95\hsize]{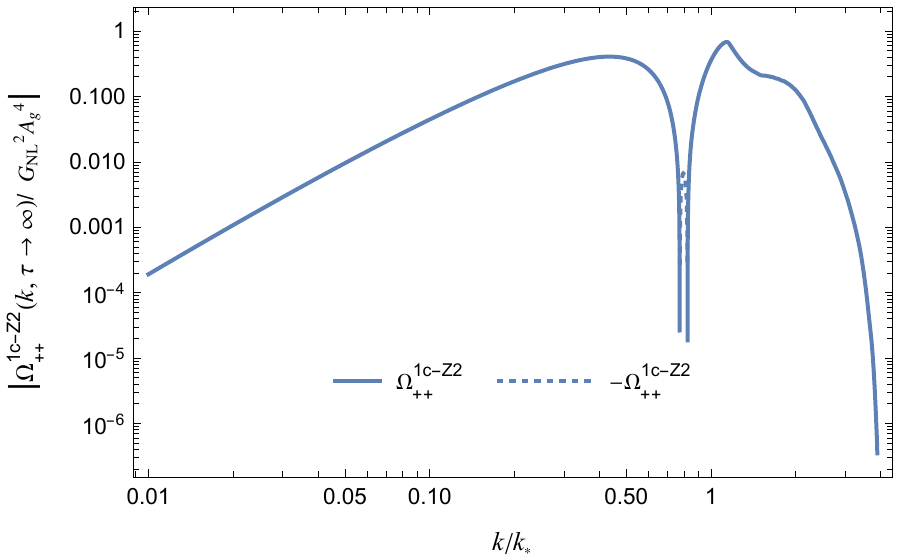}
        \end{minipage}
        \begin{minipage}{0.5\hsize}
            \centering
            \includegraphics[width=0.95\hsize]{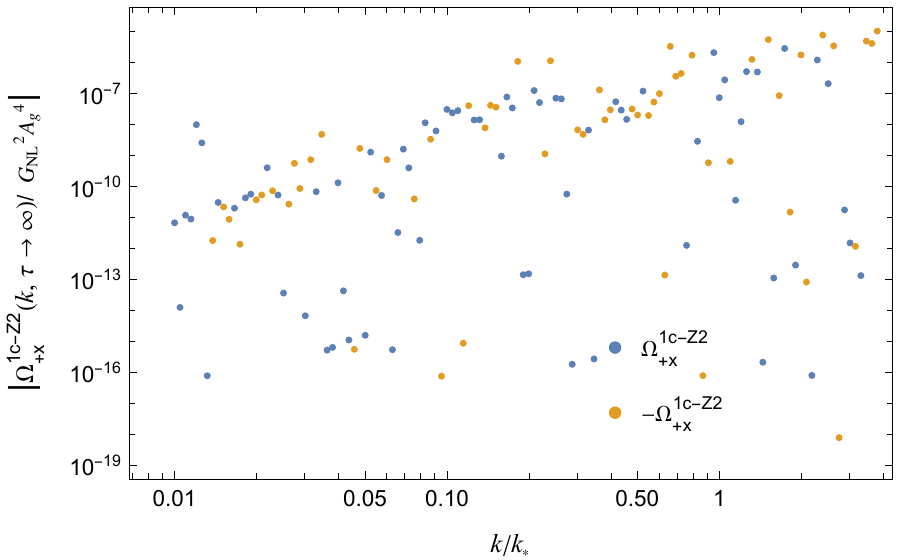}
        \end{minipage}
    \end{tabular}
    \caption{The normalized GW amplitudes $\abs{\Omega^{\text{1c-Z2}}_{++}/G_\NL^2 A_g^4}$ (left) and $\abs{\Omega^{\text{1c-Z2}}_{+\times}/G_\NL^2 A_g^4}$ (right) in the similar style to Fig.~\ref{fig: 1c-C1}}
    \label{fig: 1c-Z2}
\end{figure}

Including the deformation factor $2^2$ for the $\text{1c-Z2}$ term, the density parameter of the induced \acp{GW} can be obtained as follows.
\bae{
    \Omega^{\text{1c-Z2}}_{\lambda\lambda'}(k) &=\lim_{\tau\to\infty}\frac{2^2}{48}(k\tau)^2\overline{\mathcal{P}^{\text{1c-Z2}}_{\lambda\lambda'}(\tau,k)} \nonumber \\
    &\bmbe{=\frac{3G_\NL^2 A_g^4}{64\pi^2}\frac{k^3}{k_{*}^2}\int \dd{\cos{\theta_1}} \dd{\varphi} \dd{\cos{\theta_2}} Q^2_{\lambda\lambda'}(\bfk,\bfq_1,\bfq_2) \\
    \times \eval{\overline{J_k^2(\abs{\bfk-\bfq_1},q_1;\abs{\bfk-\bfq_2},q_2)}\frac{\Theta(2k_{*}-\abs{\bfq_1-\bfk+\bfq_2})}{\abs{\bfq_1-\bfk+\bfq_2}}}_{q_1=q_2=k_*}.}
}
The numerical resultant power spectrum is shown in Fig.~\ref{fig: 1c-Z2}.

\subsection{CZ term}

The CZ term reads
\bme{
    P^\CZ_{\lambda\lambda'}(k)=(2!F_\NL)^2 3!G_\NL \int\frac{\dd[3]{q_1}}{(2\pi)^3}\frac{\dd[3]{q_2}}{(2\pi)^3}\frac{\dd[3]{q_3}}{(2\pi)^3}Q_\lambda(\bfk,\bfq_1+\bfq_2+\bfq_3)Q_{\lambda'}(\bfk,\bfq_2) \\ \times I_k(\abs{\bfk-\bfq_1-\bfq_2-\bfq_3},\abs{\bfq_1+\bfq_2+\bfq_3},\tau)I_k(\abs{\bfk-\bfq_2},q_2,\tau) \\ 
    \times P_g(q_1)P_g(q_2)P_g(q_3)P_g(\abs{\bfk-\bfq_1-\bfq_2}).
}
Contrary to the 1c-C and 1c-Z terms, it does not have a convolved propagator and hence cannot be simplified easily. We will change integral variables several times to make the most of the momentum constraints.

First of all, the ordinary spherical coordinate takes the $z$ direction along the $\bfk$ direction, but in our case, the last momentum constraint $\abs{\bfk-\bfq_1-\bfq_2}=k_*$ can be more easily treated by defining the $z$ direction along the $\bfq\coloneqq\bfq_1+\bfq_2$ direction.
The integral variables are changed to $(q,\theta_k,\phi_k,q_2,\theta_2,\phi_2,q_3,\theta_3,\phi_3)$ where the relevant vectors are defined by
\bege{
    \bfk=k\pmqty{\sin\theta_k\cos\phi_k \\ \sin\theta_k\sin\phi_k \\ \cos\theta_k} \qc
    \bfq_2=q_2\pmqty{\sin\theta_2\cos\phi_2 \\ \sin\theta_2\sin\phi_2 \\ \cos\theta_2} \qc
    \bfq_3=q_3\pmqty{\sin\theta_3\cos\phi_3 \\ \sin\theta_3\sin\phi_3 \\ \cos\theta_3}, \\
    \bfq=\pmqty{0 \\ 0 \\ q} \qc
    \bfq_1=\bfq-\bfq_2=\pmqty{-q_2\sin\theta_2\cos\phi_2 \\ -q_2\sin\theta_2\sin\phi_2 \\ q-q_2\cos\theta_2}.
}
The Dirac deltas from the power spectra can be recast as
\bme{\label{eq: CZ deltas}
    \delta(\ln q_1-\ln k_*)\delta(\ln q_2-\ln k_*)\delta(\ln q_3-\ln k_*)\delta(\ln\abs{\bfk-\bfq}-\ln k_*) \\
    =\frac{k_*^3}{kq^2}\delta(\cos\theta_2-\bar{\mu}_{2*})\delta(\ln q_2-\ln k_*)\delta(\ln q_3-\ln k_*)\delta(\cos\theta_k-\mu_{k*}),
}
where
\bae{
    \bar{\mu}_{2*}=\frac{q}{2k_*} \qc \mu_{k*}=\frac{q^2+k^2-k_*^2}{2qk}.
}

To calculate the polarization part, we consider the rotation back of vectors to the original coordinate where $\bfk$ is in the $z$ direction.
The current coordinate is rotated back to the original one by the rotation by $-\phi_k$ around the $z$ axis followed by the rotation by $-\theta_k$ around the $y$ axis followed by the rotation back by $\phi_k$ around the $z$ axis.\footnote{The last rotation by $\phi_k$ around the $z$ axis is necessary for the Jacobian to be $q^2q_2^2q_3^2$.} Any vector $\bfp^{(q)}$ is transformed to $\bfp^{(k)}$ by these rotations as
\bae{
    \bfp^{(q)}\to\bfp^{(k)}&=R_{\phi_k}R_{\theta_k}^TR_{\phi_k}^T\bfp^{(q)} \nonumber \\ 
    &=
    \pmqty{\cos\phi_k & -\sin\phi_k & 0 \\ 
    \sin\phi_k & \cos\phi_k & 0 \\
    0 & 0 & 1}
    \pmqty{\cos\theta_k & 0 & -\sin\theta_k \\
    0 & 1 & 0 \\
    \sin\theta_k & 0 & \cos\theta_k}
    \pmqty{\cos\phi_k & \sin\phi_k & 0 \\
    -\sin\phi_k & \cos\phi_k & 0 \\
    0 & 0 & 1}\bfp^{(q)}.
}
The polarization tensors are given in the original coordinate by
\bae{
    \epsilon^{(k)\lambda}_{ij}(\bfk)=\bce{
        \frac{1}{\sqrt{2}}
        \spmqty{1 & 0 & 0 \\
        0 & -1 & 0 \\
        0 & 0 & 0} & \text{for $\lambda=+$}, \\[5pt]
        \frac{1}{\sqrt{2}}
        \spmqty{0 & 1 & 0 \\
        1 & 0 & 0 \\
        0 & 0 & 0} & \text{for $\lambda=\times$},
    }
}
and hence the projection factor $Q_\lambda(\bfk,\bfp)$ is written in terms of the $q$ coordinate expression $\bfp^{(q)}$ as
\bae{
    Q_\lambda(\bfk,\bfp)=\bfp^{(q)T}R_{\phi_k}R_{\theta_k}R_{\phi_k}^T\bm{\epsilon}^{(k)\lambda}R_{\phi_k}R_{\theta_k}^TR_{\phi_k}^T\bfp^{(q)}.
}

Now all the relevant quantities are expressed in the new integral variables. With use of the Dirac deltas~\eqref{eq: CZ deltas}, the power spectrum reduces to
\bme{
    \calP_{\lambda\lambda^\prime}^\CZ(\tau,k)=\frac{3F_\NL^2 G_\NL A_g^4}{8\pi^3}\frac{k^2}{k_*^2}\int^{\min(2,1+\tilde{k})}_{\abs{1-\tilde{k}}}\dd{\tilde{q}}\int\dd{\phi_k}\dd{\phi_2}\dd{\cos\theta_3}\dd{\phi_3}Q_\lambda(\bfk,\bfq+\bfq_3) \\ 
    \times \eval{Q_{\lambda^\prime}(\bfk,\bfq_2)I_k(\abs{\bfk-\bfq-\bfq_3},\abs{\bfq+\bfq_3},\tau)I_k(\abs{\bfk-\bfq_2},q_2,\tau)}_{q_2=q_3=k_*,\,\cos\theta_2=\bar{\mu}_{2*},\,\cos\theta_k=\mu_{k*}}.
}
Note that $\tilde{q}=q/k_*$, $\tilde{k}=k/k_*$ and the integration region for $\tilde{q}$ come from the triangle condition on $\bfq=\bfq_1+\bfq_2$ with $q_1=q_2=\abs{\bfk-\bfq}=k_*$.
Regarding the kernel part, it should be noticed that the norms $\abs{\bfk-\bfq-\bfq_3}$ and $\abs{\bfk-\bfq_2}$ depend on $\phi_k$ only through $\tilde{\phi}_3\coloneqq\phi_3-\phi_k$ and $\tilde{\phi}_2\coloneqq\phi_2-\phi_k$ respectively as can be seen in the explicit expression
\beae{
    (\bfk-\bfq-\bfq_3)^2&=k^2+q^2+q_3^2+2q_3(q-k\cos\theta_k)\cos\theta_3  
    -2k(q\cos\theta_k+q_3\sin\theta_k\sin\theta_3\cos\tilde{\phi}_3), \\
    (\bfk-\bfq_2)^2&=k^2+q_2^2-2kq_2\pqty{\cos\theta_k\cos\theta_2+\sin\theta_k\sin\theta_2\cos\tilde{\phi}_2}.
}
Therefore, by changing the integration variables as $\phi_2\to\tilde{\phi}_2$ and $\phi_3\to\tilde{\phi}_3$, the $\phi_k$ integration appears only in the projection part:
\bae{
    \bar{Q}_{\lambda\lambda^\prime}^2(\bfk,\bfq+\bfq_3,\bfq_2)\coloneqq\int\dd{\phi_k}Q_\lambda(\bfk,\bfq+\bfq_3)Q_{\lambda^\prime}(\bfk,\bfq_2),
}
which can be solved as
\bae{
    &\bar{Q}_{++}^2=\bar{Q}_{\times\times}^2 \nonumber \\
    &=\bmte{\frac{\pi q_2^2}{16}\left[\left(2 \sin^2\theta_2 \cos^2\theta_k \cos^2\tilde{\phi}_2-\sin2\theta_2 \sin2\theta_k \cos\tilde{\phi}_2+2 \cos^2\theta_2 \sin^2\theta_k-2 \sin^2\theta_2 \sin^2\tilde{\phi}_2\right) \right. \\
    \times\left(4 q^2 \sin^2\theta_k+q_3 \left(-4(q+q_3 \cos\theta_3) \sin\theta_3 \sin2\theta_k \cos\tilde{\phi}_3 \right.\right. \\
    \left.\left.+(8 q \cos\theta_3+3 q_3 \cos2\theta_3+q_3)\sin^2\theta_k +q_3(\cos2\theta_k+3) \sin^2\theta_3 \cos2\tilde{\phi}_3\right)\right) \\
    +8q_3 \sin\theta_3 \left(\sin2\theta_2 \sin\theta_k \sin\tilde{\phi}_2-\sin^2\theta_2\cos\theta_k \sin2\tilde{\phi}_2\right) \\
    \left.\times\left(2(q+q_3 \cos\theta_3) \sin\theta_k \sin\tilde{\phi}_3 -q_3 \sin\theta_3 \cos\theta_k \sin2\tilde{\phi}_3\right)\right],}
}
and
\bae{
    &\bar{Q}_{+\times}^2=-\bar{Q}_{\times+}^2 \nonumber \\
    &=\bmte{\frac{\pi q_2^2}{32}\left[\left(4 \sin ^2\theta_2 \cos\theta_k \sin2\tilde{\phi}_2-4 \sin2\theta_2 \sin\theta_k \sin\tilde{\phi}_2\right) \right. \\
    \times\left(4 q^2 \sin^2\theta_k+q_3 \left(-4(q+q_3 \cos\theta_3) \sin\theta_3\sin2\theta_k \cos\tilde{\phi}_3 \right.\right. \\
    \left.\left.+(8 q \cos\theta_3+3 q_3 \cos2\theta_3+q_3)\sin^2\theta_k+q_3 (\cos2\theta_k+3)\sin^2\theta_3 \cos2\tilde{\phi}_3\right)\right) \\
    +4 q_3 \sin\theta_3 \left((\cos2\theta_k+3)\sin^2\theta_2\cos2\tilde{\phi}_2-2 \sin2\theta_2 \sin2\theta_k \cos\tilde{\phi}_2+(3 \cos2\theta_2+1) \sin^2\theta_k\right) \\ 
    \left.\times\left(2 (q+q_3 \cos\theta_3)\sin\theta_k \sin\tilde{\phi}_3-q_3 \sin\theta_3 \cos\theta_k \sin2\tilde{\phi}_3\right)\right].}
}

\begin{figure}
    \centering
    \begin{tabular}{c}
        \begin{minipage}{0.5\hsize}
            \centering
            \includegraphics[width=0.95\hsize]{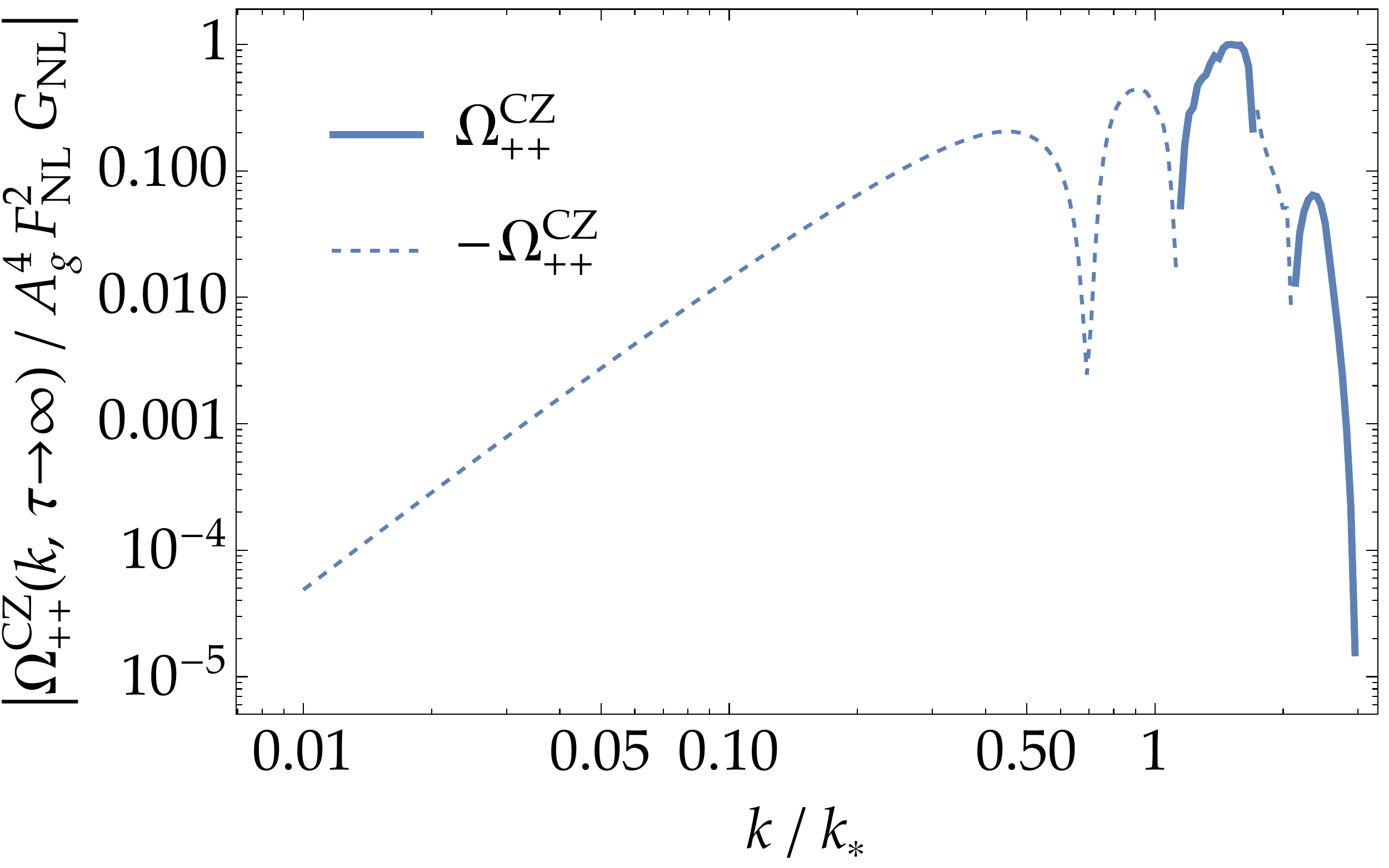}
        \end{minipage}
        \begin{minipage}{0.5\hsize}
            \centering
            \includegraphics[width=0.95\hsize]{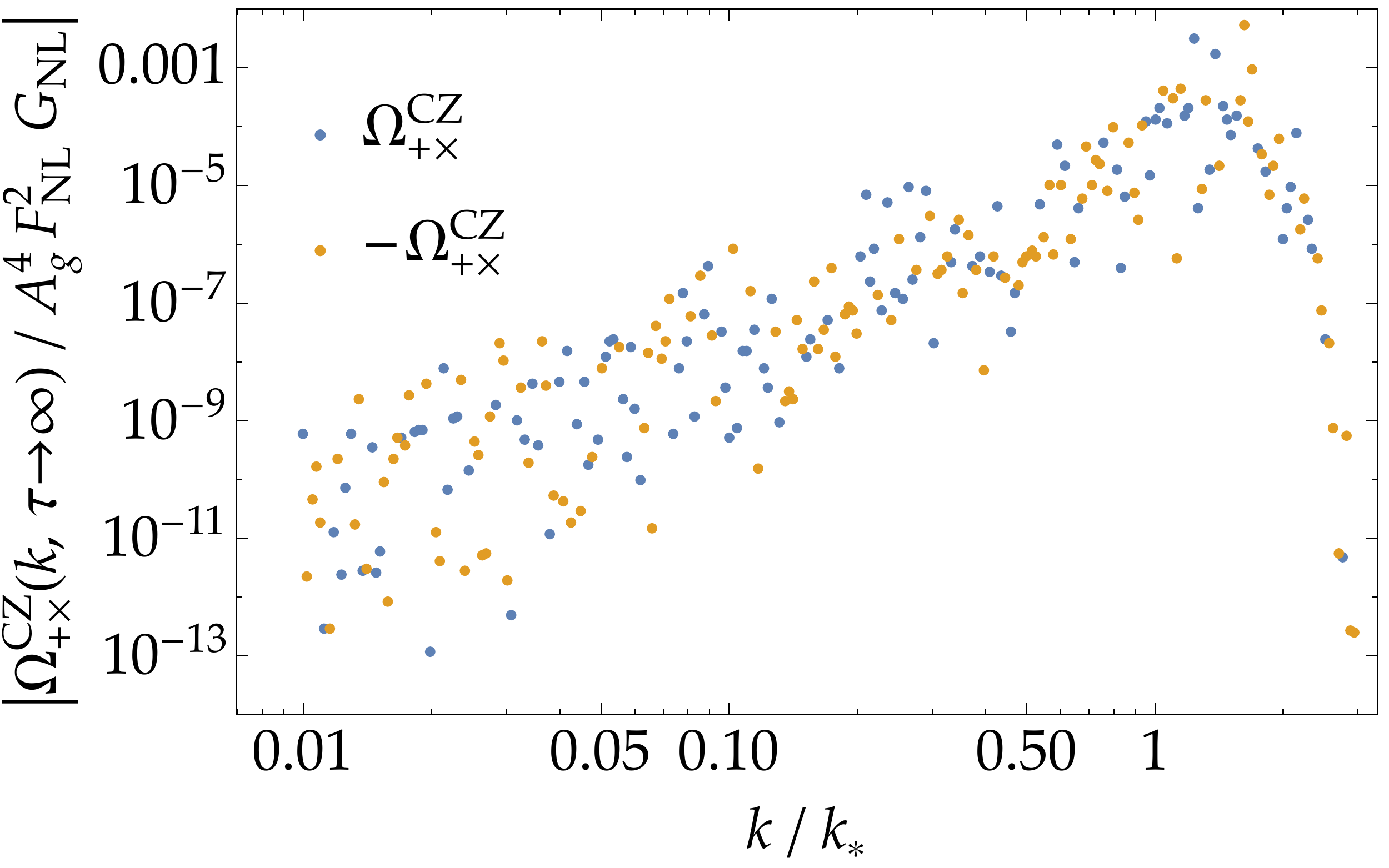}
        \end{minipage}
    \end{tabular}
    \caption{The normalized GW amplitudes $\abs{\Omega^{\text{CZ}}_{++}/F_\NL^2G_\NL A_g^4}$ (left) and $\abs{\Omega_{+\times}^\CZ}/F_\NL^2G_\NL A_g^4$ (right) in the similar style to Fig.~\ref{fig: 1c-C1}. The wavy feature around the maximum of the left panel is due to numerical errors. }
    \label{fig: CZ}
\end{figure}

Including the deformation factor $2^3$, the \ac{GW} density parameter is given by
\bae{
    \Omega_{\lambda\lambda^\prime}^\CZ(k)&=\lim_{\tau\to\infty}\frac{2^3}{48}(k\tau)^2\overline{\calP_{\lambda\lambda^\prime}^\CZ(\tau,k)} \nonumber \\
    &\bmbe{=\frac{F_\NL^2G_\NL A_g^4}{16\pi^3}\frac{k^2}{k_*^2}\int^{\min(2,1+\tilde{k})}_\abs{1-\tilde{k}}\dd{\tilde{q}}\int\dd{\phi_2}\dd{\cos\theta_3}\dd{\phi_3}\bar{Q}_{\lambda\lambda^\prime}^2(\bfk,\bfq+\bfq_3,\bfq_2) \\
    \times \eval{\overline{J_k^2(\abs{\bfk-\bfq-\bfq_3},\abs{\bfq+\bfq_3};\abs{\bfk-\bfq_2},q_2)}}_{q_2=q_3=k_*,\,\cos\theta_2=\bar{\mu}_{2*},\,\cos\theta_k=\mu_{k*}}.}
}
Its numerical result is shown in Fig.~\ref{fig: CZ}.

\bibliography{main}
\bibliographystyle{JHEP}
\end{document}